\pdfoutput=1
\documentclass[useAMS,usenatbib,usegraphicx]{mn2e}
\usepackage[english]{babel}
\usepackage{graphicx,amsmath,amssymb,multirow}
\usepackage{xspace}
\usepackage{color}
 \usepackage{url}
\usepackage[framemethod=tikz]{mdframed}
\usepackage[normalem]{ulem} 
\pdfpagewidth=\paperwidth
\pdfpageheight=\paperheight

\newcommand{\eg}{e.g.\xspace}
\newcommand{\sn}{SN\xspace}
\newcommand{\sne}{SNe\xspace}
\newcommand{\snia}{SN~Ia\xspace}
\newcommand{\sneia}{SNe~Ia\xspace}
\newcommand{\snname}[1]{SN\,#1\xspace}
\newcommand{\etal}{et~al.\xspace}
\newcommand{\ie}{i.e.\xspace}

\newcommand{\bl}{2012bl\xspace}
\newcommand{\bm}{2012bm\xspace}
\newcommand{\cg}{2012cg\xspace}
\newcommand{\cp}{2012cp\xspace}
\newcommand{\cu}{2012cu\xspace}
\newcommand{\et}{2012et\xspace}
\newcommand{\jj}{2014J\xspace}

\newcommand{\by}{2011by\xspace}

\newcommand{\fe}{2011fe\xspace}

\newcommand{\snbl}{\snname{2012bl}}
\newcommand{\snbm}{\snname{2012bm}}
\newcommand{\sncg}{\snname{2012cg}}
\newcommand{\sncp}{\snname{2012cp}}
\newcommand{\sncu}{\snname{2012cu}}
\newcommand{\snet}{\snname{2012et}}
\newcommand{\snjj}{\snname{2014J}}

\newcommand{\snby}{\snname{2011by}}

\newcommand{\snfe}{\snname{2011fe}}

\newcommand{\uvwone}{{\it uvw1}\xspace}
\newcommand{\uvwtwo}{{\it uvw2}\xspace}
\newcommand{\uvmtwo}{{\it uvm2}\xspace}

\newcommand{\wfcone}{F225W\xspace}
\newcommand{\wfctwo}{F275W\xspace}
\newcommand{\wfcthree}{F336W\xspace}
\newcommand{\wfcfour}{F125W\xspace}
\newcommand{\wfcfive}{F160W\xspace}
\newcommand{\uband}{{\it u}\xspace}
\newcommand{\Uband}{{\it U}\xspace}
\newcommand{\Bband}{{\it B}\xspace}
\newcommand{\Vband}{{\it V}\xspace}
\newcommand{\vband}{{\it v}\xspace}
\newcommand{\Rband}{{\it R}\xspace}
\newcommand{\iband}{{\it i}\xspace}
\newcommand{\Iband}{{\it I}\xspace}
\newcommand{\Jband}{{\it J}\xspace}
\newcommand{\Hband}{{\it H}\xspace}
\newcommand{\Kband}{{\it K}\xspace}
\newcommand{\Ksband}{{\it Ks}\xspace}
\newcommand{\Xband}{{\it X}\xspace}
\newcommand{\EXY}[2]{E(\textrm{#1}-\textrm{#2})}
\newcommand{\EBV}{\EXY{\Bband}{\Vband}}
\newcommand{\EXV}{E(\mathrm{\Xband}-\mathrm{\Vband})}
\newcommand{\tBmax}{t_\mathrm{\Bband}^\mathrm{max}}
\newcommand{\sB}{s_\mathrm{\Bband}}
\newcommand{\sBV}{s_\mathrm{\Bband\Vband}}
\newcommand{\mX}{m_\mathrm{\Xband}}
\newcommand{\KX}{K_\mathrm{\Xband}}
\newcommand{\mV}{\mathcal{M}_\mathrm{\Vband}}
\newcommand{\KV}{K_\mathrm{\Vband}}

\newcommand{\halpha}{H$\alpha$\xspace}
\newcommand{\NaID}{Na~\textsc{i}~D\xspace}
\newcommand{\Done}{Na~\textsc{i}~D1\xspace}
\newcommand{\Dtwo}{Na~\textsc{i}~D2\xspace}
\newcommand{\NaI}{Na~\textsc{i}\xspace}
\newcommand{\CaHK}{Ca~\textsc{ii}~H\&K\xspace}

\newcommand{\CII}{C~\textsc{ii}\xspace}
\newcommand{\SiII}{Si~\textsc{ii}\xspace}
\newcommand{\MgII}{Mg~\textsc{ii}\xspace}
\newcommand{\KI}{K~\textsc{i}\xspace}

\newcommand{\snoopy}{\texttt{SNooPy}\xspace}
\newcommand{\ccm}{CCM\xspace}
\newcommand{\ccmo}{CCM+O\xspace}
\newcommand{\ftz}{F99\xspace}
\newcommand{\RV}{R_\mathrm{\Vband}}
\newcommand{\AV}{A_\mathrm{\Vband}}
\newcommand{\AX}{A_\mathrm{\Xband}}

\newcommand{\AMW}{A^{\mathrm{MW}}}

\newcommand{\calspec}{\textsc{calspec}\xspace}
\newcommand{\synphot}{\textsc{synphot}\xspace}
\newcommand{\ctereverse}{\texttt{wfc3uv\_ctereverse}\xspace}

\newcommand{\jjpaper}{A14\xspace}

\newcommand{\swift}{{\em Swift}\xspace}
\newcommand{\uvot}{{\em UVOT}\xspace}
\newcommand{\hst}{{\em HST}\xspace}
\newcommand{\wfc}{{\em WFC}\xspace}
\newcommand{\hstwfc}{{\em HST/WFC3}\xspace}
\newcommand{\wfcuvis}{{\em WFC3/UVIS}\xspace}
\newcommand{\wfcir}{{\em WFC3/IR}\xspace}
\newcommand{\stis}{{\em STIS}\xspace}
\newcommand{\hststis}{{\em HST/STIS}\xspace}
\newcommand{\galex}{{\em GALEX}\xspace}

\newcommand{\RVjj}{1.4\pm0.1}
\newcommand{\RVcu}{2.8\pm0.1}

\title[Diversity in extinction laws for \sneia]{Diversity in extinction laws of Type Ia supernovae measured 
  between $0.2$ and $2~\mu\mathrm{m}$}
\author[R.~Amanullah \etal]{%
  R.~Amanullah,$^1$\thanks{rahman@fysik.su.se}
  J.~Johansson,$^1$
  A.~Goobar,$^1$
  R.~Ferretti,$^1$
  S.~Papadogiannakis,$^1$
  \newauthor
  T.~Petrushevska,$^1$,
  P.~J.~Brown,$^2$
  Y.~Cao,$^3$
  C.~Contreras,$^4$
  H.~Dahle,$^5$
  \newauthor
  N.~Elias-Rosa,$^6$
  J.~P.~U.~Fynbo,$^7$
  J.~Gorosabel,$^{8,9}$
  L.~Guaita,$^{10}$
  L.~Hangard,$^1$
  \newauthor
  D.~A.~Howell$^{11,12}$,
  E.~Y.~Hsiao,$^{13,4}$
  E.~Kankare,$^{14}$
  M.~Kasliwal,$^{15}$
  G.~Leloudas,$^{16,7}$
  \newauthor
  P.~Lundqvist,$^{17}$
  S.~Mattila,$^{18}$
  P.~Nugent,$^{19,20}$
  M.~M.~Phillips,$^3$
  A.~Sandberg,$^{17}$
  \newauthor
  V.~Stanishev,$^{21}$
  M.~Sullivan,$^{22}$
  F.~Taddia,$^{17}$
  G.~\"Ostlin$^{17}$
  S.~Asadi,$^{17}$\thanks{Participant of the Nordic Millimetre and Optical/NIR Astronomy Summer School 2012}
  \newauthor
  R.~Herrero-Illana,$^{8}\dagger$
  J.~J.~Jensen,$^{7}\dagger$
  K.~Karhunen,$^{23}\dagger$
  S.~Lazarevic,$^{24}\dagger$
  \newauthor
  E.~Varenius,$^{25}\dagger$
  P.~Santos,$^{5}\dagger$
  S.~Seethapuram Sridhar,$^{26,27}\dagger$
  S.~H.~J.~Wallstr\"om,$^{25}\dagger$
  \newauthor
  J.~Wiegert,$^{25}\dagger$\\
  \noindent Affiliations are listed at the end of the paper.
  }
\pagerange{\pageref{firstpage}--\pageref{lastpage}} \pubyear{2015}

\begin{document}
\label{firstpage}

\maketitle

\begin{abstract}
  We present ultraviolet (UV) observations of six nearby Type~Ia
  supernovae (\sneia) obtained with the {\em Hubble Space Telescope}, three
  of which were also observed in the near-IR (NIR) with Wide-Field Camera~3. 
  UV observations with the \swift satellite, as well as ground-based optical and
  near-infrared data provide complementary information. The combined 
  data-set covers the wavelength range $0.2$--$2~\mu$m. By also including 
  archival data of \snjj, we analyse a sample spanning observed colour excesses 
  up to $\EBV=1.4~$mag.  We study the wavelength dependent extinction of
  each individual \sn and find a diversity of reddening laws when
  characterised by the total-to-selective extinction $\RV$.  In
  particular, we note that for the two \sne with $\EBV\gtrsim1~$mag, for
  which the colour excess is dominated by dust extinction, we find
  $\RV=1.4\pm0.1$ and $\RV=2.8\pm0.1$.  Adding UV photometry reduces
  the uncertainty of fitted $\RV$ by $\sim50\,$\% allowing us to
  also measure $\RV$ of individual low-extinction objects which point to
  a similar diversity, currently not accounted for in the analyses when
  \sneia are used for studying the expansion history of the universe.

\end{abstract}



\section{Introduction}\label{sec:intro}
Studies of the cosmological expansion history using Type~Ia supernovae (\sneia) have greatly improved our understanding of the Universe. While pioneering work lead to the discovery of the accelerated expansion from a few dozen \sneia 
\citep{1998AJ....116.1009R,1999ApJ...517..565P} the present samples of several hundred \sneia out to $z\sim1.5$ 
 show to high-precision that Einstein's Cosmological Constant provides an excellent fit for this phenomenon
\citep[\eg][]{2010ApJ...716..712A,2011ApJS..192....1C,2011ApJ...737..102S,2012ApJ...746...85S,2014A&A...568A..22B}.

These findings have been possible due to the great accuracy of \sneia as distance indicators when observed 
in two or more broadband filters over several weeks, starting before light curve peak
\citep[see][for a recent review]{2011ARNPS..61..251G}.
The distance to each individual \sn is obtained by fitting a \snia template to the photometric data 
\citep[see \eg][and references therein]{%
1996ApJ...473...88R,
2002PASP..114..803N,
2007ApJ...659..122J,
2007A&A...466...11G,
2008ApJ...681..482C,2011AJ....141...19B}
in order to obtain the {\em brightness} and {\em colour} at maximum together with the {\em light curve-shape}.
The fitted properties are then typically combined using empirical linear relations to form a 
distance-dependent quantity \citep{1993ApJ...413L.105P,1998A&A...331..815T}.

The colour correction is based on the observation that red \sne are fainter than bluer ones, which could 
originate from a combination of an intrinsic \snia colour law, and extinction by dust in the host galaxy.  The
colour-brightness relation is commonly determined by minimising the scatter around the best fit cosmological 
model, but there is no a priori reason why the two effects should follow the same colour-luminosity relation, nor 
can it be expected that host galaxy dust should have the same properties in different \sn environments.  In fact, 
even in the Milky Way, a range of extinction properties have been observed for different lines-of-sight.

When \sneia are used to measure distances, these uncertainties are handled by introducing a systematic error on 
the colour-luminosity relation, which makes an important contribution to the error budget when propagated to 
the derived cosmological parameters.  Addressing the colour-brightness relation, and in particular breaking the 
degeneracy between the intrinsic colour component and dust extinction, is important for future \snia surveys to improve
beyond the current cosmological constraints.

The properties and wavelength-dependent extinction of dust in the Milky Way has been carefully studied and is
commonly characterised by the total, $\AV$, to selective,  $\EBV=A_\mathrm{\Bband}-\AV$, extinction ratio as 
$\RV=\AV/(A_\mathrm{\Bband} - \AV)$.  Lower values of $\RV$ correspond to \emph{steeper} extinction laws, since, 
for a given total extinction, $\AV$, these imply a larger reddening, $\EBV$.

The reddening of \sneia can be studied by comparing observed colours between reddened and similar 
un-reddened objects.   Several such studies of individual \sneia  
\citep[\eg][from hereon \jjpaper]{2006MNRAS.369.1880E,2008MNRAS.384..107E,2007AJ....133...58K,2010AJ....139..120F,2014ApJ...788L..21A} in nearby galaxies, point to significantly lower values of $\RV$ than has been observed in 
the Milky Way.  Since the reddening of several of these \sneia is significantly higher than what can be expected from the
Milky Way extinction along the line of sight, the observed colour excesses are likely dominated by extinction in the \sn host 
galaxies.  For low reddening 
\sne this may not be the case, but it is by nature difficult to measure $\RV$ to high precision for individual \sne with 
low $\EBV$.  However, the global reddening law for a sample of \sne can be measured, and several studies have 
obtained low $\RV$ values \citep[\eg][from hereon B14]{2006A&A...447...31A,2008A&A...487...19N,2014ApJ...789...32B} 
using this approach. 

\citet{2011A&A...529L...4C} and \citet{2014ApJ...780...37S} noted that globally derived reddening laws will be biased
if the intrinsic \snia colour dispersions are not accounted for correctly, and  \citet{2011A&A...529L...4C} 
find $\RV=2.8\pm0.3$, consistent with the Milky Way average, for their sample of 76~\sneia of which 73 had 
$\EBV<0.3~$mag.  B14 also obtain $\RV\approx3$ when they constrain their analysis to only include low-extinction 
\sneia, while highly reddened objects appear to prefer lower values ($\RV\approx1.7$) which is in agreement 
with previous studies \citep{2010AJ....139..120F}. This suggests that a single, global, reddening law derived
from a set of \sneia could depend on the colour distribution of the sample.   


The degeneracy between intrinsic \sneia colour variations and dust extinction, can also be approached by 
studying the relation between \sn colours and spectroscopic properties.  
\citet{2009ApJ...699L.139W} found $\RV=1.6$ for \sneia with "high" photospheric velocities (HV,
$>11,800$~km/s) within 5~days of maximum, while they obtained $\RV=2.4$ for objects with "normal" 
velocity (NV, $<11,800$~km/s).  The expansion velocities were quantified by measuring the velocity of the 
\SiII $6355$~\AA\  absorption feature.   \citet{2011ApJ...729...55F} confirmed their findings but also conclude 
that when \sne with $\EBV>0.35$~mag are omitted, they obtain $\RV\approx2.5$ for the two separate 
subsamples, which also have different intrinsic colours.  In other words, these findings also suggest that low 
$\RV$ values are primarily associated with high-extinction, and these could dominate when a single global 
reddening law is derived from a sample of \sneia.

One natural explanation for low $\RV$ values arises if \sneia are surrounded by circumstellar (CS) dust  
\citep{2005ApJ...635L..33W,2008ApJ...686L.103G}.
In this scenario, photons can scatter back into the line-of-sight which reduces the total 
extinction.  Photons that scatter on the CS dust will also arrive later than photons that do not interact 
which will give rise to time-dependent reddening \citep{2005ApJ...635L..33W,2011ApJ...735...20A,2015ApJ...805...74B}.  
An observation will be a superposition of the \sn spectrum at the given epoch and \sn light from earlier epochs,
but the scattering cross-section is wavelength-dependent so the fraction of delayed photons will also vary with 
wavelength.  Further, depending on the geometry, scattered photons could give rise to a plateau for the late-time 
tail of the light curve and indications of such tails have been observed in 
$\mathrm{\Bband}-\mathrm{\Vband}$ \citep{2013ApJ...772...19F}.


CS dust could also be heated by the \sn, where the temperature will depend on the dust properties and the distance 
from the explosion.  
Heated dust has been detected in mid to far-infrared for a subset of peculiar \sneia 
\citep{2011ApJ...741....7F,2012A&A...545L...7T,2013ApJ...772L...6F,2013ApJS..207....3S}, but {\em Herschel} and {\em Spitzer} observations 
\citep{2013MNRAS.431L..43J,2014arXiv1411.3332J} of a handful of normal \sneia show no signs of heated CS dust 
\citep[see also][for limits from Near-IR observations]{2014arXiv1411.3778M}.  
For two of these \sne, 2006X and 2007le, together with \sne 1999cl and the \snname{2002ic}-like PTF11kx, 
time-varying \NaID absorption features have been observed in their spectra on time-scales comparable to the 
\sn life-time which is consistent with the presence for CS material 
\citep{2007Sci...317..924P,2009ApJ...693..207B,2009ApJ...702.1157S,2012Sci...337..942D}.
Absorption from \NaI has been used as a proxy for dust extinction \citep[see \eg][]{2012MNRAS.426.1465P},
although high-resolution spectra are required \citep{2011MNRAS.415L..81P} and its validity for extinction along
lines-of-sight of \sneia has been challenged \citep{2013ApJ...779...38P}.
Time-varying \NaID can be explained by photo-ionisation of neutral sodium \citep[\eg][]{2009ApJ...699L..64B} and subsequent recombination 
by the \sn. In a recent study by \citet{2014MNRAS.444L..73S} this explanation has been 
questioned and an alternative model of photon-induced desorption of sodium from dust in planetary nebulae 
remnants was proposed.  It has also been suggested that time-variations may originate from patchy interstellar 
material along the line-of-sight \citep{2010A&A...514A..78P}. More recently, \citet{2015ApJ...801..136G} claimed 
a detection of a time-varying \KI absorption feature in \jj, while the corresponding \NaID feature remained unchanged,
which they argue is consistent with the presence of CS material at a radius of $r_\mathrm{dust}\sim10^{19}$~cm due to
the different ionisation cross-sections.

An extensive study with multi-epoch high-resolution spectra of 14 \sne \citep{2014MNRAS.443.1849S} have failed to 
reveal further examples of time varying \NaID. Nevertheless, a statistically significant preponderance of \NaID features 
which are blue shifted with respect to the local velocity have been observed in high- and mid-resolution spectra
\citep[][respectively]{2011Sci...333..856S,2013MNRAS.436..222M}, suggesting that there is outflowing material from 
the \sn progenitor system.


In this work we seek to expand our knowledge of extragalactic extinction in the line-of-sight and \snia reddening by
making the widest wavelength study of a sample of \sneia to date.  We present UV-to-NIR light curves for six \sneia based
on data mainly obtained from The Nordic Optical Telescope (NOT) and {\em The Hubble Space Telescope} (\hst).  The observations 
and data reduction is presented in \S\,\ref{sec:data} while the light curves and colours are presented in \S\,\ref{sec:color}. The 
method for deriving individual reddening laws for each \sn is described in \S\,\ref{sec:lawfitting}, and after adding \snjj 
from \jjpaper to the sample we present the results in \S\,\ref{sec:results}.  We search for CS dust in 
\S\,\ref{sec:cs} and all the results are discussed in \S\,\ref{sec:discussion}. We conclude and summarise in
\S\,\ref{sec:conclusions}.

\begin{figure*}
{\centering
\includegraphics[width=\textwidth]{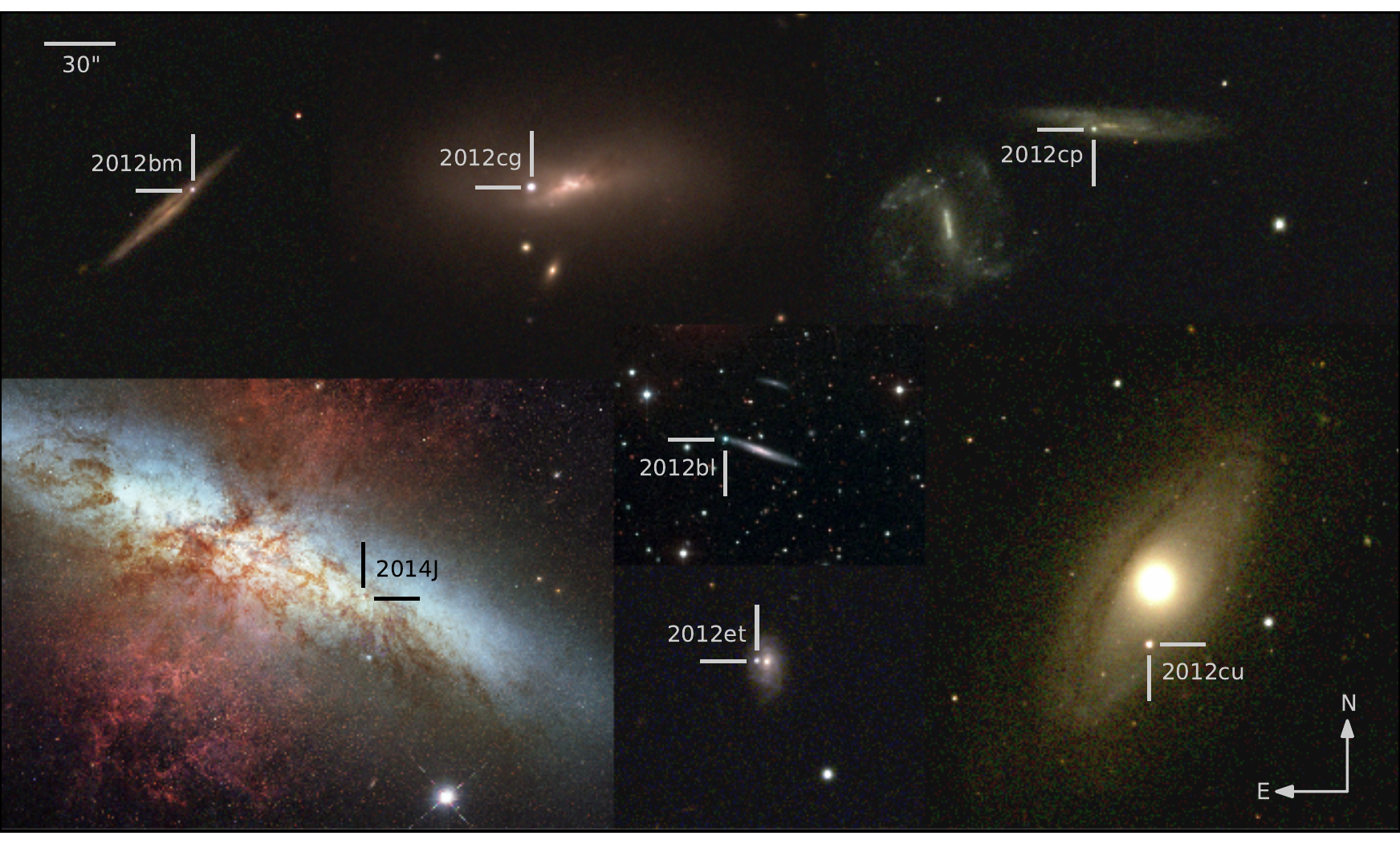}}
\caption{%
The seven \sne analysed in this work.  The patches for \sne \bm, \cg, \cp, \cu and \et were obtained from 
\Bband\Vband\Rband images from NOT, while the images for \sne \bl and \jj are obtained from the Swope telescope and \hstwfc (Program~DD-13621; PI~Goobar) images respectively.
\label{fig:snhosts}}
\end{figure*}

\begin{table*}
\begin{tabular}{l l l r l l l r r@{}l l}
\hline\hline
\multicolumn{1}{c}{\small \sn} & 
\multicolumn{1}{c}{\small $\alpha_\mathrm{\sn}$} & 
\multicolumn{1}{c}{\small $\delta_\mathrm{\sn}$} & 
\multicolumn{1}{c}{\small $r_\perp$} &
\multicolumn{1}{c}{\small Host galaxy} & 
\multicolumn{1}{c}{\small $\alpha_\mathrm{h}$} & 
\multicolumn{1}{c}{\small $\delta_\mathrm{h}$} &
\multicolumn{1}{c}{\small $v_\mathrm{h}$} & 
\multicolumn{2}{c}{\small Distance} &
\multicolumn{1}{c}{\small MW\ $A_V$}\\
& \multicolumn{1}{c}{\footnotesize{$(2000)$}} &
\multicolumn{1}{c}{\footnotesize{$(2000)$}} &
\multicolumn{1}{c}{\footnotesize{(kpc)}} &
& \multicolumn{1}{c}{\footnotesize{$(2000)$}} &
\multicolumn{1}{c}{\footnotesize{$(2000)$}} &
\multicolumn{1}{c}{\footnotesize{(km/s)}} &
\multicolumn{2}{c}{\footnotesize{(Mpc)}} &
\multicolumn{1}{c}{\footnotesize{(mag)}}\\
\hline
{\small\bl} & {\small 20:23:55.28} & {\small $-48$:21:17.3} & {\small $15.6(0.2)$} & 
	{\small ESO~234-19} & {\small 20:23:51.0} & {\small $-$48:21:32} & {\small 5608} & {\small $71$} &{\small $(1)^{a}$}& {\small 0.098}\\
{\small\bm} & {\small 13:05:45.66} & {\small $+$46:27:52.9} & {\small $7.5(0.3)$} & 
	{\small UGC~8189}      & {\small 13:05:46.6} & {\small $+$46:27:42} & {\small 7436} &  {\small $103$}&{\small $(8)^{b}$}& {\small 0.033}\\
{\small\cg} & {\small 12:27:12.83} & {\small $+$09:25:13.2} & {\small $1.3(0.2)$} & 
	{\small NGC~4424}      & {\small 12:27:11.6} & {\small $+$09:25:14} &   {\small 437} & {\small $15$}&{\small $(2)^{c}$}& {\small 0.057}\\
{\small\cp} & {\small 13:47:01.79} & {\small $+$33:53:35.0} & {\small $1.9(0.1)$} & 
	{\small UGC~8713}      & {\small 13:47:01.2} & {\small $+$33:53:37} & {\small 4956} & {\small $57$}&{\small $(3)^{d}$}& {\small 0.058}\\
{\small\cu} & {\small 12:53:29.35} & {\small $+$02:09:39.0} & {\small $5.4(1.2)$} & 
	{\small NGC~4772}      & {\small 12:53:29.1} & {\small $+$02:10:06} & {\small 1040} & {\small $41$}&{\small $(9)^{e}$}& {\small 0.074}\\
{\small\et} &  {\small 23:42:38.82} & {\small $+$27:05:31.5} & {\small $2.7(0.2)$} & 
	{\small MCG~+04-55-47} & {\small 23:42:38.4} & {\small $+$27:05:31} & {\small 7483} & {\small $105$}&{\small $(7)^{a}$}&{\small 0.221}\\
{\small\jj} & {\small 09:55:42.11} & {\small $+$69:40:25.9} & {\small $1.0(0.3)$} & 
	{\small M82} & {\small 09:55:52.7} & {\small $+$69:40:46} & {\small 203} & {\small 3.5}&{\small $(0.3)^{f}$} & {\small $0.189^f$}\\
\hline\hline
\multicolumn{11}{l}{%
$^a$\,\citet{2010Ap&SS.325..163P};\ $^b$\,Calculated based on the redshift with $H_0 = 73\pm5$~km/s/Mpc;\ $^c$\,\citet{2008ApJ...683...78C};}\\
\multicolumn{11}{l}{%
$^d$\,\citet{2009ApJS..182..474S};\ 
$^e$\,\citet{2009AJ....138..323T};\ 
$^f$\,\citet{2009ApJS..183...67D}
}
\end{tabular}
\caption{%
\sn coordinates are quoted from the discovery telegrams.  
All host galaxy data were obtained from the Nasa Extragalacic Database (NED), unless otherwise specified, 
where $v_\mathrm{h}$ is the measured recession velocity, 
and the Milky Way extinctions are from the \citet{2011ApJ...737..103S} calibration of the \citet{1998ApJ...500..525S} 
infrared-based dust maps.  The projected distances from the host galaxy nuclei, $r_\perp$, were calculated based on 
the host galaxy distances and the \sn offsets specified
in Appendix~\ref{sec:snsummary}.
\label{tb:snsummary}}
\end{table*}

All seven \sne\footnote{%
All data and figures presented in this work are available at \url{http://snova.fysik.su.se/dust/}.}
are listed in Table~\ref{tb:snsummary} and shown together with their host galaxies in Figure~\ref{fig:snhosts}.
They are also briefly summarised in Appendix~\ref{sec:snsummary}.  
 
\section{Observations and data reduction}\label{sec:data}
\subsection{Hubble Space Telescope}\label{sec:hstdata}
\begin{figure}
  \begin{center}
    \includegraphics[width=\columnwidth]{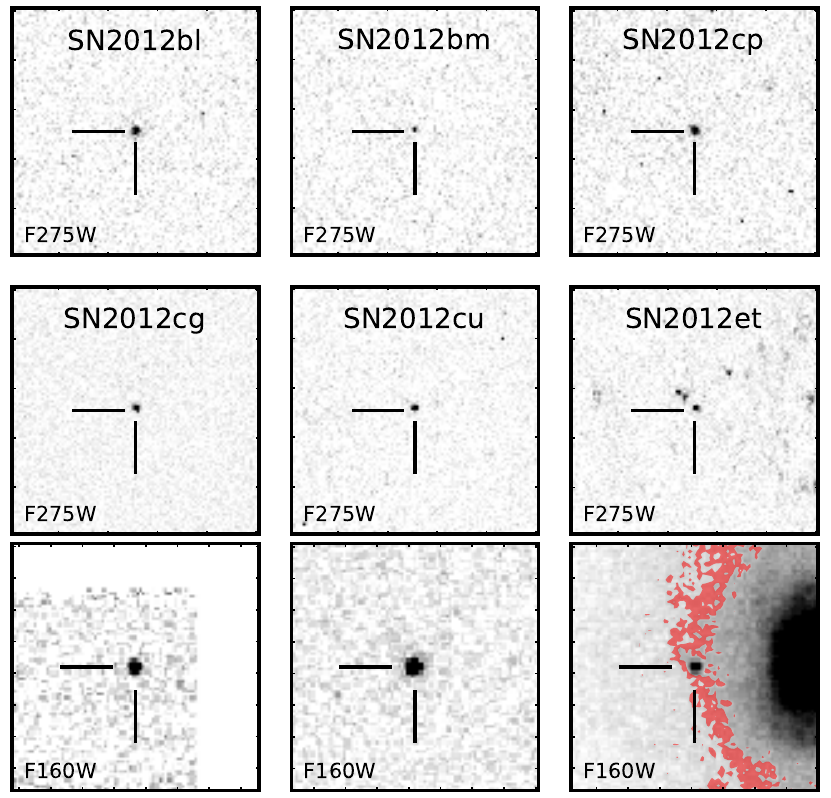}
    \caption{%
    \hstwfc patches of the six \sne observed during Cycle~19.  The patch sizes are $10''\times10''$ and in the
    upper row the first epochs obtained through the \wfctwo filter is shown for each of the three \sne \bl, \bm and \cp.  
    In the two lower rows, the corresponding patches are shown for the \sne \cg, \cu and \et, together with 
    the NIR observations in the \wfcfive filter that also were obtained for these objects.  
    The marked (red) region in the lower right panel show the isophot from which the host galaxy background 
    of \snet was estimated as explained in the text.\label{fig:wfc3patch}}
  \end{center}
\end{figure}

All previously unpublished \hst observations discussed in this work are listed in 
Table~\ref{tb:wfc3obs}. Each \sn was observed at two different epochs 
(four for \snet) with the Wide-Field Camera~3 (WFC3) during the \hst Cycle~19 under  
programme  GO-12582 (PI: Goobar).  All \sne were imaged with the \wfcuvis channel 
through the passbands \wfcone, \wfctwo and \wfcthree. In addition to this, NIR imaging in the 
\hst passbands \wfcfour and \wfcfive was obtained for the \sne \cg, \cu and \et with the \wfcir channel.  
Examples of both the \wfcuvis and \wfcir observations are shown in Figure~\ref{fig:wfc3patch}.

\wfcuvis consists of two $4096\times2051$ e2V CCD detectors with a plate scale of 0.04"/pixel, while 
the Teledyne HgCdTe infrared detector used in the \wfcir channel has a pixel scale of 0.13"/pixel.  For each 
observation, only the sub-arrays (apertures), listed in Table~\ref{tb:wfc3obs}, of the detectors were read-out.  
All WFC3 data were reduced using the standard STScI reduction pipeline and calibrated using CALWF3.  

\begin{table}
  \centering
  \begin{tabular}{c c l l}
    \hline\hline
    Civil date & MJD      & \multicolumn{1}{c}{Aperture}  & \multicolumn{1}{c}{\sn} \\
    \hline
    2012-04-09 & 56026.4 & UVIS1-2K2A-SUB & \bm \\
    2012-04-09 & 56026.5 & STIS 52X0.2 G230LB & \bm\\   
    2012-04-13 & 56030.6 & UVIS1-2K2A-SUB & \bm \\
    2012-04-13 & 56030.7 & STIS 52X0.2 G230LB & \bm\\   
    \hline
    2012-04-16 & 56033.1 & STIS 52X0.2 G230LB & \bl\\     
    2012-04-16 & 56038.2 & UVIS1-2K2A-SUB & \bl \\    
    2012-04-20 & 56037.4 & STIS 52X0.2 G230LB & \bl\\     
    2012-04-21 & 56038.2 & UVIS1-2K2A-SUB & \bl \\
    \hline
    2012-06-04 & 56082.4 & UVIS2-M1K1C-SUB & \cg \\
    2012-06-04 & 56082.4 & IRSUB64 & \cg \\
    2012-06-04 & 56082.5 & STIS 52X0.2 G230LB & \cg\\    
    2012-06-04 & 56082.5 & STIS 52X0.1 G430L & \cg\\       
    2012-06-18 & 56096.4 & STIS 52X0.2 G230LB & \cg\\     
    2012-06-18 & 56096.4 & STIS 52X0.1 G430L & \cg\\       
    2012-06-19 & 56097.5 & UVIS2-M1K1C-SUB & \cg \\
    2012-06-19 & 56097.5 & IRSUB64 & \cg \\
    \hline
    2012-06-04 & 56082.6 & UVIS1-2K2A-SUB & \cp \\
    2012-06-04 & 56082.7 & STIS 52X0.2 G230LB & \cp\\     
    2012-06-16 & 56094.6 & UVIS1-2K2A-SUB & \cp \\
    2012-06-17 & 56095.6 & STIS 52X0.2 G230LB & \cp\\     
    \hline
    2012-07-02 & 56110.8 & UVIS2-M1K1C-SUB & \cu \\
    2012-07-02 & 56110.8 & IRSUB512 & \cu \\ 
    2012-07-07 & 56115.8 & UVIS2-M1K1C-SUB & \cu \\
    2012-07-07 & 56115.8 & IRSUB512 & \cu \\
    \hline
    2012-10-01 & 56201.5 & UVIS2-M1K1C-SUB & \et \\
    2012-10-01 & 56201.5 & IR & \et \\
    2012-10-01 & 56201.8 & STIS 52X0.2 G230LB & \et\\      
    2012-10-01 & 56201.8 & STIS 52X0.2 G430L & \et\\        
    2012-10-05 & 56205.6 & UVIS2-M1K1C-SUB & \et \\
    2012-10-05 & 56205.6 & IR & \et \\
    2012-10-09 & 56209.8 & UVIS2-M1K1C-SUB & \et \\
    2012-10-09 & 56209.8 & IR & \et \\
    2012-10-13 & 56213.9 & UVIS2-M1K1C-SUB & \et \\
    2012-10-13 & 56213.9 & IR & \et \\
    \hline\hline
  \end{tabular}
  \caption{Wide-Field Camera-3 and STIS observations together with the used apertures, where UVIS1 and UVIS2 are 
    the two chips of the \wfcuvis channel.  The observations with the \wfcir and \wfcuvis channels were obtained through the
    \wfcfour, \wfcfive and \wfcone, \wfctwo, \wfcthree broadband passbands respectively. \label{tb:wfc3obs}}
\end{table}

The \wfcuvis CCDs, like all \hst CCDs, are plagued with charge transfer inefficiencies (CTI), i.e. 
degradations of the detector performance over time due to damage in the silicon lattice from cosmic rays.  
The CTI can be reverse corrected at the pixel level using 
\ctereverse\footnote{\url{http://www.stsci.edu/hst/wfc3/tools/cte_tools}} for all frames except for images 
obtained with the UVIS2-M1K1C-SUB sub-array.  For this aperture we instead used the recipe suggested by
\citet{2014wfcSTAN}.  The individual flat-fielded \wfcuvis and \wfcir images were then 
resampled and combined using \texttt{astrodrizzle} \citep{2010bdrz.conf..382F}.


Aperture photometry was carried out on the combined images using an aperture radius of $0.4"$, although in some
cases a smaller aperture of radius $0.16"$ had to be used for the \wfcuvis data.  This was the case for the low
signal-to-noise measurements of \snbm, observations of \sncg that were plagued by nearby coincidence cosmic rays, 
and all epochs of \snet, where a nearby object contaminated the larger
aperture (see the mid-right panel of Figure~\ref{fig:wfc3patch}). The smaller radius measurements had to be aperture 
corrected to the $0.4''$ radius for which STScI provides zeropoints\footnote{We used the zeropoints from 
March~6th, 2012.} for WFC3. Since there were no bright stars present in the fields we used the general enclosed 
energy tables provided by STScI for this purpose.  To account for the time and wavelength dependence of the 
\wfcuvis aperture corrections we adopt an additional uncertainty of 0.05~mag for all measurements using the 4~pixel 
radius, which was derived by comparing the enclosed energy tables with photometry from different aperture radii of 
the high signal-to-noise data. 

\begin{figure*}
	\centering
    \includegraphics[width=\textwidth]{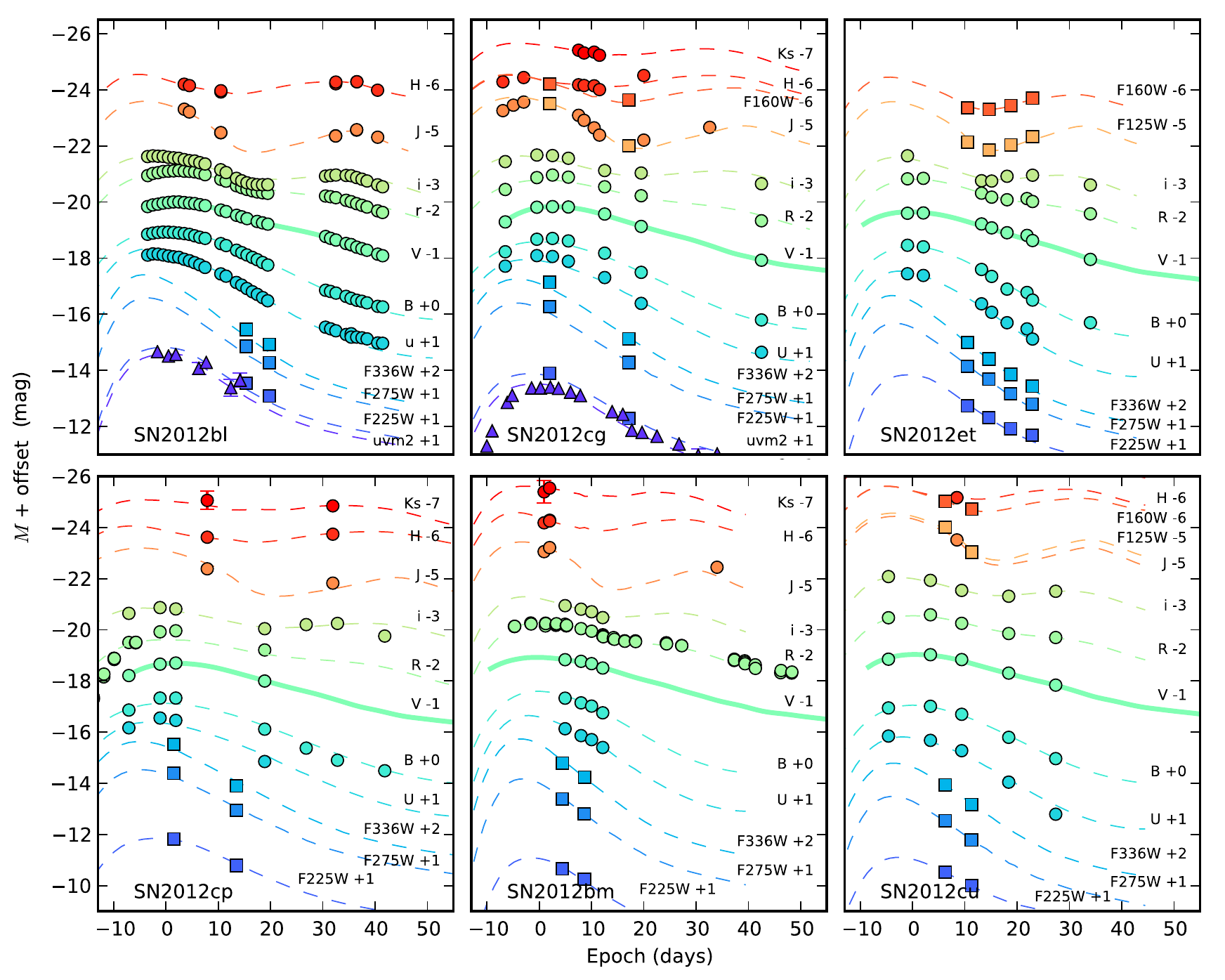}
    \caption{%
    		Observed light curves for the SN sample. Our analysis is based on optical %
    		\Uband\Bband\Vband\Rband\Iband and NIR \Jband\Hband\Kband photometry from ground based 
		observatories (cirles), UV (\wfcone,\wfctwo,\wfcthree) and NIR (\wfcfour,\wfcfive) photometry from the 
		\hst (squares). For the \sne \bl and \cg we also use \swift/\uvot photometry (triangles).
		The data are shown in their instrumental system but have been shifted using the distance moduli based on the
		distances in Table~\ref{tb:snsummary} and offset for clarity. 
		The thick, solid, lines show the \snoopy \Vband-models used to calculate colours as described in the text, while
		the dashed lines show synthesised light curves of \snfe to guide the eye.  When a \sn has observations in 
		a similar filter from multiple instruments only one of the light curves is shown. The epoch is the observed date with 
		respect to the fitted day of maximum in the \Bband-band from Table~\ref{tb:fitlawresults}.
     \label{fig:lcs}}
\end{figure*}

Only the \snet NIR measurements suffered from host galaxy contamination.  The host contribution was estimated 
by placing apertures of the same radius used for the \sn photometry along the isophot of the galaxy that intersects
with the position of \snet as illustrated in the lower right panel of Figure~\ref{fig:wfc3patch}.  
The background was then estimated as the median of these measurements, and the root-mean-square of the measurements 
was added in quadrature to the photometric uncertainty.
All the {\em WFC3} photometry is presented in Table~\ref{tb:photometry} and shown in Figure~\ref{fig:lcs}.  
Here we also present updated photometry of \snjj using the method described here which is consistent with the results 
presented in A14. 

\begin{table*}
\begin{tabular}{lrlrrrcrrrrr}
\hline\hline
\multicolumn{1}{c}{\small MJD} & \multicolumn{1}{c}{\small Phase} & & \multicolumn{1}{c}{\small\it X} & {\small $A^\mathrm{MW}_\mathrm{\it X}$} & {\small $K_\mathrm{\it X}$} & & \multicolumn{1}{c}{\small {\it V}} & {\small $A^\mathrm{MW}_\mathrm{\it V}$} & {\small $K_\mathrm{\it V}$} & \multicolumn{1}{c}{\small $(\mathrm{\it V}-\mathrm{\it X})_0$}\\
\multicolumn{1}{c}{\small (days)} & \multicolumn{1}{c}{\small (days)} &  \multicolumn{1}{c}{\small Filter} & \multicolumn{1}{c}{\small (mag)} & {\small (mag)} & {\small (mag)} &  {\small Match} & \multicolumn{1}{c}{\small (mag)} & {\small (mag)} & {\small (mag)} & \multicolumn{1}{c}{\small (mag)}\\
\hline
\multicolumn{11}{c}{\small SN2012cg}\\
\hline
{\small $56082.4$} & {\small $0.2$} & WFC3 F336W & {\small $11.74(0.06)$} & {\small $0.09$} & {\small $0.01$} & M & {\small $12.04(0.04)$} & {\small 0.06} & {\small 0.00} & {\small\ \ \ $0.61$\ \ \ }\\
{\small $56097.5$} & {\small $14.0$} & WFC3 F336W & {\small $13.75(0.01)$} & {\small $0.09$} & {\small $0.02$} & M & {\small $12.53(0.07)$} & {\small 0.06} & {\small 0.00} & {\small\ \ \ $-0.90$\ \ \ }\\
{\small $56082.4$} & {\small $0.2$} & WFC3 F125W & {\small $12.37(0.01)$} & {\small $0.01$} & {\small $0.05$} & M & {\small $12.04(0.04)$} & {\small 0.06} & {\small 0.00} & {\small\ \ \ $-0.55$\ \ \ }\\
{\small $56097.5$} & {\small $14.1$} & WFC3 F125W & {\small $13.88(0.03)$} & {\small $0.02$} & {\small $-0.07$} & M & {\small $12.53(0.07)$} & {\small 0.06} & {\small 0.00} & {\small\ \ \ $-1.91$\ \ \ }\\
{\small $56082.4$} & {\small $0.2$} & WFC3 F225W & {\small $16.38(0.05)$} & {\small $0.13$} & {\small $0.02$} & M & {\small $12.04(0.04)$} & {\small 0.06} & {\small 0.00} & {\small\ \ \ $-2.55$\ \ \ }\\
{\small $56097.5$} & {\small $14.0$} & WFC3 F225W & {\small $17.60(0.08)$} & {\small $0.13$} & {\small $0.02$} & M & {\small $12.53(0.07)$} & {\small 0.06} & {\small 0.00} & {\small\ \ \ $-3.68$\ \ \ }\\
{\small $56082.4$} & {\small $0.2$} & WFC3 F160W & {\small $12.67(0.02)$} & {\small $0.01$} & {\small $0.01$} & M & {\small $12.04(0.04)$} & {\small 0.06} & {\small 0.00} & {\small\ \ \ $-0.88$\ \ \ }\\
{\small $56097.5$} & {\small $14.1$} & WFC3 F160W & {\small $13.24(0.03)$} & {\small $0.01$} & {\small $0.55$} & M & {\small $12.53(0.07)$} & {\small 0.06} & {\small 0.00} & {\small\ \ \ $-0.69$\ \ \ }\\
\multicolumn{11}{c}{\small $\cdots$}\\
\hline
\end{tabular}

\caption{%
  The photometry of all \sne.  All magnitudes are in the natural Vega system.  The rest-frame magnitude and colour can be 
  obtained from Columns 4--6 using Equation~\eqref{eq:colour}. Columns 5 and 9 are the Galactic extinctions and Columns 6 and 10 are  $K_\mathrm{\Xband}$ and 
  $K_\mathrm{\Vband}$ corrections for the two bands, respectively.  All corrections were calculated after the \snfe template 
  had been reddened with the best fitted \ftz law, shown in Table~\ref{tb:fitlawresults}, for each \sn.  The \Vband magnitude is 
  only shown for data points used in the colour analysis, i.e. with phases between $-10$ and $+35$ days.
  Column 2 show the effective light curve-width-corrected phase, while Column~7 specifies whether the \Vband magnitude 
  was measured for the same epoch (D) or if it was calculated using the \snoopy model (M).  The corresponding 
  intrinsic colour for the \snfe template is shown when available in Column~11.   The \swift magnitudes of \snjj are from
  \citet{2015ApJ...805...74B}.
  (This table is available in its entirety in a machine-readable form in the online journal. A portion is shown here 
   for guidance regarding its form and content.)
\label{tb:photometry}}
\end{table*}

For most \hst visits we also obtained long-slit spectroscopy with the Space Telescope Imaging Spectrograph (STIS) where 
the $1024\times1024$~pixel SITe CCD detector was used with the G230LB grating covering the wavelength range 
$1600$--$3100$~\AA, and for two \sne (\cg, \et)  G430L, $2900$--$5700$~\AA. The data were reduced using the \texttt{calstis}
pipeline which is part of the \texttt{STSDAS} package.  The pipeline was ran for all the spectra up to the point where the 
calibrated 2D spectra were created.  Only the spectra of \sncg and the red G430L spectrum of \snet contained any significant 
signal and were extracted using the \texttt{calstis} \texttt{x1d} routine with a 4~pixel aperture.

\subsection{\swift/\uvot}
UV photometry was also obtained with the Ultra-Violet/Optical Telescope \citep[\uvot,][]{2005SSRv..120...95R} on the 
\swift spacecraft \citep{2004ApJ...611.1005G} for the \sne \bl, \cg, \cp and \cu in the \uvwone, \uvwtwo and \uvmtwo filters.  
However, the \uvwone and \uvwtwo filters are not well suited for extinction studies due to the significant ``red-tails'' of 
these filters. \cite{2010ApJ...721.1627M} used the UV spectra of \snname{1992A} \citep{1993ApJ...415..589K} and 
estimated that 52\,\% and 44\,\% of the light in \uvwone and \uvwtwo respectively, originate from wavelengths redder 
than $3000$~\AA.  \citet{2015ApJ...805...74B} show how the observed photons and the corresponding effective 
wavelengths shift dramatically as reddening increases.

The \uvmtwo filter, on the other hand, is much better constrained (1\,\% of the light comes from $>3000$~\AA) and
is used for the extinction studies in this work.  
The \uvmtwo magnitudes were measured using the pipeline from the \swift Optical/Ultraviolet Supernova Archive 
\citep[SOUSA,][]{2014Ap&SS.354...89B}, including correction for the time-dependent sensitivity and revised zeropoints 
from \citet{2011AIPC.1358..373B}.
The result is presented in Table~\ref{tb:photometry} and shown in Figure~\ref{fig:lcs}.
Only the \sne \bl and \cg showed a significant \sn flux in the \uvmtwo band after subtraction of the underlying host galaxy flux.
The flux measured for \sncp was consistent with the host galaxy brightness and no signal was measured for \sncu. 


\subsection{Ground based observations}
The ground based spectroscopic observations are listed in Table~\ref{tb:spec} while the photometric observations are 
summarised together with the measured magnitudes in Table~\ref{tb:photometry} and shown in Figure~\ref{fig:lcs}.  
All \sne except for \snbl were observed with
the 2.56~metre Nordic Optical Telescope (NOT) under programmes 45-009 and~46-018 (PI: Amanullah).  Imaging and 
spectroscopy was obtained with the $6.4'\times6.4'$ Andalucia Faint Object Spectrograph and Camera (ALFOSC) using the 
filter set \Uband (\#7)\footnote{NOT filter ID}, \Bband (\#74), \Vband (\#75), \Rband (\#76), \iband (\#12) and the $R=360$ 
grism~(\#4)\footnote{NOT grism ID}.  The data were reduced using standard IRAF routines and the QUBA pipeline 
\citep{2011MNRAS.416.3138V}.   The photometry was measured by first fitting the point-spread-function (PSF) using DAOPHOT 
\citep{1987PASP...99..191S} to stars in the fields.  The calibrated magnitudes were obtained either from Landolt fields \citep{1992AJ....104..340L} observed during the night, or, when these were not available, by comparing the fluxes to stars in
the fields that were then calibrated against Landolt fields for the photometric nights.

\begin{table*}
\centering
\begin{tabular}{l l r l c r l}
\hline\hline
\multicolumn{1}{c}{Civil date} & 
\multicolumn{1}{c}{MJD} & 
\multicolumn{1}{c}{Epoch} &
\multicolumn{1}{c}{Instrument} & $\lambda$ & 
\multicolumn{1}{c}{Exp.} & \multicolumn{1}{c}{\sn}\\
& & \multicolumn{1}{c}{(days)} & & \multicolumn{1}{c}{(\AA)} & \multicolumn{1}{c}{(s)}\\
\hline
2012-05-26 & 56073.8 & $+260$ & WHT/ISIS & 3500$-$9500 & 600 & \fe\\
\hline
2012-03-30 & 56016.3 &  $-2$ &  DuPont/WFCCD/WF4K-1 & 3600$-$9200 & 700 & \bl\\
2012-03-31 & 56017.4 &  $-1$ & DuPont/WFCCD/WF4K-1 & 3600$-$9200 & 700 &\bl\\
2012-05-01 & 56048.3 &  $+29$ & DuPont/WFCCD/WF4K-1 & 3600$-$9200 & 900 &\bl\\
\hline
2012-03-28 & 56014.0 &  $-4$ & ASIAGO/AFOSC & 3500$-$8200& 2700 & \bm\\
2012-04-09 & 56027.0 &  $+9$ & NOT/ALFOSC & 3200$-$9000 & 2400 & \bm\\
2012-04-16 & 56034.1 &  $+16$ & NOT/ALFOSC & 3200$-$9000 & 2400 & \bm\\
\hline
2012-05-24 & 56071.9 &  $-9$ & NOT/ALFOSC & 3700$-$7200 & 5477 & \cg\\
2012-05-26 & 56073.9 &  $-7$ & NOT/FIES   & 3640$-$7360  & 1800 & \cg\\
2012-05-28 & 56075.9 &  $-5$ & NOT/ALFOSC & 3700$-$7200 & 6588 & \cg\\
2012-06-02 & 56080.9 &  $+1$ & NOT/FIES   & 3640$-$7360 & 900 & \cg\\ 
2012-06-04 & 56082.9 &  $+3$ & NOT/ALFOSC & 3200$-$9000 & 300 & \cg\\
2012-06-14 & 56092.9 &  $+13$ & NOT/FIES   & 3640$-$7360 & 1200 & \cg\\
2013-03-09 & 56360.4 &  $+280$ & Keck/LRIS & 3075$-$10300 & 600 & \cg\\
\hline
2012-05-26 & 56074.0 &  $-7$ & NOT/ALFOSC & 3200$-$9000 & 2700 & \cp\\
2012-06-04 & 56083.0 &  $+2$ & NOT/ALFOSC & 3200$-$9000 & 1800 & \cp\\
\hline
2012-07-05 & 56113.9 &  $+9$ & NOT/ALFOSC & 3200$-$9000 & 1200 & \cu\\
\hline
2012-09-13 & 56183.8 &  $-7$ & ASIAGO/AFOSC & 3500$-$8200 & 2700 & \et\\
2012-09-22 & 56193.1 &  $+2$ & NOT/ALFOSC & 3200$-$9000 & 600 & \et\\
\hline
2014-11-09 & 56970.0 &  $+282$ & APO/DIS & 3300$-$9800 & 750 & \jj\\
\hline\hline
\end{tabular}
\caption{Ground based spectroscopic observations.  The Asiago spectra were obtained through the
Asiago Transient Classification Program \citep{2014AN....335..841T}.  The epochs are shown with
respect to the date of $\Bband$-maximum shown in Table~\ref{tb:fitlawresults}.
\label{tb:spec}}
\end{table*}

With the NOT we also obtained high-resolution spectroscopy of \sncg using the FIbre-fed Echelle Spectrograph 
\citep[FIES,][]{2014AN....335...41T} in its high-resolution mode, $R=67000$.  Simultaneous wavelength reference (Thorium-Argon)
spectra were obtained and the data were reduced using the software \texttt{FIEStool} which is provided by the observatory.

NIR observations were carried out for \sne \bm, \cg and \cp with the $4'\times4'$ NOTCam instrument in the \Jband, \Hband 
and \Ksband bands under the programmes 45-007 (PI: Kankare) and 46-020 (PI: Mattila).  We used the wide field imaging 
option of the $1024\times1024$ pixel HgCdTe NOTCam detector with a plate scale of  $0\farcs234$/pixel.  The observations 
were carried out using either 5 or 9-point dithering patterns. For the \sne with extended host galaxies, beam-switching was used 
to guarantee a successful sky subtraction. 
%
%
We used the NOTCam Quick-Look reduction package based on IRAF.  Bad pixels (which includes two dead columns) were 
masked and we used master differential skyflats which were obtain from bright and faint skyflats.  Further, the images were 
corrected for geometric distortions\footnote{Magnus G\aa lfalk, private communication.}  before the individual images were 
aligned and coadded.  

\sncg was also observed in \Jband\Hband\Ksband with the $4.3'\times4.3'$ CAIN instrument on the 1.52~metre 
Carlos Sanchez Telescope at Observatorio del Teide on Tenerife.  CAIN III is a $256\times256$ pixel HgCdTe 
NIR detector,  where we used the wide field option with a plate scale of $1.0''$/pixel.  
The data were reduced using a dedicated IRAF package  provided by J. Pullido and A. Barrena. 

The two \sne \bm and \cp were also observed by the {\em Palomar Transient Factory} with the Palomar 48-inch telescope
\citep{2009PASP..121.1334R} in the Mould $\Rband$. We use their photometry which was obtained by first determining the 
PSF prior to subtractions, and then carrying out PSF photometry on the subtracted frames \citep{2015MNRAS.446.3895F}. 
The calibration of this data has been described in \citet{2012PASP..124...62O}.

Photometric follow-up of \snbl was carried out by the {\em Carnegie Supernova Project} using the Swope telescope for the
optical and the du~Pont telescope for the NIR.  The reduction and photometry have been described in detail in
\citet{2006PASP..118....2H} and \citet{2010AJ....139..519C}. PSF photometry was performed with respect to a local sequence 
of standard stars calibrated to the \citet{1992AJ....104..340L} and \citet{2002AJ....123.2121S} standard fields for the 
optical observations and the \citet{1998AJ....116.2475P} for the NIR.

Further, we here also present spectroscopic observations of \sne \fe, \cg and \jj\ obtained using the ISIS instrument at
the William Herschel Telescope, the LRIS instrument at Keck and the DIS instrument at the Apache Point Observatory.  These
data were reduced in a standard manner.

\section{\sn spectra and colours}\label{sec:color}
\begin{figure}
  \begin{center}
    \includegraphics[width=1.0\columnwidth]{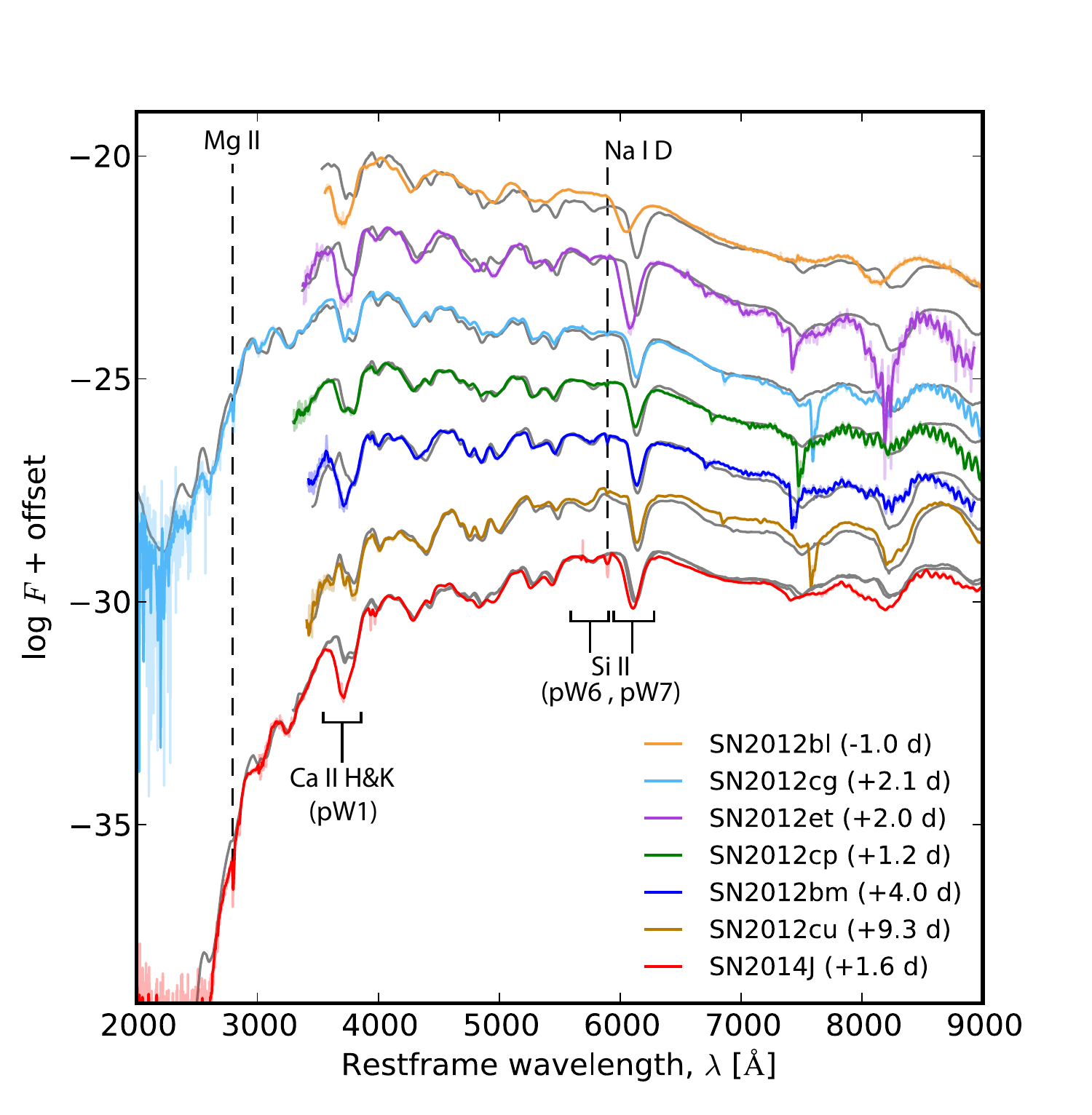}
    \caption{%
    		Spectra near maximum, when available, for the \sne in our sample. The grey spectra are for \snfe
     	at similar epochs \citep{2013A&A...554A..27P,2014MNRAS.439.1959M} reddened by the extinction laws 
	    that we derive in \S\,\ref{sec:lawfitting}.  For \sne \cg and \jj we have combined the optical spectra with \stis spectra 
	    \citep[the \snjj spectrum was taken from][]{2014MNRAS.443.2887F}. Spectroscopic features discussed in 
	    \S\,\ref{sec:cs}, \S\,\ref{sec:fetemplate} and \S\,\ref{sec:highres} have been marked.
      \label{fig:spectra}}
  \end{center}
\end{figure}

The spectra for the seven \sne around maximum are shown in Figure~\ref{fig:spectra}, where we also
added the classification spectrum of \sne \bm and \et \citep{2014AN....335..841T} to the set of 
spectroscopic observations already described above. For \sncg, both the \hststis G250LB 
and G430L spectra, and the ALFOSC spectrum, that was obtained the same day, are shown.
We used SNID \citep{2007ApJ...666.1024B} to type the spectra and all \sne could be classified
as ''normal'', although \snet also provided acceptable matches to \snname{1999aa}.
For each spectrum in Figure~\ref{fig:spectra} we also show the spectrum of  the
normal \snfe for the matching epoch \citep{2014MNRAS.439.1959M} for comparison.

All photometric data are shown in Figure~\ref{fig:lcs} and for each measurement in each filter, \Xband, where 
$\mathrm{\Xband} \in \left\{\mathrm{\uvmtwo},\mathrm{\wfcone},\ldots,\mathrm{\Ksband}\right\}$,
we calculated the colour \Xband$-$\Vband.  When \Xband and \Vband-band measurements are not available for the 
same dates, and the colour could not be obtained directly, a \Vband-band model was used to calculate the colours.   
When applicable, we used a smoothed spline model but for the \sne \bm and \cp the \Vband-band data 
were too sparsely sampled to allow reliable spline fits.  For these objects we instead fitted the \snia template
from \citet[][H07 from hereon]{2007ApJ...663.1187H} to the observed data. Both the spline and template-based models 
were obtained using the \snoopy light curve fitter \citep{2011AJ....141...19B} and are shown as thick lines in 
Figure~\ref{fig:lcs}.   Further, for all model-based colours we add $0.05$~mag in quadrature to uncertainties in 
order to account for the inaccuracy of the models.

All colours were corrected for Galactic extinction, $\AMW$, using the extinction law by 
\citet[][\ccm from hereon]{1989ApJ...345..245C} with $\RV=3.1$ and the $\AV^{\mathrm{MW}}$ values shown in the 
last column of Table~\ref{tb:snsummary}.   The values are from the \citet{2011ApJ...737..103S} recalibration of 
the \citet{1998ApJ...500..525S}  dust maps,  except for \snjj where we use a 
value obtained from studying neighbouring regions \citep{2009ApJS..183...67D}.

The colours were also $K$ corrected \citep[\eg][]{2002PASP..114..803N} and $S$ corrected
\citep[\eg][]{2000AIPC..522...65S,2002AJ....124.2100S,2003AJ....125..166K} to a common rest-frame filter 
system that will be used for the remaining analysis throughout the paper.  The analysis was carried out using the 
\wfcuvis and \swift/\uvot filters for the UV, the ALFOSC filter set for the optical and the NOTCam filters for the NIR, 
since the bulk of the data were obtained using these filters.  The combined $K$ and $S$ corrections ($\KX$ from now 
on) are calculated synthetically using the SED of \snfe as described in \S\,\ref{sec:intrinsic} and the filter transmissions 
provided by the different observatories.  

To summarise, all colours were obtained as
\begin{equation}
\mathrm{\Xband}-\mathrm{\Vband} = (\mX - \AMW_\mathrm{\Xband} - \KX) -  (\mV - \AMW_\mathrm{\Vband} - \KV)\, ,
\label{eq:colour}
\end{equation}
where $\mX$ are the measured instrumental magnitudes in filter \Xband  and $\mV$ is the \Vband magnitude for the 
same date either measured directly, or obtained from a spline or template model.  All values are presented in 
Table~\ref{tb:photometry} where we have also added the \swift photometry of \snjj from \citet{2015ApJ...805...74B} to 
the measurements described in \S\,\ref{sec:data}.
As will be further discussed in \S\,\ref{sec:lawfitting}, not only the observed colours, but also, $\AMW$ and $\KX$ depend on 
the reddening of each \sn and were therefore obtained iteratively during the extinction law fitting.  The values shown 
in Table~\ref{tb:photometry} were
obtained after the  \snfe SED had been reddened with the \citet[][hereafter \ftz]{1999PASP..111...63F} law using the parameters
shown in Table~\ref{tb:fitlawresults} for each \sn.

\begin{figure*}
  	\centering
    \includegraphics[width=\textwidth]{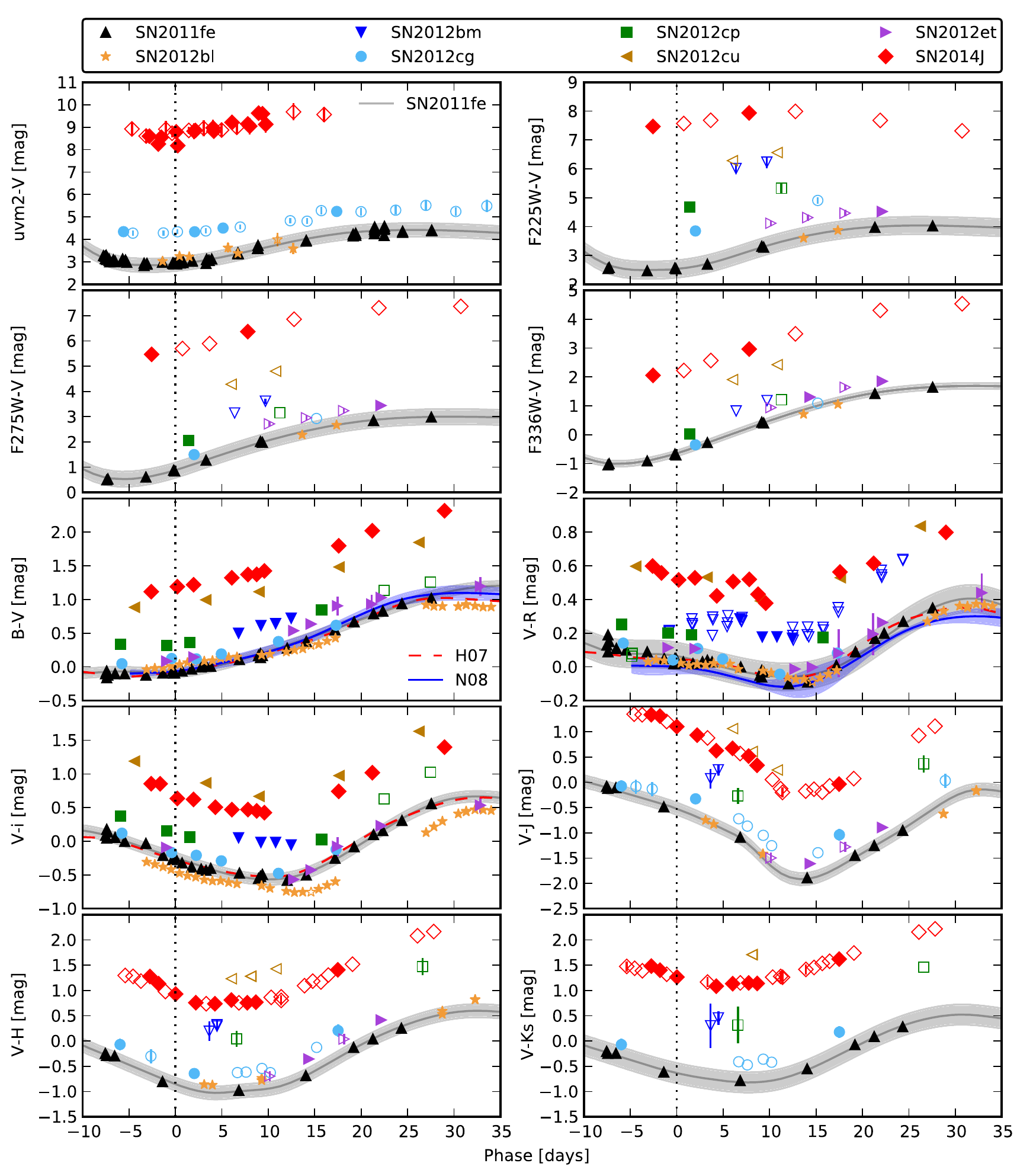}
    \caption{A selection of measured colours from UV to NIR together with literature \sne \by and \fe.  The phases,
      $p = (t -\tBmax)/\sB$, were obtained using the values of $\tBmax$ and $\sB$ from Table~\ref{tb:fitlawresults}. 
       The colours from the H07 spectral template (dashed red) and 
       \citet{2008A&A...487...19N} (N08, solid blue with dispersion region)
       are also shown together with the derived colour model of \snfe (solid grey) with the adopted dispersions (grey region).
       Filled symbols show colours where a \Vband-band measurement was obtained for the same date, while a 
       \Vband-band model was used to obtain the measurements that are shown with open symbols. 
       Errorbars are only plotted for data points where the uncertainty exceeds 0.1~mag. See text for details.
      \label{fig:colors}}
\end{figure*}

A subset of the colours are also shown in Figure~\ref{fig:colors} where the colours obtained 
using measurements and the \Vband-band model are shown with filled and open symbols respectively.

\sneia are intrinsically bright in the rest-frame \Bband-band and this filter is commonly used to measure 
\snia distances.   \snia light curves are typically quantified by a few parameters such as the brightness and 
time of maximum, $\tBmax$, for \eg the \Bband-band, and the light curve shape.  The light curve shape can be 
quantified using different methods, where perhaps the most straight-forward is to measure the brightness decline for 
the first 15~days past $\tBmax$.  \cite{1993ApJ...413L.105P} showed that fast declining \sneia are typically intrinsically 
fainter than slow declining objects,  a discovery that lay the foundation for using \sneia as distance indicators for precision 
cosmology.  An alternative approach to quantify the light curve shape is to introduce a \emph{stretch} parameter 
\citep{1997ApJ...483..565P}, $\sB$, which can be defined as the value needed to match the time-evolution of an 
observed \Bband light curve to a standard \snia template.  



The time-evolution of \snia $\mathrm{\Xband}-\mathrm{\Vband}$ colours also depends on light curve shape. 
For example, \citet[from hereon N08]{2008A&A...487...19N} present colour-light curve shape relations in the optical 
while B14 show that the \sneia colours can be standardised over a wide range of decline rates by introducing 
\emph{colour-stretch},  $\sBV = t_\mathrm{\Bband\Vband}/30~\mathrm{days}$, where $t_\mathrm{\Bband\Vband}$ is the 
time between $\tBmax$ and the maximum of the \Bband$-$\Vband colour.  However, this approach requires
$\mathrm{\Bband}-\mathrm{\Vband}$ coverage up to $t\sim\tBmax+40$ in order to accurately determine $\sBV$.

We here standardise the measured colours using $\sB$. All colours will be studied and compared as a function 
of  \emph{phase}, $p$, defined as $p = (t - \tBmax)/\sB$, where $t$ are the observing dates.   The phases for the colours 
shown in Figure~\ref{fig:colors} have been determined by using the values of $\tBmax$ and $\sB$ from Table~\ref{tb:fitlawresults},
which are fitted simultaneously with the reddening law as described in \S\,\ref{sec:lawfitting}.

\newcommand{\XVO}{\left(\mathrm{\Xband} - \mathrm{\Vband}\right)_0}
\section{Fitting reddening laws}\label{sec:lawfitting}
We will use a similar method to \jjpaper to fit different parametrised reddening laws to the 
measured colour excesses,
\begin{equation}
\EXV = \left(\mathrm{\Xband} - \mathrm{\Vband}\right) - \XVO\,,
\label{eq:excess}
\end{equation}
where $\XVO$ is the assumed $\mathrm{\Xband} - \mathrm{\Vband}$ colour for the unreddened objects.

In this work we will study different extinction laws and how well they describe the observed \snia reddening. Each extinction law,
$A(\lambda;\bar{a})$, depends on a set of parameters $\bar{a}$ (\eg $\EBV$ and $\RV$), that will be fitted to the observed colour
excesses, $\EXV_p$, by minimising
\begin{equation}
	\chi^2 = \sum_{X}\sum_{p} \frac{\left[ E(\mathrm{\Xband}-\mathrm{\Vband})_p - (A_{\mathrm{\Xband}_p} - A_{\mathrm{\Vband}_p})\right]^2}{\sigma_{\mathrm{\Xband}_p\mathrm{\Vband}_p}^2}\, .
	\label{eq:chi2}
\end{equation}
Here $A_{\mathrm{\Xband}_p}$ and $A_{\mathrm{\Vband}_p}$ are the predicted extinctions in the \Xband and \Vband
filters for phase $p$ and can be calculated as
\begin{equation}
	A_{\mathrm{\Xband}_p} = -2.5\log_{10}\left(\frac{\int T_\mathrm{\Xband}(\lambda)\cdot
		10^{-0.4\cdot A(\lambda;\bar{a})}\cdot S_0(\lambda;p)\lambda\,d\lambda}{%
		\int T_\mathrm{\Xband}(\lambda)S_0(\lambda;p)\lambda\,d\lambda}\right)\,,
	\label{eq:extinction}
\end{equation}
if the effective filter transmission, $T_\mathrm{\Xband}(\lambda)$, and the spectral energy distribution (SED),
$S_0(\lambda;p)$, of the unreddened source are assumed to be known.  Further, 
$\sigma_{\mathrm{\Xband}_p\mathrm{\Vband}_p}^2$ are the uncertainties added in quadrature.  This includes the measurement
errors, but will in most cases be dominated by the intrinsic colour uncertainties discussed in \S\,\ref{sec:intrinsic}.

In addition to the extinction law parameters $\bar{a}$, we will also minimise Equation~\eqref{eq:chi2} with respect to
the parameters $\tBmax$ and $\sB$ that together determine the phase, $p$. The {\em observed} values of these 
parameters are extinction dependent \citep[see \eg][\jjpaper]{1988PhDT.......171L,1993ApJ...413L.105P,2002PASP..114..803N}
since broadband measurement of an object suffering from extinction will effectively probe redder wavelengths 
than the observation of the same unreddened object would have.  The \snia light curve decline rate varies 
with wavelength, and slower decline rates are in general expected to be observed for reddened
objects.

Similarly, the Galactic extinction, $\AMW_\mathrm{\Xband}$, as well as the $\KX$ corrections, are also properties measured 
through broadband filters and therefore also depend on the SED of the source, as seen from 
\eg Equation~\eqref{eq:extinction}.  Since the observed SED in the Milky Way is affected by the extinction in the \sn 
host galaxy, these properties will also depend on the reddening law.  The effect will be particularly significant in the UV, where
the wavelength dependence for the extinction laws is steep.  For example, the Galactic extinction in the \swift/\uvmtwo band
for the line of sight to \snjj would have been $\AMW_\mathrm{\Xband}\approx0.3$ for an unreddened source while it is 
$\AMW_\mathrm{\Xband}\approx0.1$ for \snjj.
%

We take this effect into account by fitting each reddening law, $A(\lambda;\bar{a})$, iteratively and 
update $\AMW_\mathrm{\Xband}$ and $\KX$ using the fit $\bar{a}$ values in each iteration.  That is, after each iteration,
we re-calculate the observed colour and colour excesses using Equations~\eqref{eq:colour} and~\eqref{eq:excess} before 
re-fitting the reddening law using Equation~\eqref{eq:chi2}. The procedure is repeated until the change in the fitted 
parameters is less than $1\,\%$ between iterations. 

As mentioned above we use the average value of $\RV=3.1$ when correcting for the Galactic extinction. 
For the majority of the \sne the Galactic extinction is negligible compared to the host reddening, with the 
exception of \snet.  For this \sn we also tried carrying out the fits using the extreme values of 
$\RV=2.2$ and $\RV=5.8$ observed in the Milky Way \citep[\eg CCM,][]{1999PASP..111...63F} 
and concluded that this could impact the fit values with up to one statistical standard deviation.

\subsection{\snia intrinsic colours and SED}\label{sec:intrinsic}
In order to minimise Equation~\eqref{eq:chi2}, we need to make assumptions of the intrinsic colours, $\XVO$, 
and SED, $S_0(\lambda;p)$ of the unreddened objects. The reddening laws for \sneia are typically derived by 
either comparing them to individual objects that show similarities in light curve properties and spectral evolution 
\citep[see \eg][\jjpaper]{2006AJ....131.1639K}, or to \snia colour and SED templates. 

In Figure~\ref{fig:colors} the H07 template is plotted (dashed red) for the optical colours.  Although the H07 template 
does extend from the UV to NIR, the data it is based on are sparse at the endpoints of this range. N08 studied the intrinsic 
optical colours between $-10$ and $+50$~days from \Bband-band maximum. Two of their colour laws are 
shown (in blue) for normal, $\sB=1$, \sneia in Figure~\ref{fig:colors} together with the colour dispersions they
derive.   Although these templates provide excellent coverage at optical wavelengths none of them cover, to high 
accuracy, the full wavelength range required for the UV-NIR analysis in this work.

The best studied unreddened \snia to date is \snfe, discovered by the Palomar Transient Factory in the nearby spiral 
galaxy M101 \citep[PTF11kly,][]{2011Natur.480..344N}. Its close proximity 
allowed detailed spectroscopic and photometric observations over a broad wavelength range from the UV 
\citep{2012ApJ...753...22B,2014MNRAS.439.1959M}, through the optical 
\citep[\eg][]{2013NewA...20...30M} to the near- \citep{2012ApJ...754...19M,2013ApJ...766...72H}, and mid-
\citep{2013ApJ...767..119M}, infrared.  The \sn was also targeted in the far-IR \citep*{2013MNRAS.431L..43J} and 
radio \citep{2012ApJ...750..164C}, but was not detected at these wavelengths.  

\snfe is a normal \snia in both the optical and the NIR  \citep{2013NewA...20...30M,2012ApJ...754...19M} 
and does not show any spectroscopic peculiarities \citep{2013A&A...554A..27P}.  
Further, the low Galactic and host galaxy reddening along the line-of-sight, 
$\EBV_{\rm MW} = 0.011 \pm 0.002$~mag and $\EBV_{\rm host} = 0.014 \pm 0.002$~mag 
\citep[deduced from the integrated equivalent widths of the \NaID lines;][]{2013A&A...549A..62P}, makes it
an excellent comparison object for studying reddening of \sneia.   The colours of \snfe shown in 
Figure~\ref{fig:colors} (solid black lines) have been obtained by combining the available UV--NIR data,
as described in Appendix~\ref{sec:femodel} based on the measurements (black triangles).  We have also combined 
the spectroscopic data to create a daily sampled SED, shown in Figure~\ref{fig:sedmodel}.  From the excellent 
$\mathrm{\Bband}-\mathrm{\Vband}$ match between \snfe and H07 we can also conclude that any potential
discrepancies introduced by using the \snoopy model for the \Vband-band is not likely to have any significant
effect on the derived reddening laws when the colours are compared to the corresponding \snfe colours.

In order to properly compare the colours of reddened \sneia to \snfe, and fit reddening laws, we also need 
accurate estimates of the expected intrinsic colour dispersions.  It is desirable to take into account
how these vary in time, and how they are correlated both in time and with different colours.  However, since we lack 
this information for all colours involved in the analysis, we adopt a simplified approach where all colours 
are treated equally with phase-independent dispersions for each colour.  We further follow the procedure from
\jjpaper and assume that the colour uncertainties between \sn phases are completely correlated and that there
is no correlation between different colours.

\citet[][M13 from hereon]{2013ApJ...779...23M} following the work of \citet{2010ApJ...721.1608B} and 
\citet{2010ApJ...721.1627M}, studied the UV$-$optical colours of 23~\sneia observed by \swift.  They reported a dispersion of 
$\sim0.3$~mag in UV$-$\vband for their $\EBV<0.2$~mag sample after correcting the data for extinction and the result does 
not seem to be affected significantly by the choice of $\RV$.  We adopt this dispersion, shown as grey bands around the \snfe
colour template in Figure~\ref{fig:colors}, for the 
$\mathrm{\uvmtwo}-\mathrm{\Vband}$, 
$\mathrm{\wfcone}-\mathrm{\Vband}$, and $\mathrm{\wfctwo}-\mathrm{\Vband}$ colours.  
For the $\mathrm{\wfcthree}-\mathrm{\Vband}$ and $\mathrm{\Uband}-\mathrm{\Vband}$ we adopt a dispersion 
of $0.1$~mag based on the $\mathrm{\Uband}-\mathrm{\Vband}$ dispersion from N08.  Note that since these dispersions 
were derived for unreddened objects, and the dispersion of \sneia colours decrease with wavelength, they can be considered
as conservative for highly reddened \sne.  When reddened \sneia are studied using the \wfcuvis filters, the observations will 
effectively probe redder wavelengths compared to similar studies of unreddened objects as illustrated in \eg Figure~\ref{fig:leak}, 
and as a consequence the intrinsic dispersion of the former observations can be expected to be lower than the latter.
On the other hand, even larger dispersions have been also observed in the UV, and we will use these intrinsic uncertainties 
under the assumption that all the \sne we study are similar to the reference \snfe, which we will discuss further in 
\S\,\ref{sec:fetemplate}. 

B14 derive a dispersion for the pseudo-colour $\mathrm{\Bband}_\mathrm{max}-\mathrm{\Vband}_\mathrm{max}$ of 
$0.06$~mag using a Cauchy Prior for the colour distribution of their \sneia observed with the Carnegie Supernova 
Project.  This is consistent with the phase-dependent results of N08 who derive a  
$\mathrm{\Bband}-\mathrm{\Vband}$ dispersion of $\lesssim0.1$~mag, which we use here, within the 
range $-10$ to $+35$ days.  For the $\mathrm{\Vband}-\mathrm{\Rband}$ and $\mathrm{\Vband}-\mathrm{\iband}$  colours
we adopt the N08 dispersions of $0.08$~mag.
 
The NIR colours of \sneia typically show low dispersion \citep[see \eg][]{2000MNRAS.314..782M} and for the NIR-optical 
$\mathrm{\Vband}-\mathrm{\Jband}$ and $\mathrm{\Vband}-\mathrm{\Hband}$ colours we can estimate the dispersion 
in our phase range by using all \sne with $\EBV<0.2$~mag from the ''Gold Sample'' in \citet{2015arXiv150507707S} and find $0.12$~mag 
and $0.13$~mag respectively.  For $\mathrm{\Vband}-\mathrm{\Ksband}$ we conservatively adopt $0.2$~mag based on the
studies carried out at maximum (B14) and the dispersion of the \Ksband-band \citep{2012PhDT........42F,2014arXiv1408.0465F}.



\begin{figure*}
    \centering
    \includegraphics[width=\textwidth]{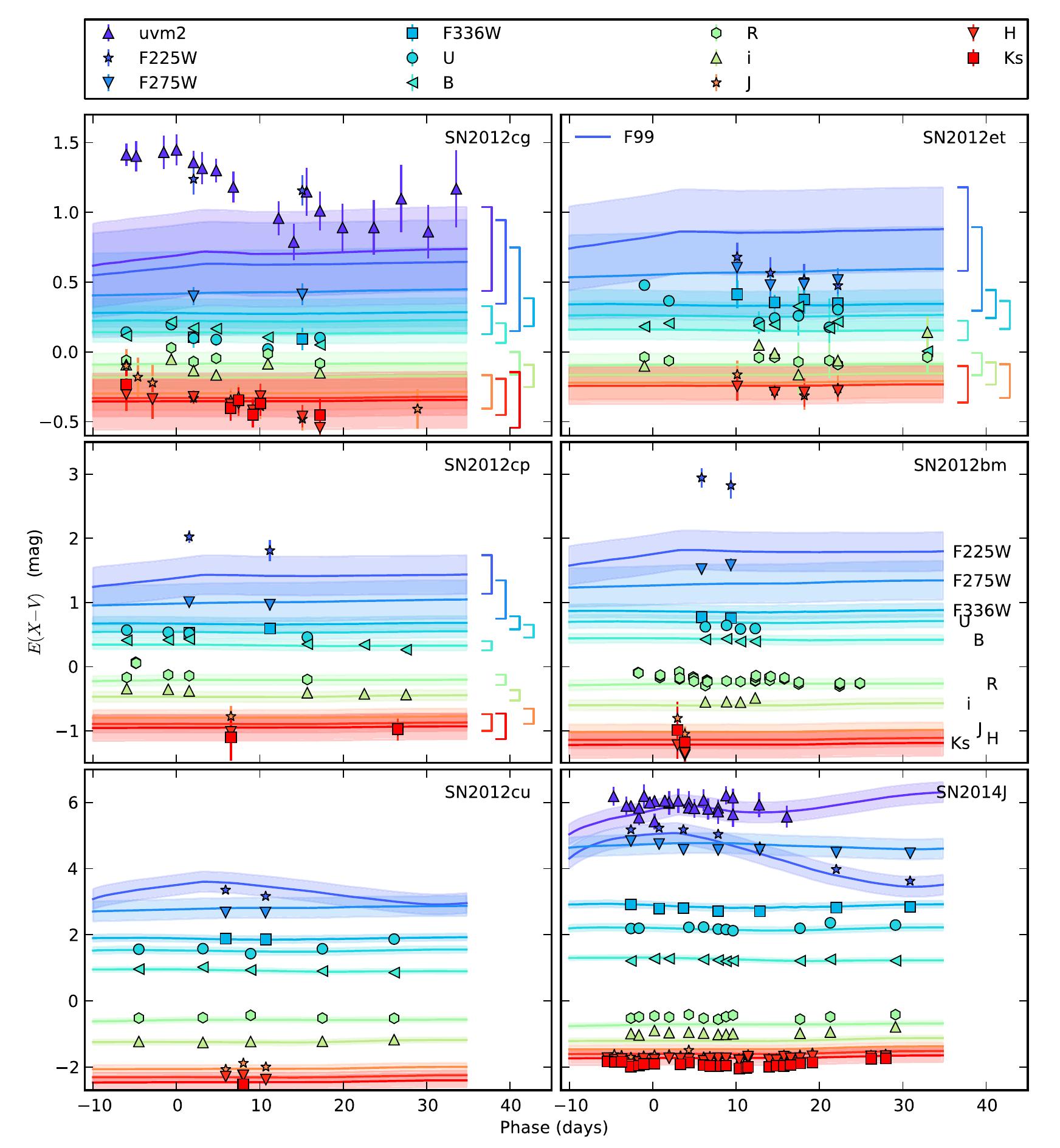}
    \caption{%
    	The observed colour excess, $E(\mathrm{\Xband}-\mathrm{\Vband})$, against phase 
	for all \sne and filters.  The predicted colour excess from the \snfe template and the fitted \ftz law, presented in 
	Table~\ref{tb:fitlawresults}, is shown with solid lines and the regions surrounding them mark the adopted intrinsic 
	colour uncertainties.  The uncertainties are also indicated by open brackets for the three \sne with the least reddening.  
	The phase, $p=(t-\tBmax)/\sB$, was calculating based on the fitted $\tBmax$ and $\sB$ values from 
	Table~\ref{tb:fitlawresults}. The apparent $\EXV$ time-dependence of the \wfcone filter for \eg \snjj originates from the red 
	tails of these filters which causes the observed colour excess to vary with the intrinsic colour of the source.
    	\label{fig:afits}}
\end{figure*}

The observed colour excesses, $\EXV$, for each reddened \sn, together with the adopted colour dispersions, shown 
as background regions, are plotted in Figure~\ref{fig:afits}.  \snbl has not been plotted in the figure.  It will be omitted 
from the extinction analysis due to its lack of reddening.

If the colour excess is observed at single wavelengths, \ie $E(\lambda_1-\lambda_2) = A_{\lambda_1} - A_{\lambda_2}$, 
it will be time-independent for interstellar dust extinction.  However, for broad-band photometry,  we can expect $\EXV$ to 
vary with time which is illustrated by \eg the UV colours in Figure~\ref{fig:afits}.
As the intrinsic colour of the source varies with time, the effective wavelengths of all broad-band filters will change 
as well.  Further, since the first term on the right-hand side in Equation~\eqref{eq:excess} is also affected by a 
reddening law, any change in the effective wavelength of the filters \Xband and \Vband will affect this term more 
than the second term, which will induce a time-variability for $\EXV$.  For most filters, the time-variability will 
be negligible except in the UV (where the extinction has a steep wavelength dependence), and in particular for filters with 
\emph{red tails}, \eg \wfcone (see~Appendix~\ref{sec:redtail}), the effect is significant.  This is illustrated by the 
bottom panel of Figure~\ref{fig:leak} where the effective wavelength of three \wfcuvis UV filters have been plotted for a 
\snia at maximum for different extinctions.

\subsection{Reddening laws}
We test three different extinction law parametrisations. In addition to the widely used CCM law modified by 
\citet[][hereafter \ccmo]{1994ApJ...422..158O}, which has been derived from studying different lines-of-sight in 
the Milky Way,  we also fit the parametrisation from \ftz that has been derived in a similar manner.  For each of 
these we fit both the colour excess $\EBV$, which relates to the optical depth, and the ratio of the total-to-selective 
extinction, $\RV$.  In the Milky Way the value of $\RV$ typically varies between $\RV=2.2$--$5.8$, for different 
lines-of-sight \citep[\eg CCM,][]{1999PASP..111...63F}, despite this we allow extrapolations of $\RV$ within the 
range $\RV=0.5$--$8$ when we fit the laws to the \sn colours.


We also test a simple power-law model \citep{2008ApJ...686L.103G},
\[
A(\lambda;A_\mathrm{\Vband},a,\beta) =  A_\mathrm{\Vband}\left[1 - a + a\left(\frac{\lambda}{\lambda_\mathrm{\Vband}}\right)^\beta\right]\, ,
\]
where the reference wavelength, $\lambda_\mathrm{\Vband}$, will be chosen as $\lambda_\mathrm{\Vband} = 0.55~\mu$m.
The \ccmo and \ftz laws can be approximated by a power-law in the optical and NIR range, and \citet{2008ApJ...686L.103G} 
showed that the observed reddening law of an object embedded in circumstellar dust can be approximated by the expression
above with $a = 0.8$ ($a = 0.9$) and $\beta=-2.5$ ($\beta=-1.5$) for Milky Way (LMC) like dust.  The reason
why the original extinction laws are not preserved in the CS scenario is due to the geometry where multiply 
scattered photons on the CS dust will reach the observer while this is extremely unlikely for an ordinary 
interstellar dust geometry.  Note that the values in \citet{2008ApJ...686L.103G} were obtained under the assumption that 
the light source is constant in flux and colour.  For time-dependent light sources the observed reddening is expected to 
vary in time which has been investigated for \sneia by \eg \citet{2011ApJ...735...20A}.

When the power-law is fitted to colour excesses, $\EXV = \AX - \AV$, $\AV$ and $a$ are almost completely degenerate,
and we are in fact only sensitive to the product $a\cdot\AV$. We will for this reason fix $a$ to $a\equiv1$ in all fits below, 
which also has the implication that the resulting $\AV$ can no longer be interpreted as solemnly being the extinction in the 
$\Vband$-band.  We have further checked that fixing $a$ does not affect the fitted values of $\beta$, and that we obtain 
consistent values of the product $a\cdot\AV$ when $a$ is allowed as a free parameter, although the individual values of 
$a$ and $\AV$ are in this case of course poorly constrained.

In A14, the SALT2 \citep{2007AA...466...11G} law from \citet{2014A&A...568A..22B} was also studied.  This is a colour law and 
not an extincton law, and is further only defined for wavelengths $\lambda<8000$~\AA.  It can therefore not be compared on an
equal footing to the other laws and will not be studied in detail in this work.  We will however compare it to the reddening or our 
low extinction \sne in \S\,\ref{sec:origin}.

\subsection{Results}\label{sec:results}
The fitted parameters for the three extinction laws are presented in Table~\ref{tb:fitlawresults} for each individual \sn.  
The extinction laws were fitted to all available colours between phases $-10$ and $+35$ days and the predicted colour 
excesses using the best fitted \ftz laws are also shown in Figure~\ref{fig:afits}.
The revised photometry of \snjj and updated intrinsic colour model with respect to \jjpaper only had a minor impact
on the fitted results for this highly reddened \sn, and the values presented here are consistent with \jjpaper.
\begin{table*}
\centering
\begin{tabular}{ccr@{}lr@{}lr@{}lr@{}lr@{}lr@{}l}
\hline\hline
 &  & \multicolumn{2}{c}{\small 2012cg} & \multicolumn{2}{c}{\small 2012et} & \multicolumn{2}{c}{\small 2012cp} & \multicolumn{2}{c}{\small 2012bm} & \multicolumn{2}{c}{\small 2012cu} & \multicolumn{2}{c}{\small 2014J}\\
\hline\\[-2ex]
\multirow{6}{*}{\rotatebox[origin=c]{90}{\small F99}}
  & {\small $\EBV$ } & {\small $0.15$}&{\small$\,(0.02)$} & {\small $0.17$}&{\small$\,(0.03)$} & {\small $0.35$}&{\small$\,(0.03)$} & {\small $0.46$}&{\small$\,(0.04)$} & {\small $0.99$}&{\small$\,(0.03)$} & {\small $1.36$}&{\small$\,(0.02)$}\\[0.5ex]
  & {\small $\RV$ } & {\small $2.7$}&{\small$^{+0.9}_{-0.7}$} & {\small $1.7$}&{\small$^{+0.6}_{-0.5}$} & {\small $3.0$}&{\small$^{+0.4}_{-0.4}$} & {\small $3.0$}&{\small$^{+0.5}_{-0.4}$} & {\small $2.8$}&{\small$^{+0.1}_{-0.1}$} & {\small $1.4$}&{\small$^{+0.1}_{-0.1}$}\\[0.5ex]
  & {\small $\tBmax$ } & {\small $56080.6$}&{\small$\,(0.3)$} & {\small $56191.0$}&{\small$\,(0.5)$} & {\small $56081.1$}&{\small$\,(0.3)$} & {\small $56018.0$}&{\small$\,(1.9)$} & {\small $56104.5$}&{\small$\,(0.4)$} & {\small $56687.9$}&{\small$\,(0.2)$}\\[0.5ex]
  & {\small $\sB$ } & {\small $1.12$}&{\small$\,(0.02)$} & {\small $1.01$}&{\small$\,(0.03)$} & {\small $1.18$}&{\small$\,(0.02)$} & {\small $1.24$}&{\small$\,(0.12)$} & {\small $1.04$}&{\small$\,(0.03)$} & {\small $1.14$}&{\small$\,(0.01)$}\\[0.5ex]
  & {\small $\chi^2/\nu$ } & {\small $1.67$}& & {\small $1.50$}& & {\small $1.60$}& & {\small $2.83$}& & {\small $0.63$}& & {\small $2.38$}&\\[0.5ex]
\hline\\[-2ex]
\multirow{6}{*}{\rotatebox[origin=c]{90}{\small CCM$+$O}}
  & {\small $\EBV$ } & {\small $0.15$}&{\small$\,(0.02)$} & {\small $0.17$}&{\small$\,(0.03)$} & {\small $0.36$}&{\small$\,(0.03)$} & {\small $0.48$}&{\small$\,(0.04)$} & {\small $1.00$}&{\small$\,(0.03)$} & {\small $1.40$}&{\small$\,(0.02)$}\\[0.5ex]
  & {\small $\RV$ } & {\small $2.6$}&{\small$^{+0.8}_{-0.7}$} & {\small $1.4$}&{\small$^{+0.8}_{-0.7}$} & {\small $3.0$}&{\small$^{+0.4}_{-0.4}$} & {\small $2.8$}&{\small$^{+0.4}_{-0.4}$} & {\small $2.8$}&{\small$^{+0.2}_{-0.2}$} & {\small $1.2$}&{\small$^{+0.1}_{-0.1}$}\\[0.5ex]
  & {\small $\tBmax$ } & {\small $56080.6$}&{\small$\,(0.3)$} & {\small $56191.1$}&{\small$\,(0.5)$} & {\small $56081.1$}&{\small$\,(0.3)$} & {\small $56019.1$}&{\small$\,(1.7)$} & {\small $56104.7$}&{\small$\,(0.4)$} & {\small $56688.1$}&{\small$\,(0.2)$}\\[0.5ex]
  & {\small $\sB$ } & {\small $1.12$}&{\small$\,(0.02)$} & {\small $1.01$}&{\small$\,(0.03)$} & {\small $1.18$}&{\small$\,(0.02)$} & {\small $1.18$}&{\small$\,(0.10)$} & {\small $1.03$}&{\small$\,(0.03)$} & {\small $1.13$}&{\small$\,(0.01)$}\\[0.5ex]
  & {\small $\chi^2/\nu$ } & {\small $1.60$}& & {\small $1.34$}& & {\small $1.42$}& & {\small  $2.14$}& & {\small $0.91$}& & {\small $1.42$}&\\[0.5ex]
\hline\\[-2ex]
\multirow{6}{*}{\rotatebox[origin=c]{90}{\small Power-law}}
  & {\small $\AV$ } & {\small $0.44$}&{\small$\,(0.19)$} & {\small $0.30$}&{\small$\,(0.13)$} & {\small $1.51$}&{\small$\,(0.40)$} & {\small $1.47$}&{\small$\,(0.43)$} & {\small $3.59$}&{\small$\,(0.29)$} & {\small $1.97$}&{\small$\,(0.10)$}\\[0.5ex]
  & {\small $\beta$ } & {\small $-1.2$}&{\small$\,(0.4)$} & {\small $-1.7$}&{\small$\,(0.5)$} & {\small $-0.8$}&{\small$\,(0.2)$} & {\small $-1.2$}&{\small$\,(0.4)$} & {\small $-1.0$}&{\small$\,(0.1)$} & {\small $-2.0$}&{\small$\,(0.1)$}\\[0.5ex]
  & {\small $\tBmax$ } & {\small $56080.6$}&{\small$\,(0.3)$} & {\small $56191.1$}&{\small$\,(0.5)$} & {\small $56081.2$}&{\small$\,(0.3)$} & {\small $56020.4$}&{\small$\,(2.7)$} & {\small $56104.8$}&{\small$\,(0.4)$} & {\small $56688.1$}&{\small$\,(0.2)$}\\[0.5ex]
  & {\small $\sB$ } & {\small $1.12$}&{\small$\,(0.02)$} & {\small $1.00$}&{\small$\,(0.03)$} & {\small $1.18$}&{\small$\,(0.02)$} & {\small $1.16$}&{\small$\,(0.15)$} & {\small $1.02$}&{\small$\,(0.03)$} & {\small $1.13$}&{\small$\,(0.01)$}\\[0.5ex]
  & {\small $\chi^2/\nu$ } & {\small $1.79$}& & {\small $1.17$}& & {\small $1.95$}& & {\small $2.13$}& & {\small $1.78$}& & {\small $1.65$}&\\[0.5ex]
\hline\hline\\[-2ex]
\end{tabular}

\caption{%
   The best fitted parameters to all measured colours between phases $-10$ and $+35$ days for the reddening laws investigated 
   in this work.  For the power-law we have fixed $a$ to $a\equiv1$ to break the degeneracy between this parameter
   and $\AV$. For each value we quote the $68\,\%$ level fitted uncertainty when the parameters are considered individually.  
   The degrees of freedoms were calculated under assumptions that the colours are independent measurements, which is
   not the case.  \label{tb:fitlawresults}}
\end{table*}

Further, all three reddening laws capture the general wavelength dependence of the observed colour excesses with the 
possible exception of \sncg which will be discussed below.  This shows that the observed \sn colours from UV to NIR can 
indeed be described by the colours of \snfe, with the adopted intrinsic dispersions, together with an extinction law with only 
two free parameters.

\begin{figure*}
  \centering
    \includegraphics[width=\textwidth]{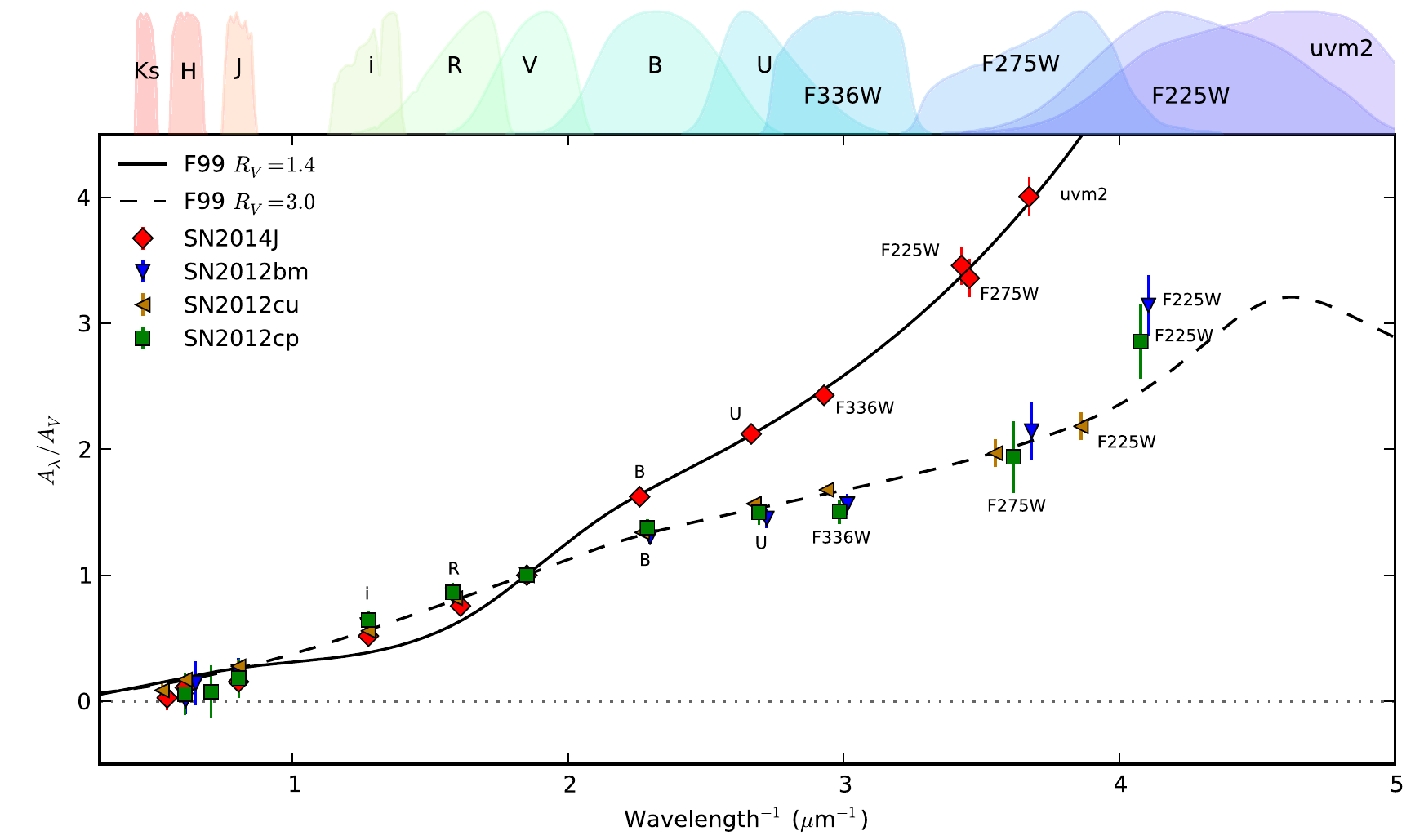}
    \caption{%
         The normalised extinction, $\AX/\AV$ for the four \sne with $\EBV>0.2$~mag. Here,
         $\AX/\AV = 1 + \EXV/\AV$ where $\EXV$ is calculated as the weighted (with the measurement uncertainties) 
         averages for all epochs, and $\AV$ is obtained from the fitted \ftz parameters in Table~\ref{tb:fitlawresults}.  
         Following \jjpaper, each data 
         point has been plotted at the wavelength where the residual to the extinction law is matching the corresponding 
         residual between $\EXV$ and the fitted reddening law in Figure~\ref{fig:afits}. Note that all fitted parameters 
         described in this work were obtained by minimising Equation~\eqref{eq:chi2}, and {\em not by fitting the extinction 
         laws to the data points in this plot.}  The filter passbands for the filters used for the minimisation are shown at the top.
         \label{fig:extdiversity}}
\end{figure*}

The six \sne span a broad range both in reddening, $\EBV=0.2$--$1.4$~mag, and in $\RV=1.4$--$3.0$ which confirms 
the findings from previous optical and NIR studies (\eg B14), that point to a diversity of observed reddening laws of 
\sneia.  The significant difference in the derived extinction laws is also illustrated in Figure~\ref{fig:extdiversity} where 
the relative extinction,  $A_\lambda/\AV$, for the \ftz law has been plotted for two different values of 
$\RV$. The diversity is particularly striking blue-wards of the \Uband-band ($\lambda^{-1} > 2.5~\mu\textrm{m}^{-1}$), 
emphasising the power of using UV data to study diversity in extinction.

In the Figure~\ref{fig:extdiversity} we also show the relative extinction, $\AX/\AV = 1 + \EXV/\AV$, for the four \sne 
with $\EBV>0.2$~mag against inverse wavelength.  Here, $\EXV$ is calculated as the weighted average for all phases 
and $\AV$ was obtained from the fitted \ftz parameters as $\AV=\RV\cdot\EBV$.  Because of this, the derived values for 
$\AX/\AV$ are model-dependent and will be different for the different extinction laws.   Further, since the extinction $\AX$ is 
obtained from comparing observations in the filter \Xband of reddened and unreddened sources,  and the measurements 
correspond to different effective wavelengths, it is non-trivial to derive a specific wavelength corresponding to $\AX/\AV$.  
For these reasons, we follow the procedure from \jjpaper and plot the $\AX/\AV$ points in a manner where we chose to
preserve the residuals between the fitted model and the data points from Figure~\ref{fig:afits}.  This is obtained by plotting
each point, $\AX/\AV$, at the wavelength where the residual to the extinction law is the same as the corresponding weighted 
average residual of $\EXV$ over all epochs. We emphasise that this plot is only a way of visualising the data and the reddening
laws and that all extinction parameters were obtained by minimising Equation~\eqref{eq:chi2}, \ie not by fitting the extinction 
laws to the data points in Figure~\ref{fig:extdiversity}.

\begin{figure}
  \centering
    \includegraphics[width=\columnwidth]{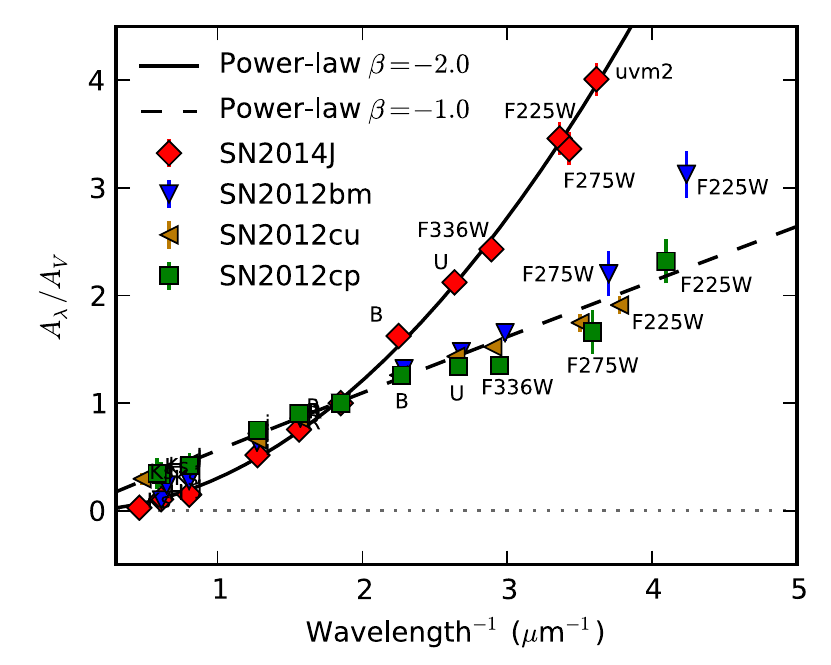}
    \caption{%
    	 	 Similar to Figure~\ref{fig:extdiversity} but here the power-law fits from Table~\ref{tb:fitlawresults} are shown with
		 the observed reddening.  The wavelengths of the data points will differ from Figure~\ref{fig:extdiversity} due to
		 the way these plots are created, as explained in the text.  Note that all fitted parameters 
         described in this work were obtained by minimising Equation~\eqref{eq:chi2}, and {\em not by fitting the extinction 
         laws to the data points in this plot.} 
         \label{fig:powerdiv}}
\end{figure}

One observation that can be made by comparing the results in Table~\ref{tb:fitlawresults} and Figure~\ref{fig:afits} is that 
although all extinction laws appear to be general enough to describe all \sne, the \ftz law shows tension for 
\sne \bm and \jj.  These \sne have in common that they both show a fair amount of reddening ($\EBV > 0.5$~mag).
The more dust along the line-of-sight to a \sn, the more sensitive the fit will be to the extinction law, and less to 
possible intrinsic \sn colour variations.  In other words, we can expect any deviation between the assumed 
parametrisation and the actual observed extinction law to become more significant the redder the \sne are.
It has already been pointed out that since the \ftz law is empirical in nature, it may have features that do not properly 
extrapolate to low values of $\RV$.   This could be the case for \snjj, where the \ftz law with $\RV=1.4$
leads to a more prominent feature around $\sim1.5~\mu\mathrm{m}^{-1}$ ($\sim6700$~\AA) than what is supported by 
the data.   In Figure~\ref{fig:powerdiv} we compare the power-law fits for the four most reddened \sne, and here we can 
see how this is consistent with all the data points over the full wavelength range for \snjj.

From the reduced $\chi^2$ in Table~\ref{tb:fitlawresults} we can conclude that the power-law model provides better fits 
than \ftz for both low-$\RV$ \sne, \et and \jj, while the opposite is true for the $\RV\lesssim3$ \sne \cg, \cp and \cu.
In particular, the \ftz and \ccmo law seem to provide nearly perfect fits to the observations of \sncu, which is also seen 
in Figure~\ref{fig:extdiversity}.  \snbm shows tension for all fitted laws. 

The \ccmo law performs better in terms of goodness-of-fit for several cases, \eg for \snjj.  However, this law does not extrapolate 
well to low $\RV$ values which give rise to a sharp, discontinuous, knee when the NIR and optical parametrisations of the law 
are merged.

\section{Searching for circumstellar dust}\label{sec:cs}
All the \sne studied here show a colour excess evolution that is within the adopted dispersion with the exception of
\sncg. This \sn reveals a minimum in its NIR colour evolution around two weeks past max which is possibly accounted to 
the diversity observed in NIR \snia light curves  between the first and second peaks.  This is within 
the assumed intrinsic uncertainty although the assumption that these are fully correlated between epochs does not hold 
in this case. What is perhaps even more striking are the $\sim2\sigma$ 
variations of the $\mathrm{\uvmtwo}-\mathrm{\Vband}$ evolution which suggests that this colour evolve slower 
than the corresponding \fe colour, resulting in an apparent decrease with time of the colour excesses.

The fact that the observed colour evolution of \sncg is not consistent with a single extinction law is also illustrated by
Table~\ref{tb:extphase} where the \ftz law has been fitted for two phase intervals: around maximum and between 
10 and 20~days past $\tBmax$.  The fits were carried out for the three \sne that had UV data in both these intervals 
while keeping the fitted values of $\tBmax$ and $\sB$ fixed to the values given in Table~\ref{tb:fitlawresults}.
The fitted parameters of \sne \cp and \jj are within errors between the two epochs while this is not the case for \sncg.
\begin{table}
\centering
\begin{tabular}{clr@{}lr@{}lc@{}l}
\hline\hline\\
\multicolumn{1}{c}{\small \sn} & \multicolumn{1}{c}{\small Phase} &\multicolumn{2}{c}{\small $\EBV$} &\multicolumn{2}{c}{\small $\RV$} &\multicolumn{2}{c}{\small $\chi^2/\nu$}\\
\hline\\[-2ex]
\multirow{2}{*}{\small SN2012cg} & $[-5,+5]$ &{\small $0.16$}&{\small$\,(0.02)$} & {\small $1.7$}&{\small$^{+0.8}_{-0.6}$} & {\small $1.51$}\\[0.5ex]
 & $[+10,+20]$ &{\small $0.13$}&{\small$\,(0.03)$} & {\small $3.6$}&{\small$^{+1.5}_{-1.0}$} & {\small $1.34$}\\[0.5ex]
\hline\\[-2ex]
\multirow{2}{*}{\small SN2012cp} & $[-5,+5]$ &{\small $0.33$}&{\small$\,(0.03)$} & {\small $1.8$}&{\small$^{+0.6}_{-0.5}$} & {\small $1.63$}\\[0.5ex]
 & $[+10,+20]$ &{\small $0.34$}&{\small$\,(0.03)$} & {\small $2.2$}&{\small$^{+0.8}_{-0.6}$} & {\small $0.55$}\\[0.5ex]
\hline\\[-2ex]
\multirow{2}{*}{\small SN2014J} & $[-5,+5]$ &{\small $1.35$}&{\small$\,(0.03)$} & {\small $1.5$}&{\small$^{+0.1}_{-0.1}$} & {\small $1.81$}\\[0.5ex]
 & $[+10,+20]$ &{\small $1.32$}&{\small$\,(0.03)$} & {\small $1.6$}&{\small$^{+0.1}_{-0.1}$} & {\small $1.56$}\\[0.5ex]
\hline\hline\\[-2ex]
\end{tabular}

\caption{%
   Fitted \ftz parameters using only data between  $-5$--$+5$ and $+10$--$+20$~days from maximum, respectively. 
   While \sne \cg and \jj both have UV--NIR observations for both intervals, the fits to \sncp are only based on UV--optical
   data. The parameters $\tBmax$ and $\sB$ were fixed to the values given in Table~\ref{tb:fitlawresults}.
   \label{tb:extphase}}
\end{table}

Evolving colour excesses, or a reddening law that changes with time, is predicted if circumstellar dust 
is present in the \sn environment. 
\citet{2011ApJ...735...20A} studied this effect at optical wavelengths and we can extend that analysis, using the 
same tools, to also cover UV and NIR in order to test the observed colour evolution of \sncg against the 
expectations for a CS scenario.

\begin{figure}
\centering
\includegraphics[width=\columnwidth]{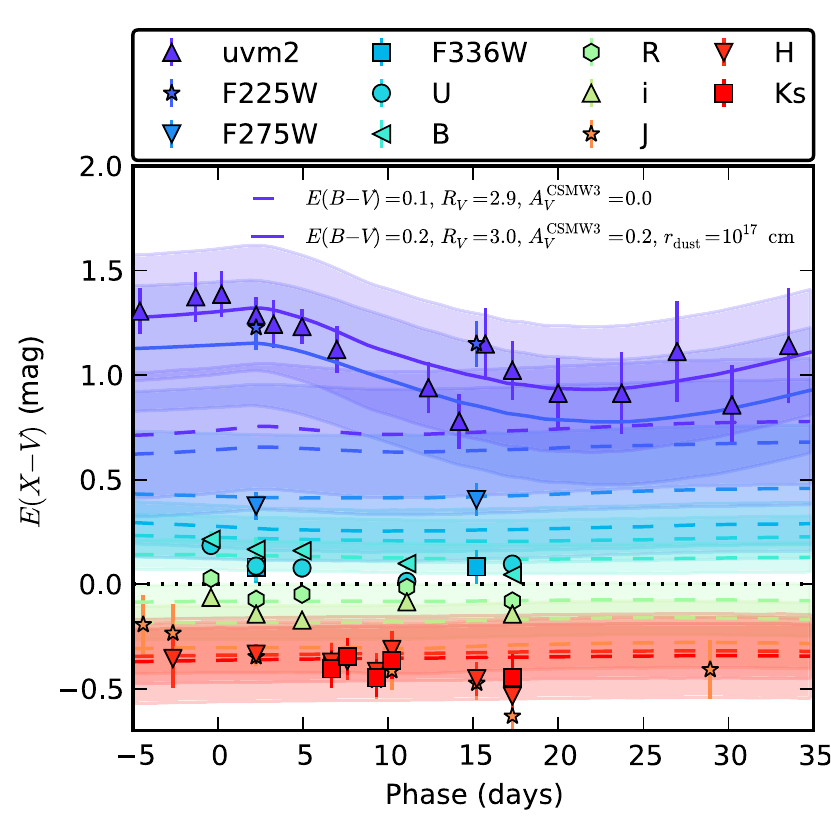}
\caption{
  The measured colour excesses for different phases for \sncg are taken from Figure~\ref{fig:afits}.  Overplotted are the
  best fits of a combined circumstellar and \ftz dust law, where the former has been adopted from \citet{2011ApJ...735...20A}
  where the dust is distributed in a thin spherical shell.  The solid lines shows the best fit when only the 
  $\mathrm{\uvmtwo}-\mathrm{\Vband}$ and $\mathrm{\wfcone}-\mathrm{\Vband}$ colours are considered, while the dashed 
  lines show the fit to all colours.  The former is obtained with a CS dust shell at radius $r_\mathrm{dust}\sim10^{17}$~cm from
  the \sn that give rise to an extinction of $\AV\sim0.2$, while latter fit is consistent with no CS dust.
  \label{fig:cgcs}}
\end{figure}

In Figure~\ref{fig:cgcs} the same colour excesses of \sncg that are plotted in the upper left panel of Figure~\ref{fig:afits} are 
shown.  The solid lines represent the result when we fit a combined \ftz and CS dust law to the bluest colours, 
$\mathrm{\uvmtwo}-\mathrm{\Vband}$ and $\mathrm{\wfcone}-\mathrm{\Vband}$, while keeping $\tBmax$ and $\sB$ fixed 
to the values from Table~\ref{tb:fitlawresults}.  The best fit values for the \ftz law are $\EBV\approx0.2$~mag and 
$\RV\approx3.0$.  The CS dust in this case consists of dust modeled to match the average extinction properties in the Milky 
Way \citep{2003ApJ...598.1017D} and distributed in a thin spherical shell.  Here the best fit yields an extinction of
 $\AV^\mathrm{CSMW3}\approx0.2$ for a shell at a radius of $r_\mathrm{dust}\sim10^{17}$~cm. 

Although adding Milky Way-like CS dust to the picture can give an explanation for the time-evolution of the bluest colours, it does 
not provide an improvement beyond a single \ftz law when all colours are considered simultaneously, which is shown by the 
dashed lines in Figure~\ref{fig:cgcs}.  Here the best fit is obtained for $\AV^\mathrm{CSMW3}\approx0.0$.  The explanation for
this is that the specific CS dust model we test also predicts evolution of the redder colours  which in this case is not supported 
by the data.  If CS dust is present, the total extinction, $A_\lambda$, can only decrease with time due to late-arriving 
scattered photons.  From the results in Table~\ref{tb:extphase}, we see that this is not the case for $\AV$,  although the value 
of $\AV$ is in this case model dependent.


It is also possible to detect CS dust by studying the time-evolution of multi-epoch high-resolution spectroscopy.  
This was mentioned in \S\,\ref{sec:intro} and is the observational signature that has been used to claim detections of 
CS dust around \sneia.  

\begin{figure*}
  \begin{center}
    \includegraphics[width=\textwidth]{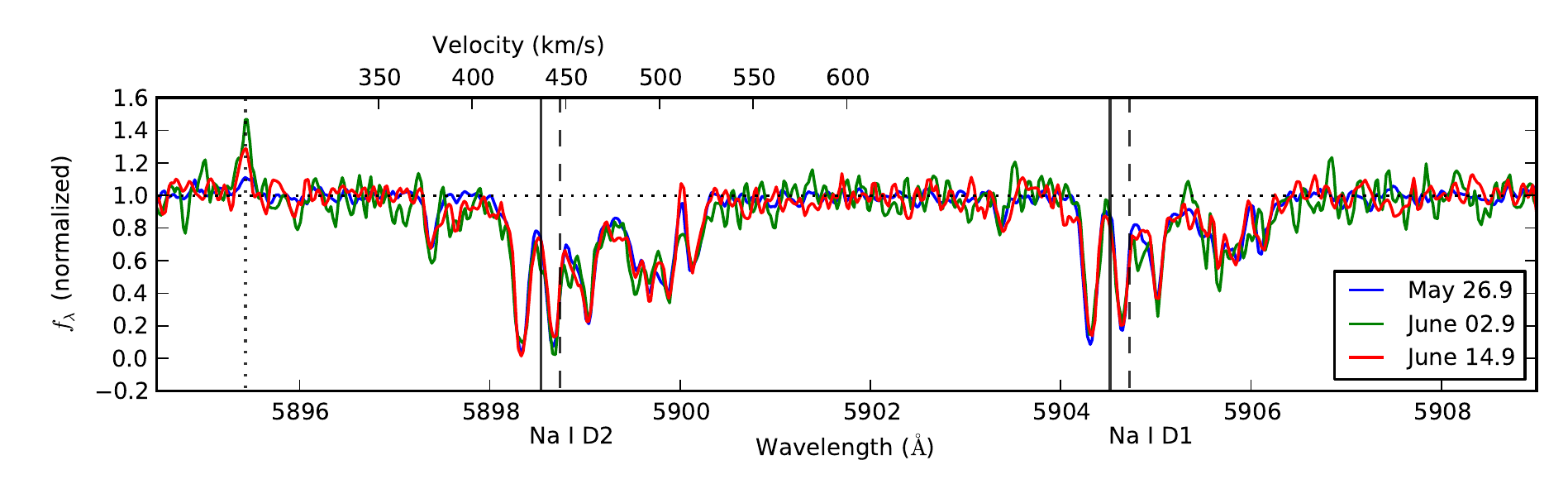}
    \caption{\NaID absorption doublet of SN 2012cg. The rest frame wavelength of \NaID in NGC~4424 
      (solid lines) and for the approximate stellar velocity along the line-of-sight (dashed) are indicated. The 
      dotted vertical line indicates a sky emission line. In Section~\ref{sec:cs} the profile of \NaID is discussed in detail. 
      \label{fig:cgnaid}}
  \end{center}
\end{figure*}

For \sncg we obtained high-resolution FIES spectra for three epochs which are listed in Table~\ref{tb:spec} and 
correspond to phases $-7$, $1$ and $13$ days with respect to $\tBmax$.  All spectra contain well resolved 
unsaturated \NaID features (Figure~\ref{fig:spectra}), which are shown in Figure~\ref{fig:cgnaid}. The bulk of the 
distinguishable features are likely part of the visible interstellar medium (ISM), but it can not be excluded that \NaI 
associated with CS medium could contribute to the observed profile.  A full description of the \NaID profile and summary 
of other interstellar absorption features is presented \S\,\ref{sec:highres}.


The FIES spectra were normalised by fitting 3rd-order polynomials to the continua bracketing \NaID. No telluric 
corrections were performed on the spectra, due to the lack of standard star observations on some of the epochs. 
We fitted Gaussian profiles to the weighted average of the three epochs, and find that the \NaID line ratios 
(D1$/$D2) range from $2$--$1.2$ for individual features.  The deepest features have low ratios, indicating that 
they are not optically thin. Based on a standard star spectrum taken on the second epoch, we identify telluric 
features overlapping with \Dtwo, where as \Done appears to be located at a less affected part of the spectrum. 
We therefore focus most of the time-variation analysis on \Done, because it is less contaminated by telluric 
features and it is more likely to be optically thin due to the lower absorption cross-section of this transition.
We measure the total equivalent width of \Done of the respective epochs to be, $702\pm9$, $711\pm32$ and 
$685\pm20$ m\AA.

Since no significant time-variations of \NaID absorption between different epochs could be detected beyond the 
noise level, we attempt to use the non-detections to constrain possible CS dust models assuming a thin sphere
of CS material shell distribution around the \sn. Using the \NaI photo-ionisation model described by 
\citet{2009ApJ...699L..64B} 
and recently applied to \snjj in \citet{2015ApJ...801..136G}, we can exclude CS material, and thus also CS dust, at 
certain radii from the \sn. If 
the \NaI is optically thin, a change in equivalent width of \NaID is directly proportional to a decrease in column density. 
Furthermore, the product of the integrated photon count flux above the ionisation energy of \NaI ($2400$ \AA) and its 
corresponding ionisation cross-section is proportional to the fractional decrease in column density. We can thus compute 
the fractional decrease in equivalent width of \NaID expected due to photo-ionisation of CS material at a given radius 
from a \sn. In this model, CS \NaI can be excluded which is far enough away from the \sn to not have been fully ionised 
by the first epoch and close enough to see changes by the following epochs. The diversity of \sneia spectra in UV 
implies that the results will be sensitive to the spectral templates used. Noting that the \stis spectra of \sncg resemble 
those of \snfe, we use the \snfe spectral template described in Appendix~\ref{sec:femodel}.
However, the H07 template extends to shorter wavelengths and earlier times than the \snfe template. 
We therefore extrapolate our spectral template by H07, whereby the flux is rescaled for continuity. 
This slightly increases the photo-ionising flux.

\begin{figure}
  \begin{center}
    \includegraphics[width=\columnwidth]{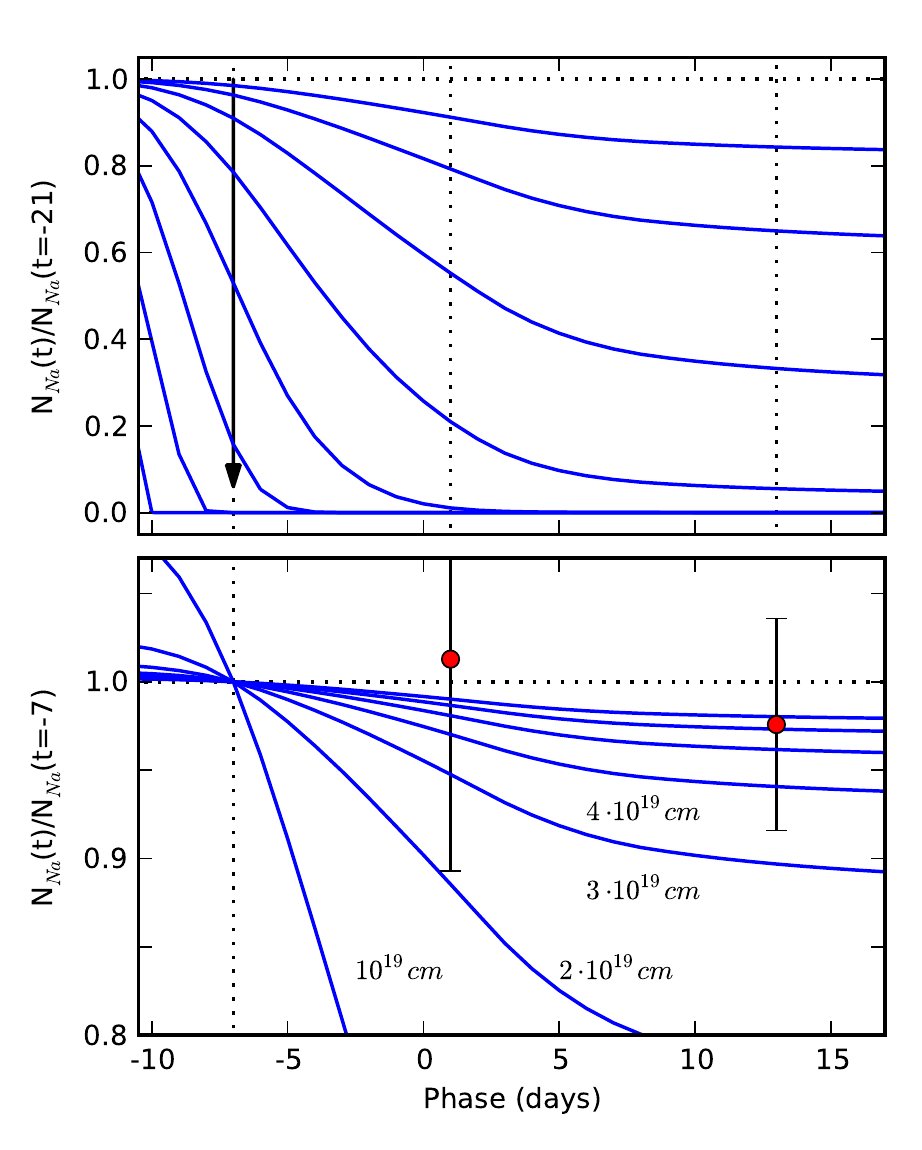}
    \caption{%
      	\NaI ionisation models for thin shells of CS material at different distances from a \sn, suggesting 
        the absence of circumstellar material at distances between $r_\mathrm{dust}=2\cdot10^{18}$ -- $3\cdot10^{19}$~cm 
        around \sncg. The upper panel shows the fractional decrease of \NaI due to ionization as a function of time at radii 
	  	ranging from $r_\mathrm{dust}=10^{18}$ -- $3\cdot10^{19}$~cm. The arrow indicates an approximate upper 
		limit for Na~I around \cg to be ionised beyond detection level by $-8$ days from \Bband maximum based on 
		the S/N around the \NaID doublet of the first FIES spectrum. The lower panel shows the fractional change 
		of \NaI with respect to the first epoch. The fractional change of equivalent width of \Done 
		with $3\sigma$ error bars set an approximate lower radius limit at which CS material could be present 
		while the ionisation effects remain undetected. 
    	\label{fig:cg_ion}}
  \end{center}
\end{figure}

In Figure~\ref{fig:cg_ion} the model fractional decrease of \NaI at different radii from a \sn are shown.  We determined 
the inner radius of excluded CS \NaI of \cg by assuming that an absorption feature deeper than three times the 
root-mean-square of the noise would be detectable in the earliest spectrum. This yields a distance of 
$r_\mathrm{dust}<2\cdot10^{18}$ cm within which the \NaI column density would have decreased below detection 
levels since explosion.  With respect to the first epoch, the fractional change of the equivalent width of \Done is considered 
to set an upper limit on the change in column density. Within $3\sigma$ errors, the last epoch excludes the models in 
which the dust is located closer than $r_\mathrm{dust}>3\cdot10^{19}$ cm from \sncg. Using the total equivalent width 
of \Done of course has the caveat that it considers the entire measured column density to be situated at one radius 
from the \sn, which we know cannot be true based on the profile. We therefore also consider how much a single feature 
must change to be detected given the signal-to-noise of the spectra. The limits obtained by the absence of change 
beyond three times the root-mean-square noise are comparable to those set by the fractional change of the total 
equivalent width. Lastly we note that the limits are not valid if there is CS material that is not optically thin, since the 
equivalent width then does not change linearly with column density. 
 
In summary, although the existence of Milky Way-like CS dust at $r_\mathrm{dust}\sim10^{17}$~cm is consistent 
with the observed time-evolution of $E(\mathrm{\uvmtwo}-\mathrm{\Vband})$ and non-detection of time-varying 
\NaID absorption for \sncg, the minimal $E(\mathrm{\Xband}-\mathrm{\Vband})$ evolution of the redder 
wavelengths disfavours this explanation.   
This suggests that CS dust can only be present either in very small amounts, $\AV^\mathrm{CSMW3}<0.05$ 
or if it is located at larger radii $r_\mathrm{dust}\sim10^{19}$~cm for which the time-delays are much larger than 
the time-scales studied here and are not expected to affect the observed colours \citep{2011ApJ...735...20A}.
Further, non-detections in radio \citep{2012ATel.4453....1C} and far-IR \citep{2013MNRAS.431L..43J} are
other observational signatures suggesting that \sncg exploded in an environment free from CS material. Here
the far-IR observations are sensitive to pre-existing CS dust that would be heated by the explosion.

\section{Discussion}\label{sec:discussion}

\newcommand{\pW}{{\it pW}}
\subsection{Using \snfe as a colour template}\label{sec:fetemplate}
One possible explanation for the time-evolution of the \sncg $UV-V$ colour excess could be that it is
intrinsically different from \snfe. We have already argued that all the \sne in our sample, are normal \sneia by 
comparing their overall spectra (see Figure~\ref{fig:spectra}) and their SNID classification,  but we can also 
study individual features which has proven to be a useful approach for sub-classifying normal \sneia.

The properties of individual features can be quantified by \eg measuring their velocities and pseudo-equivalent 
widths (\pW), where pseudo refers to the fact that the equivalent widths are obtained using a pseudo continuum 
since proper continua are absent in \snia spectra. The pseudo continuum can be defined with a straight line 
between the two flux peaks surrounding an absorption feature and the \pW\ is then calculated as the integral of 
the spectrum flux relative to the continuum.  Using this method, the error of the measurement will typically be 
dominated by the systematic uncertainty introduced by the determination of the pseudo-continuum.  

\citet{2006PASP..118..560B} suggested a classification scheme based on pseudo-equivalent width measurements 
of the absorption features near $5750$~\AA\ and $6100$~\AA\ (following the naming convention from 
\citet{2007A&A...470..411G} we refer to these as \pW6 and \pW7 respectively), that can be associated with 
\SiII\,$\lambda5972$~\AA\ and \SiII\,$\lambda6355$~\AA\ as shown in Figure~\ref{fig:spectra}.  They identify four 
different groups when considering \pW6\ and \pW7\ for their sample where in particular the "core normal" (CN) \sne, 
to which \snfe belongs, are tightly clustered and show a high degree of general spectral homogeneity.

In Table~\ref{tb:pwv} we show measurements of \pW6 and \pW7 for spectra close to maximum
after correcting our spectra for host galaxy reddening \citep[although this is not expected to have any major
impact on the results as discussed in][]{2011A&A...526A.119N} using the fitted reddening laws from 
\S\,\ref{sec:results}.  We also present the \pW\ for \CaHK \citep[\pW1 following][]{2007A&A...470..411G} 
and note that all \sne have $\pW1 < 150$~\AA.  It has been shown that $\pW1$ correlates with the 
intrinsic $\mathrm{\Uband}-\mathrm{\Bband}$ colour \citep{2013ApJ...773...53F,2014ApJ...789...32B} 
where \sne with $\pW1 > 150$~\AA\ are intrinsically redder than the bulk of normal \snia.

\begin{table}
\centering
\begin{tabular}{l r r@{\,}l r@{\,}l r@{\,}l c}
\hline\hline
 & \multicolumn{1}{c}{Phase} & \multicolumn{2}{c}{\pW1} & \multicolumn{2}{c}{\pW6} & \multicolumn{2}{c}{\pW7} \\
 \multicolumn{1}{c}{\sn} & \multicolumn{1}{c}{(days)} & \multicolumn{2}{c}{(\AA)} & \multicolumn{2}{c}{(\AA)} & \multicolumn{2}{c}{(\AA)} &
 \multicolumn{1}{c}{Type}\\
\hline
\fe   & +0 &   94&(1)  & 14&(1) &  94&(1) & CN\\
\bl   & -1  & 110&(2)  &  1&(1) &   89&(2) & CN\\
\bm & +9 & 125&(2)  & 13&(1) &  76&(2) & CN\\
\cg  & +2 & 103&(4)  &   9&(3) &  75&(4) & CN\\
\cp  & +2 &   98&(1)  &  10&(1) &  74&(1) & CN\\
\cu  & +9 &    70&(1) &  24&(1) &  99&(1) & CN\\
\et   & +2 & 123&(2)  &  11&(1) & 133&(2) & BL\\
\jj     & -1 & 155&(2)  & 14&(1) & 100&(3) & CN\\
\hline\hline
\end{tabular}
\caption{%
	Calculated pseudo-equivalent widths (\pW) based on the optical spectra.  The \pW\ naming convention 
	from \citet{2007A&A...470..411G} is followed with \pW1 corresponding to \CaHK, and \pW6 and \pW7 
	corresponding to \SiII\,$\lambda5972$~\AA\ and \SiII\,$\lambda6355$~\AA\ respectively. 
	The quoted uncertainties are the statistical errors.  Systematic uncertainties from estimating the
	pseudo continuum could be dominating in some cases.
	The classification as "core normal" (CN) or "broad-line" (BL) is based on the criteria from 
	\citet{2012AJ....143..126B}.  Given that these criteria were derived based on spectra around maximum,
	the classification of \sncu should be considered less certain. The measurement of \snjj was carried out
	on a spectrum published in \citet{2014ApJ...784L..12G}.
	See the text for details.%
	\label{tb:pwv}}
\end{table}

Using the classification criteria from \citet{2012AJ....143..126B} we can classify all \sne as CN except for 
\snet which lands in range that \citet{2006PASP..118..560B} defines as the "broad-line" (BL) group.  These 
\sneia have broader and deeper $6100$~\AA\ absorption but are in most aspects not very different from 
CN \sne.  When the observed spectra from the two groups are compared with synthetic spectra 
generated with SYNOW \citep{2003AJ....126.1489B} similar photospheric velocities and excitation temperatures 
can be used to describe both the CN and BL groups  \citep{2006PASP..118..560B}.

Given these results, there is no evidence for any significant discrepancy between the optical 
properties of our \sne and \snfe, with the exception for the 
additional source of luminosity observed in the very early optical lightcurve
of SN \jj \citep{2015ApJ...799..106G}.  
  However, several studies have shown that normal \sne in the optical still show 
dispersion in the UV and M13 argued that their sample could be divided in up to four groups.  Here, \snfe 
belongs to what they define as the ''near-UV (NUV) blue'' group which they conclude is on average 0.44~mag bluer in 
$\mathrm{\uband}-\mathrm{\vband}$ than the reddest, "NUV-red",  group.  They consider \sne with 
$\EBV<0.25$~mag, and this constraint could still allow extinction to contribute with up to $0.9$--$1.3$~mag 
(depending on the extinction law) to the $\mathrm{\uvmtwo}-\mathrm{\vband}$ colour, which may offer an 
explanation for the significant colour dispersion.  However, M13 show that the two groups are distinguishable 
even after the individual \sne have been corrected for reddening. 

Further, they also conclude that the groups, on average, show different spectroscopic properties. All NUV-blue 
\sne have ''normal'' photospheric expansion velocities around maximum when quantified based on the 
\SiII~$\lambda6355$~\AA\ feature, while the remaining groups are populated with both ''normal'' (NV) and 
''high'' (HV) velocity \sne.  They use the definition from \citet{2009ApJ...699L.139W}, where \sne with 
\SiII velocities of $v_\mathrm{\SiII} > 11,800~\mathrm{km}\,\mathrm{s}^{-1}$ at maximum are classified as HV and the 
remaining as NV.   Although there may not be a sharp distinction between the two classes 
\citep[see \eg][]{2012MNRAS.425.1819S},  the M13 results are consistent with previous studies showing that an optical 
colour-velocity dependence exists with HV \sne being on average intrinsically redder 
\citep{2011ApJ...729...55F,2012AJ....143..126B,2013ApJ...773...53F,2014ApJ...797...75M}.
On the other hand, exceptions from these findings is, \snname{2011de}, the brightest \snia observed \citep{2014ApJ...796L..18B},
and the results by \citet{2012ApJ...749..126W},  showing, using the \hst Advanced Camera for Surveys,  that the 
UV-optical colours of the HV object \snname{2004dt} are significantly bluer than the NV \snname{2005cf}.

\begin{figure}
\centering
\includegraphics[width=\columnwidth]{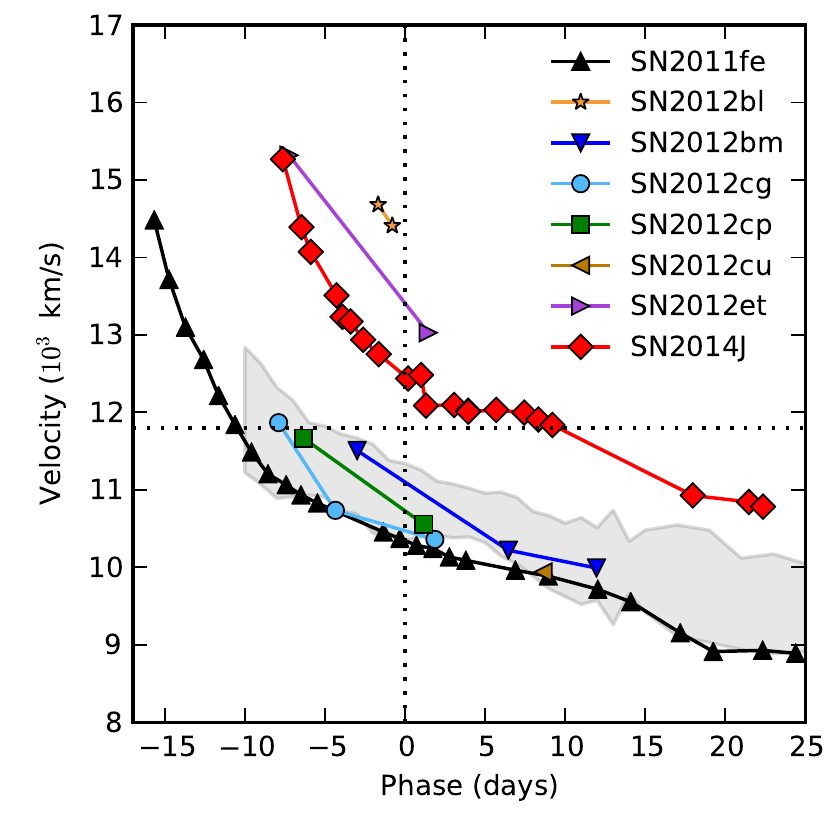}
\caption{%
  Measured velocities of \SiII$\lambda6355$ for different phases of the \sne studied in this work together with
  reference objects. For \sne \fe and \jj the velocities were calculated based on the spectra published by
  \citet{2013A&A...554A..27P}, \citet{2014ApJ...784L..12G} and \citet{2015ApJ...798...39M}.
  The grey region is marking the average \snia velocity range \citep{2012ApJ...745...74F,2013ApJ...773...53F}.  
  The dashed line marks the difference used to separate high and low velocity objects by \citet{2009ApJ...699L.139W} at 
  maximum.
\label{fig:velocity}}
\end{figure}

We have measured the \SiII velocities, which are presented in Figure~\ref{fig:velocity}, for all spectra we have 
obtained of the \sne at $t<+25$~days from $\tBmax$ by fitting Gaussian profiles.  
From the figure we conclude that \sne \fe, \bm, \cg and \cp show  similar velocity evolutions, and the single
measurement we have of \sncu is also consistent with \snfe.  The \sne \bl, \et and \jj, on the other hand,
are classified as HV. Note however that the velocity range for the full sample is still small enough 
for the scatter induced by the optical colour-velocity relation to land within the adopted uncertainty of 
$0.1$~mag for $\mathrm{\Bband}-\mathrm{\Vband}$ \citep{2014ApJ...797...75M}.  
\citet{2014ApJ...797...75M} further find no evidence for such a relation for the 
$\mathrm{\Vband}-\mathrm{\Rband}$ and $\mathrm{\Vband}-\mathrm{\iband}$. 

\snet is both classified as BL (although the \pW6 and \pW7 values put its close to the border of
the BL and CN groups) and HV, and such objects have been shown to be redder than the bulk  
of normal \sneia \citep{2013ApJ...773...53F,2014ApJ...789...32B}. However, the value of \pW1 for 
\snet at $-8$~days
is \pW1$=82\pm7$\AA, which is well below \pW1$\sim150$~\AA\ where the BL HV start to deviate in 
colour \citep[see Figure~17 in][]{2014ApJ...789...32B}.   
%

M13 also determine that NUV-blue \sne show evidence of unburned \CII in their optical spectra while
the remaining groups consist of objects both with and without unburned \CII detections. These results are 
also consistent with optical spectroscopic studies where \sneia with signatures of unburned carbon are bluer in their 
optical colours \citep[\eg][]{2011ApJ...743...27T,2012ApJ...745...74F,2012MNRAS.425.1917S}. 
Unburned carbon can be detected through absorption attributed to the most prominent line in the optical, 
\CII~$\lambda6580$, but the prospects of detecting this line decreases for epochs approaching maximum light
\citep[see \eg Figure~11 in][]{2012ApJ...745...74F}.
The \sne \fe, \cg and \jj have all evidence for unburned material in their spectra 
\citep{2011Natur.480..344N,2012ApJ...756L...7S,2015ApJ...798...39M}, while the first spectra of the remaining objects
were all obtained $>-7$~days with respect to $\tBmax$.

\begin{figure}
\centering
\includegraphics[width=\columnwidth]{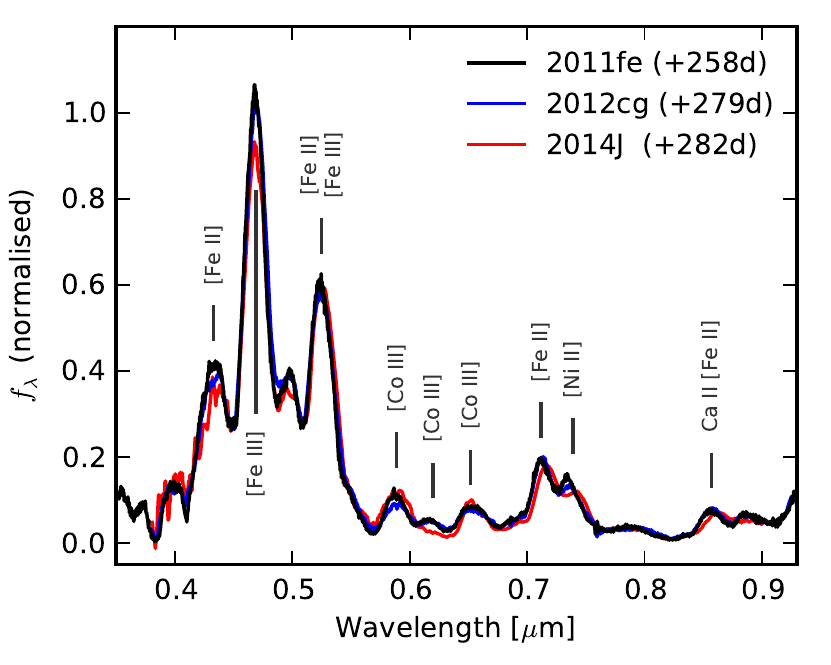}
\caption{%
Nebular spectra of \sne \fe, \cg and \jj obtained with the William Herschel Telescope, the Keck and the Apache Point
Observatory.   The epochs of the spectra with respect to maximum is specified in the legend.  The spectra of \sne 
\cg and \jj were corrected for reddening and smoothed using a 5~pixel boxcar, and then scaled to match the spectrum 
of \snfe. The main features of the nebular spectra have been marked.
\label{fig:nebular}}
\end{figure}

Finally, we can also compare the optical spectra of \sne \fe, \cg and \jj at the late-time nebular phase when the material
in the \sn ejecta is optically thin.  In Figure~\ref{fig:nebular} we present spectra of the three \sne at $260$--$280$~days past 
maximum after they have been corrected for reddening.  The spectra are dominated by forbidden lines from the decay
chain $^{56}\mathrm{Ni}\rightarrow$$^{56}\mathrm{Co}\rightarrow$$^{56}\mathrm{Fe}$.  The spectra of \sne \cg and \fe are
remarkably similar while for \snjj the [Co~\textsc{iii}] line at $6200$~\AA\ is suppressed and the features at $7000$--$7400$~\AA\ 
are redshifted with respect to the others.   The two peaks identified as [Fe~\textsc{ii}] $\lambda$7155 and 
[Ni~\textsc{ii}] $\lambda$7378 \citep[\eg][]{2010Natur.466...82M,2010ApJ...708.1703M}, are consistently blueshifted with 
respect to their rest frame by $800$-$1000$~km/s in \sne \fe and \cg. In contrast, these lines appear redshifted in \snjj by 
a similar amount. This is in line with the findings of \citet{2010Natur.466...82M} that supernovae showing a HV gradient 
\citep{2005ApJ...623.1011B}, generally associated with HV supernovae \citep{2009ApJ...699L.139W}, demonstrate 
exclusively redshifted nebular velocities, while LV supernovae preferentially demonstrate blue shifted nebular profiles.

In conclusion, based on the optical spectroscopy available both around maximum light and in the nebular phase, 
there are no observational signatures that would disfavour comparing the colours of the \sne to \snfe for studying 
reddening together with the adopted uncertainties.  In particular, while \sne \fe and \cg have slightly different light curve 
shapes, they show remarkable similarities in spectroscopical features and evolution.  
However, the \sne \by and \fe have also been found to be almost identical in the optical while showing
significant discrepancies in the UV \citep{2013ApJ...769L...1F,2015MNRAS.446.2073G}.  Next we investigate the 
impact on the fitted reddening laws if we instead compare the measured colours to \snby.

\subsection{\snby as colour template}
\begin{figure}
  \begin{center}
    \includegraphics[width=\columnwidth]{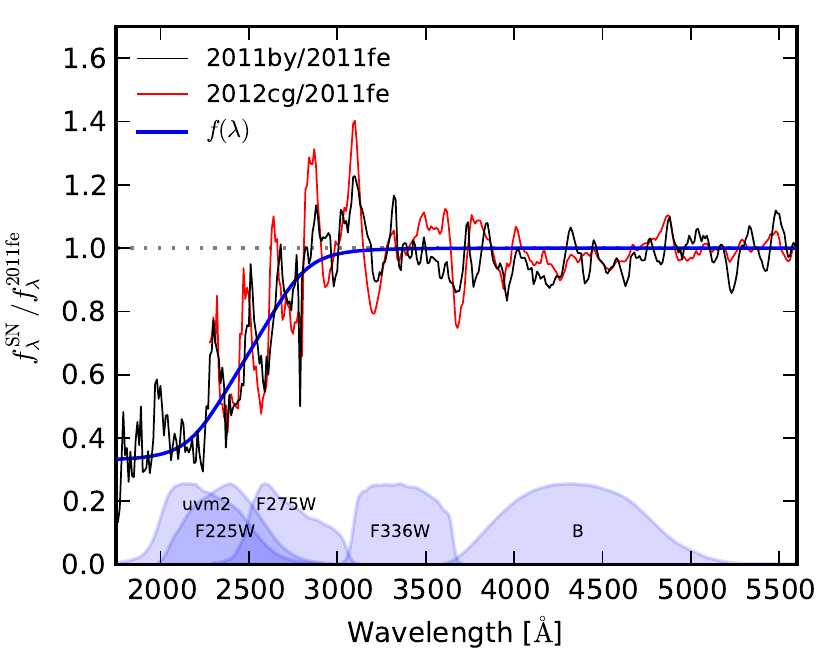}
    \caption{%
    	The flux ratio between \sne \by and \fe, and \cg and \fe at maximum after they have been corrected for Galactic 
	extinction, corrected for reddening (in the case of \sncg), and normalised between $5000$--$5500$~\AA. We use the 
	reddening parameters derived for maximum for \sncg from Table~\ref{tb:extphase} and both spectra have been 
	smoothed using boxcar smoothening. These spectra are only shown for wavelengths where the signal-to-noise 
        per spectral element exceeds 2.5.  Some of the passbands have been plotted in the background
	to illustrate which filters are affected by the difference in spectroscopic shape.  The blue line shows 
	Equation~\eqref{eq:ratiofunc} that is used to model the flux ratio between \sne \by and \fe as described in
	Appendix~\ref{sec:bytemplate}.
	\label{fig:nuvratio}}
  \end{center}
\end{figure}

\snby was discovered \citep{2011CBET.2708....1J} in NGC~3972 and found to be a spectroscopically normal \snia 
\citep{2013MNRAS.430.1030S} with minimal reddening \citep{2012MNRAS.426.2359M}.  \hststis UV observations 
were obtained at maximum (Program GO-12298; PI:~Ellis) and using these data \citet{2013ApJ...769L...1F} and 
\citet{2015MNRAS.446.2073G} have shown that while \snby is almost identical to \snfe
in its optical properties it is intrinsically fainter in the UV.  The two \sne are compared at maximum in 
Figure~\ref{fig:nuvratio}, where the ratio between the spectra are plotted.   Comparing this to the ratio between \sne
\cg and \fe at maximum, also shown in the figure, suggests that \snby may be an even better reference \sn than
\snfe in the UV. 

Both \sne \by and \cg are fainter than \snfe in the UV, which could explain why \sncg is marginally redder in its UV colours
than predicted by the fitted reddening laws.  Two other \sne in the sample, \bm and \cp, show a similar behaviour 
for the bluest colour \wfcone$-$\Vband.  

\hststis spectroscopy of \snby was obtained at phases $-10$ and $-1$~days, which unfortunately does not 
allow us to construct a SED covering the full range of our observations as we did for \snfe.  However, we
can use the available spectrum at maximum and the \swift observations of this \sn to model the difference
between \sne \by and \fe in the UV.  In Appendix~\ref{sec:bytemplate} we describe this procedure and construct a 
model SED of \snby under the assumption that it is identical to \snfe at wavelengths $\lambda>3000$~\AA.

\begin{figure}
  \centering
  \includegraphics[width=\columnwidth]{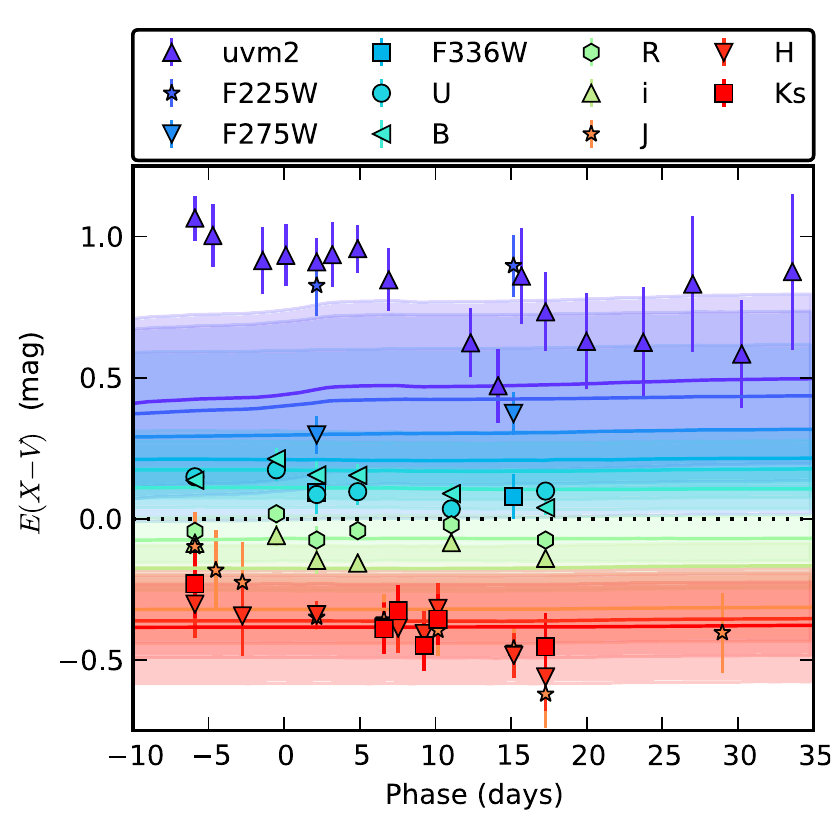}
  \caption{%
    The colour excess for \sncg, similar to the upper left panel of Figure~\ref{fig:afits}, with the exception that the 
    measured and predicted colour excesses are based on the SED model described in 
    \S\,\ref{sec:bytemplate}, which has been derived to mimic \snby in the UV.  The solid lines represent the best fitted 
    \ftz law with the parameters given in Table~\ref{tb:fitlawresults11by}.
    \label{fig:exvcgby}}
\end{figure}

Using this as the comparison \sn, we can refit the reddening laws for the \sne \cg, \cp and \bm for which
we have reason to believe that \snby may provide a better comparison object.  The fitted \ftz parameters at maximum 
are presented in Table~\ref{tb:fitlawresults11by}.  The fitted parameters for \sne \cp and \bm change marginally compared to
Table~\ref{tb:fitlawresults} but we do note that reduced $\chi^2$ has decreased.  However, the most 
interesting result here is for \sncg. Not only the fitted reddening law parameters change, but also the 
time-evolution of $E(\mathrm{\uvmtwo}-\mathrm{\Vband})$ becomes less significant, which can be seen in
Figure~\ref{fig:exvcgby}.  This can be explained from Figure~\ref{fig:exvbyfe}, where the colour excesses of \snby are
shown with respect to \snfe, and a similar time-evolution for $E(\mathrm{\uvmtwo}-\mathrm{\Vband})$ is shown. We
note that the there is still a tension between the bluest colours and the fitted reddening law which may suggest that
the spectrum of \sncg drops faster than \snby at $\lambda<2500$~\AA.

\begin{table}
\centering
\begin{tabular}{cr@{}lr@{}lr@{}l}
\hline\hline
 & \multicolumn{2}{c}{\small 2012cg} & \multicolumn{2}{c}{\small 2012cp} & \multicolumn{2}{c}{\small 2012bm}\\
\hline\\[-2ex]
{\small $\EBV$ } & {\small $0.12$}&{\small$\,(0.02)$} & {\small $0.34$}&{\small$\,(0.03)$} & {\small $0.45$}&{\small$\,(0.04)$}\\[0.5ex]
{\small $\RV$ } & {\small $3.7$}&{\small$^{+1.4}_{-1.0}$} & {\small $3.2$}&{\small$^{+0.5}_{-0.4}$} & {\small $3.0$}&{\small$^{+0.4}_{-0.4}$}\\[0.5ex]
{\small $\tBmax$ } & {\small $56080.5$}&{\small$\,(0.3)$} & {\small $56081.1$}&{\small$\,(0.3)$} & {\small $56018.3$}&{\small$\,(1.1)$}\\[0.5ex]
{\small $\sB$ } & {\small $1.13$}&{\small$\,(0.02)$} & {\small $1.19$}&{\small$\,(0.02)$} & {\small $1.24$}&{\small$\,(0.08)$}\\[0.5ex]
{\small $\chi^2/\nu$ } & {\small $0.88$}& & {\small $1.09$}& & {\small $2.13$}&\\[0.5ex]
\hline\hline\\[-2ex]
\end{tabular}

\caption{%
   Best fitted parameters of the \ftz law using the colour and SED model that has been adapted to match
   \snby in the UV as described in \S\,\ref{sec:bytemplate}.  The quoted uncertainties are the statistical
   $1\sigma$ errors from the $\chi^2$ fit.
   \label{tb:fitlawresults11by}}
\end{table}

From the spectroscopic similarities between \sne \by and \cg as shown in Figure~\ref{fig:nuvratio} and the similar
UV-optical colours we find that \snby appears to be a better comparison \sn for \sncg than \snfe.   This study also
show how studying the UV-colour evolution of \sneia provides a powerful tool for comparing predictions of CS
dust scenarios with intrinsic colour variations.


\subsection{The impact of the UV-data}\label{sec:nuvdata}
Another approach for testing the robustness of the reddening laws is to study how the fitted parameters are 
affected by the data at different wavelengths.  We do this by comparing the fitted \ftz parameter values for two 
different cases, where we use all measured colours (UV-NIR) and where we omit the UV data (OPT-NIR).  
A tension between the two cases could be a sign of either assuming an incorrect reddening law or colour 
reference \sn.  We keep $\tBmax$ and $\sB$ fixed for this study and use \snby as reference for the \sne
with red UV colours.  The results are presented in Table~\ref{tb:extwaveftz}. 
\begin{table}
\centering
\begin{tabular}{clr@{}lr@{}lc@{}l}
\hline\hline\\
\multicolumn{1}{c}{\small \sn} & \multicolumn{1}{c}{\small $\lambda$-range} &\multicolumn{2}{c}{\small $\EBV$} &\multicolumn{2}{c}{\small $\RV$} &\multicolumn{2}{c}{\small $\chi^2/\nu$}\\
\hline\\[-2ex]
\multirow{2}{*}{\small SN2012cg$^\dagger$}
 & {\small UV-NIR} &{\small $0.11$}&{\small$\,(0.02)$} & {\small $3.8$}&{\small$^{+1.5}_{-1.0}$} & {\small $0.71$}\\[0.5ex]
 & {\small OPT-NIR} &{\small $0.11$}&{\small$\,(0.05)$} & {\small $4.0$}&{\small$^{+3.2}_{-1.5}$} & {\small $0.56$}\\[0.5ex]
\hline\\[-2ex]
\multirow{2}{*}{\small SN2012et}
 & {\small UV-NIR} &{\small $0.16$}&{\small$\,(0.02)$} & {\small $1.8$}&{\small$^{+0.6}_{-0.5}$} & {\small $1.06$}\\[0.5ex]
 & {\small OPT-NIR} &{\small $0.15$}&{\small$\,(0.09)$} & {\small $<$}&{\small$1.9$} & {\small $1.72$}\\[0.5ex]
\hline\\[-2ex]
\multirow{2}{*}{\small SN2012cp$^\dagger$}
 & {\small UV-NIR} &{\small $0.34$}&{\small$\,(0.03)$} & {\small $3.1$}&{\small$^{+0.5}_{-0.4}$} & {\small $0.83$}\\[0.5ex]
 & {\small OPT-NIR} &{\small $0.36$}&{\small$\,(0.04)$} & {\small $3.0$}&{\small$^{+0.6}_{-0.5}$} & {\small $1.03$}\\[0.5ex]
\hline\\[-2ex]
\multirow{2}{*}{\small SN2012bm$^\dagger$}
 & {\small UV-NIR} &{\small $0.45$}&{\small$\,(0.03)$} & {\small $3.0$}&{\small$^{+0.4}_{-0.3}$} & {\small $1.68$}\\[0.5ex]
 & {\small OPT-NIR} &{\small $0.41$}&{\small$\,(0.05)$} & {\small $3.5$}&{\small$^{+0.6}_{-0.5}$} & {\small $0.70$}\\[0.5ex]
\hline\\[-2ex]
\multirow{2}{*}{\small SN2012cu}
 & {\small UV-NIR} &{\small $0.99$}&{\small$\,(0.03)$} & {\small $2.8$}&{\small$^{+0.1}_{-0.1}$} & {\small $0.47$}\\[0.5ex]
 & {\small OPT-NIR} &{\small $1.00$}&{\small$\,(0.05)$} & {\small $2.7$}&{\small$^{+0.2}_{-0.2}$} & {\small $0.43$}\\[0.5ex]
\hline\\[-2ex]
\multirow{2}{*}{\small SN2014J}
 & {\small UV-NIR} &{\small $1.36$}&{\small$\,(0.02)$} & {\small $1.4$}&{\small$^{+0.1}_{-0.1}$} & {\small $2.01$}\\[0.5ex]
 & {\small OPT-NIR} &{\small $1.28$}&{\small$\,(0.04)$} & {\small $1.6$}&{\small$^{+0.1}_{-0.1}$} & {\small $2.29$}\\[0.5ex]
\hline\hline\\[-2ex]
\multicolumn{8}{l}{{\small $^\dagger$\sn colours were compared to \snby}}\\ 
\end{tabular}

\caption{%
   The best fitted \ftz parameters using either only the optical--NIR colours or all colours (UV--NIR) while the
    $\tBmax$ and $\sB$ parameters were fixed to the values given in Tables~\ref{tb:fitlawresults} 
    and~\ref{tb:fitlawresults11by} respectively. All colour data between phases $-10$ and $+35$~days from $\tBmax$ were 
    used for the fits.
   \label{tb:extwaveftz}}
\end{table}

The results are consistent for all \sne except for \snjj, where the addition of the UV data shifts the 
fitted values by $\sim2$--$3\sigma$.  We fit a low value of $\RV$ for this \sn and, as shown in 
Figure~\ref{fig:powerdiv},  a power-law appears to provide a better fit in this case.  In 
Table~\ref{tb:extwavepower} we present the results on the two colour sets when using a power-law 
relation for the two low-$\RV$ \sne.  While there is a slight, $\sim1\sigma$, tension for \snet, the results 
are fully consistent between the two cases for \snjj.
\begin{table}
\centering
\begin{tabular}{clr@{}lr@{}lc@{}l}
\hline\hline\\
\multicolumn{1}{c}{\small \sn} & \multicolumn{1}{c}{\small $\lambda$-range} &\multicolumn{2}{c}{\small $\AV$} &\multicolumn{2}{c}{\small $\beta$} &\multicolumn{2}{c}{\small $\chi^2/\nu$}\\
\hline\\[-2ex]
\multirow{2}{*}{\small SN2012et}
 & {\small UV-NIR} &{\small $0.30$}&{\small$\,(0.13)$} & {\small $-1.7$}&{\small$\,(0.5)$} & {\small $0.63$}\\[0.5ex]
 & {\small OPT-NIR} &{\small $0.16$}&{\small$\,(0.11)$} & {\small $-3.3$}&{\small$\,(1.5)$} & {\small $0.82$}\\[0.5ex]
\hline\\[-2ex]
\multirow{2}{*}{\small SN2014J}
 & {\small UV-NIR} &{\small $1.97$}&{\small$\,(0.10)$} & {\small $-2.0$}&{\small$\,(0.1)$} & {\small $1.40$}\\[0.5ex]
 & {\small OPT-NIR} &{\small $1.93$}&{\small$\,(0.10)$} & {\small $-2.1$}&{\small$\,(0.1)$} & {\small $1.02$}\\[0.5ex]
\hline\hline
\end{tabular}

\caption{%
   The best fitted power-law parameters for the two \sne with low $\RV$ values.  The fits were carried out using either 
    only the optical--NIR colours or all colours (UV--NIR) while the $\tBmax$ and $\sB$ parameters were fixed to the values 
    given in Table~\ref{tb:fitlawresults}.  All colour data between phases $-10$ and $+35$~days from $\tBmax$ were used for the fits.
   \label{tb:extwavepower}}
\end{table}

\begin{figure}
  \begin{center}
    \includegraphics[width=\columnwidth]{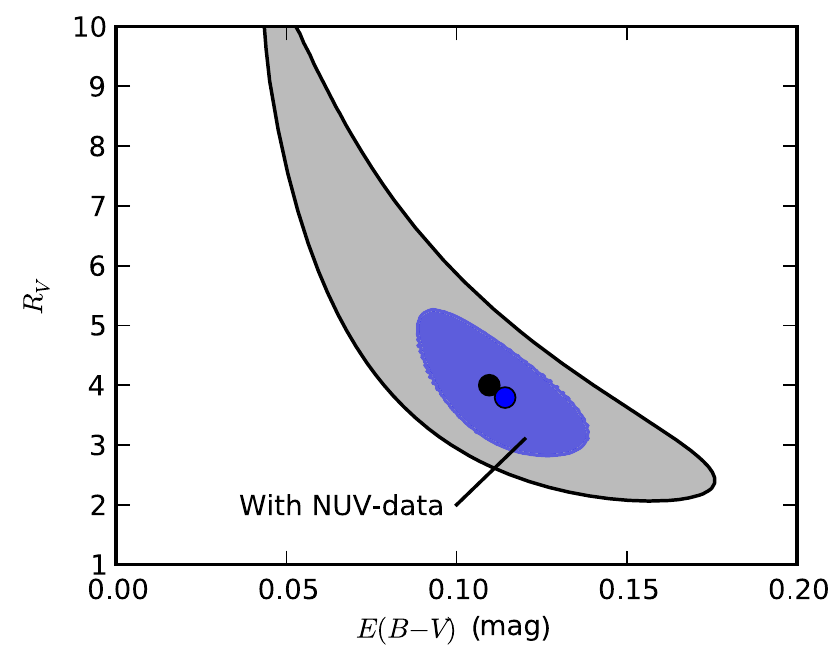}
    \caption{%
    	Contours showing the joint $68\,\%$ uncertainty region for the fitted parameters $\RV$ and $\EBV$ 
	for a \ftz extinction laws for \sncg in comparison to \snby.  The two contours show the fit results when the full 
	data set (UV--NIR) is used (dark blue contour region) and when the colours based on \hst and \swift UV 
	observations are omitted (light grey contour region).  All data between phases $-10$ and $+35$~days with 
	respect to $\tBmax$ were used, and the best fit values for the two cases have been marked by black and 
	blue dots respectively.
    	\label{fig:contour}}
  \end{center}
\end{figure}

The major impact of adding the UV-colours is, however, on the uncertainties of the fitted values, in particular for 
low-reddening \sne. This is illustrated in Figure~\ref{fig:contour} where the grey contour shows the 
$68\,\%$ confidence region for $\EBV$ and $\RV$ when the \ftz law is fitted to the optical--NIR data of \sncg, 
while the blue contour shows the uncertainties from the full UV--NIR fit.
%
From the figure we conclude that extending the wavelength range significantly improves the constraints on 
$\RV$ despite the high intrinsic dispersion adopted for the \swift and \hst colours.


\subsection{Extinction from interstellar absorption features}\label{sec:highres}
In \S\,\ref{sec:cs} we discussed the constraints on the existence of circumstellar dust that can be derived from non-detection 
of time-varying \NaID absorption lines.  \NaID lines in high-resolution spectra are also commonly used as a proxy for reddening 
\citep{2011MNRAS.415L..81P} and the equivalent width of the lines have been shown to correlate well with extinction in the 
Milky May \citep[see][]{1997A&A...318..269M,2012MNRAS.426.1465P}.  However, \citet{2013ApJ...779...38P} found that of 
their sample of \sneia $\sim25\,\%$, of which most have blue shifted \NaID features, have unusually high \NaI column 
densities compared to the values expected from the \citet{2012MNRAS.426.1465P} relation.  It is currently unclear whether the 
high abundance of sodium is intrinsic to the environment of some \snia or if the Milky Way reference value is anomalous compared 
to other galaxies. Other species, such as diffuse interstellar bands, have been shown to be useful for studying the ISM in \sneia
host galaxies \citep{2005A&A...429..559S} and in particular the diffuse interstellar band (DIB) feature at $5780$~\AA\, may provide 
a  better proxy for reddening of \sneia than \NaID \citep{2013ApJ...779...38P}.  In Table~\ref{tb:ewnaid} we present equivalent 
width values of \NaID and the DIB at $5780$~\AA\ with corresponding reddening estimated by the  empirical relations
\citep{2012MNRAS.426.1465P} and \citep{2013ApJ...779...38P}, respectively, for the \sne of which we have obtained spectra or
 published data exists.  Further empirical reddening relations exist for ISM features \CaHK \citep{2015arXiv150302697M}
 and \MgII \citep{2008MNRAS.385.1053M}.  These relations have so far not been extensively studied for their validity in 
estimating the reddening of \sne.

\begin{table*}
\centering
  \begin{tabular}{l r@{\,}l r@{\,}l r@{}l r@{\,}l r@{\,}l r@{\,}l r@{\,}l r@{\,}l}
    \hline\hline
    \multicolumn{1}{c}{\small \sn}  &  
    \multicolumn{2}{c}{\small \Done} & \multicolumn{2}{c}{\small \Dtwo} & \multicolumn{2}{c}{\small $\EBV$} & 
    \multicolumn{2}{c}{\small DIB~5780} & \multicolumn{2}{c}{\small $\AV$} & 
    \multicolumn{2}{c}{\small \MgII $2796\,$} & \multicolumn{2}{c}{\small $\EBV$} & \multicolumn{2}{c}{\small $\EBV$}\\

    & \multicolumn{2}{c}{\footnotesize{\small (m\AA)}} & \multicolumn{2}{c}{\footnotesize{\small (m\AA)}} & \multicolumn{2}{c}{\footnotesize{\small (\NaID)}} & 
    \multicolumn{2}{c}{\footnotesize{\small (m\AA)}} & \multicolumn{2}{c}{\footnotesize{\small (DIB)}} & \multicolumn{2}{c}{\footnotesize{\small (\AA)}} & 
    \multicolumn{2}{c}{\footnotesize{\small (\MgII)}} & \multicolumn{2}{c}{\footnotesize{\small (Phot.)}}\\
    \hline
    {\small \jj} & {\small $2558$}&{\small $(6)$} & {\small $2831$}&{\small $(9)$} & \multicolumn{2}{c}{\small -} & 
    {\small $344$}&{\small $(4)$} & {\small $1.8$}&{\small $(0.9)$} & {\small $\leq4.6$}&{\small $(0.2)$} & {\small $\leq0.14$}&{\small $(0.08)$} &
    {\small $1.36$}&{\small $(0.02)$}\\[0.5ex]
    {\small \cg} & {\small $700$}&{\small $(9)$} & {\small $1018$}&{\small $(8)$} & {\small $1.6$}&{\small $^{+0.4}_{-0.3}$} & 
    {\small $85$}&{\small $(5)^\dagger$} & {\small $0.4$}&{\small $(0.2)$} & {\small $\leq3.5$}&{\small $(0.2)$} & {\small $\leq0.08$}&{\small $(0.05)$} &
    {\small $0.11$}&{\small $(0.02)$}\\[0.5ex]
    {\small \cu} & {\small $849$}&{\small (3)$^\ddagger$} & {\small $925$}&{\small $(3)^\ddagger$} & {\small $1.7$}&{\small $^{+0.4}_{-0.3}$} & 
    \multicolumn{2}{c}{\small -} & \multicolumn{2}{c}{\small -} & \multicolumn{2}{c}{\small -} & \multicolumn{2}{c}{\small -} &
    {\small $0.99$}&{\small $(0.03)$}\\[0.5ex]
    {\small \et} & \multicolumn{2}{c}{\small -}   & {\small $650$}&{\small $(40)^\star$}   & {\small $0.3$}&{\small $^{+0.3}_{-0.1}$} & 
    \multicolumn{2}{c}{\small -} & \multicolumn{2}{c}{\small -} & \multicolumn{2}{c}{\small -} & \multicolumn{2}{c}{\small -} &
    {\small $0.16$}&{\small $(0.02)$}\\[0.5ex]
    \hline
    {\small \fe} & {\small $27.4$}&{\small $(1.1)^\diamond$}  & {\small $47.1$}&{\small $(0.8)^\diamond$}   & {\small $0.017$}&{\small $^{+0.004}_{-0.003}$} & 
    \multicolumn{2}{c}{\small -} & \multicolumn{2}{c}{\small -} & {\small $2.9$}&{\small $(0.2)$} & {\small $0.06$}&{\small $^{+0.03}_{-0.02}$} &
    \multicolumn{2}{c}{\small -}\\[0.5ex] 
    {\small \by} & \multicolumn{2}{c}{\small -}   & \multicolumn{2}{c}{\small -} & \multicolumn{2}{c}{\small -} & 
    \multicolumn{2}{c}{\small -} & \multicolumn{2}{c}{\small -} & {\small $\leq2.3$}&{\small $(0.2)$} & {\small $\leq0.04$}&{\small $(0.02)$} &
    \multicolumn{2}{c}{\small -}\\[0.5ex] 
    \hline\hline
    \multicolumn{16}{l}{\small Equivalent width values retrieved from $^\dagger\,$\citet{2013ApJ...779...38P}, 
    \small $^\ddagger\,$\citet{2014MNRAS.443.1849S}, $^\star\,$\citet{2013MNRAS.436..222M}}\\
    \multicolumn{10}{l}{and $^\diamond\,$\citet{2013A&A...549A..62P}.} \\
  \end{tabular}
  \caption{Equivalent width values of \NaID, the DIB feature at $\lambda\lambda 5780$~\AA\ and \MgII at $\lambda\lambda 2796$~\AA\ for which 
reddening parameters can be inferred from \citet{2012MNRAS.426.1465P}, \citet{2013ApJ...779...38P} and \citet{2008MNRAS.385.1053M}, respectively. 
Upper limits are determined for \MgII measurements, since the doublet is blended in the \stis spectra.
In the case of \snjj, the \NaID equivalent width extrapolates the relation by \citet{2012MNRAS.426.1465P} beyond any reasonable value for $\EBV$. 
Best photometric $\EBV$ values have been taken from Table~\ref{tb:extwaveftz} for comparison and the $\EBV$ for \fe and \by is assumed to be negligible.
\label{tb:ewnaid}}
\end{table*}

Besides \NaID, narrow \CaHK features, with respective equivalent widths of $770\pm30$ and $579\pm35$ m\AA\ 
could only be measured from the first FIES epoch of \sncg. These values are well above the range in which the relations in 
\citep{2015arXiv150302697M} are valid. 
\citet{2013ApJ...779...38P} also quote the detection of the DIB $5780$~\AA\ which
is consistent with the non-detection in the individual FIES epochs within the signal-to-noise. 
The most prominent absorption features in \NaID, shown in Figure~\ref{fig:cgnaid} 
roughly agree with $v_h$ from Table~\ref{tb:snsummary} and 
are slightly blueshifted with respect to the apparent stellar velocity of $\sim448$~km/s along the line-of-sight of 
\sncg \citep[see][]{2006AJ....131..747C}. Additional features of the \NaID and \CaHK profiles
span a range from $\sim370$--$530$~km/s, whereby a number of features are blended on the redshifted side of the profile.
\sncg is included in the analysis of \citet{2013ApJ...779...38P} and is not counted in the group of \sne with anomalously 
high \NaI column densities. However, the best fit $\EBV$ in Table~\ref{tb:extwaveftz} for \cg highly disagrees with the value
computed from the \citet{2012MNRAS.426.1465P} relation. 


\begin{figure}
  \begin{center}
    \includegraphics[width=\columnwidth]{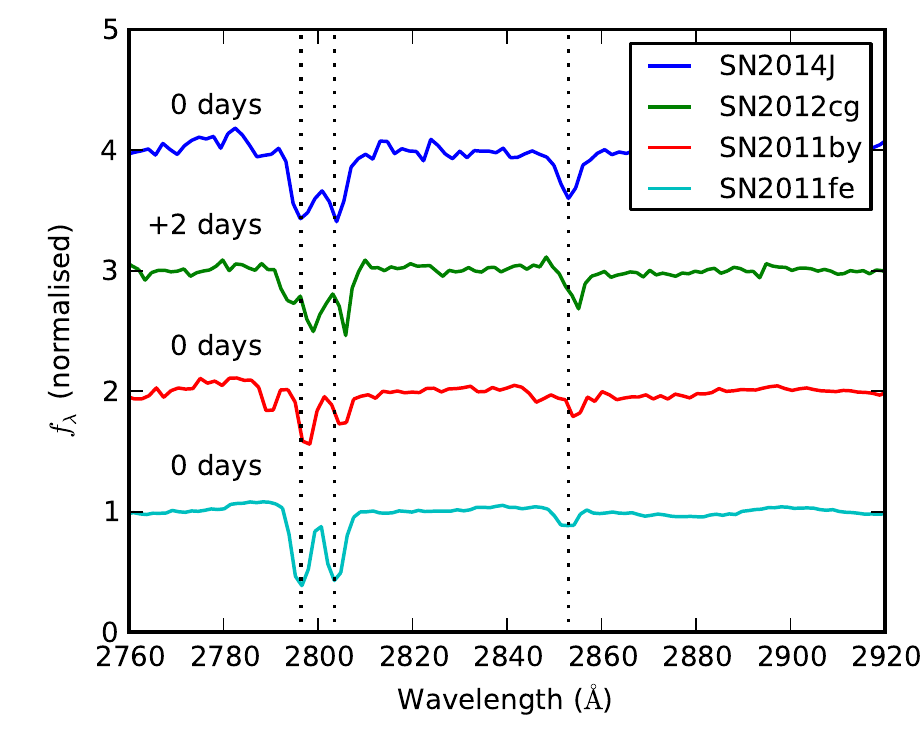}
    \caption{\MgII at $\lambda\lambda2796, 2803$~\AA\ and Mg~I at $\lambda\lambda2853$~\AA\ marked by dotted lines in 
    \stis spectra of four \sne with different reddening. The continua of the unsmoothed spectra are normalised and offset for clarity. 
    Above the respective spectra, the phases are indicated at which the spectra were taken. 
    	\label{fig:Mg}}
  \end{center}
\end{figure}

Four epochs of high-resolution spectra of \sncu have been described in \citet{2014MNRAS.443.1849S}. 
No time-variability is detected in the deep \NaID absorption features, which likely consist of several 
blended components. The velocity of the \NaID lines of $\sim1130$~km/s are in agreement with the H~I 
velocity along th line-of-sight of \sncu in NGC~4772 \citep{2000AJ....120..703H}. 
We perform the photo-ionisation analysis described in Section~\ref{sec:cs} on the \NaID doublet of \cu. 
The non-variation of \Done indicates that there is no \NaI between $6\cdot10^{18}$ -- $2\cdot10^{19}$~cm. 
However, these constraints are weak since the absorption lines are close to saturation and not optically thin.
The $\EBV$ computed from the \citet{2012MNRAS.426.1465P} relation
is $\sim2\sigma$ greater than the photometric value. Furthermore, this value must be considered a lower limit
since the \NaID doublet of \sncu appears to be close to saturation.

\snet is included in the sample of \citet{2013MNRAS.436..222M} in which an \Dtwo equivalent width is quoted from 
a mid-resolution spectrum. The corresponding \citet{2012MNRAS.426.1465P} relation $\EBV$ is $\sim1\sigma$ greater 
than expected. 

It is clear from \citet{2013ApJ...779...38P} that \NaID do not serve as good reddening proxies for \sneia and 
that the scatter in the \citet{2012MNRAS.426.1465P} relation is underestimated. This appears to be the case for 
\sne with high reddening, such as \jj, as well as low reddening such as \cg.
Furthermore, many of the \NaID profiles do not appear to be optically thin, implying that the column densities 
are rather underestimated, which would further increase the discrepancy between the \citet{2012MNRAS.426.1465P} 
relation and observations. 
There is thus motivation to investigate other ISM species, such as \eg \MgII, which are 
more difficult to observe and have not been studied in large samples so far. 

\stis spectra of \sne \jj \citep[from][]{2014MNRAS.443.2887F}, \cg, \fe and \by\footnote{\stis spectra of SNe \fe and \by were taken as part of Program GO-12298; PI: Ellis}
around the \MgII doublet at $\lambda\lambda2796, 2803$~\AA\ are shown in Figure~\ref{fig:Mg}. 
The absorption features can be used to test the relation 
\[
\EBV =  (0.008\pm0.001)W^{(1.88\pm0.17)}
\]
of \citet{2008MNRAS.385.1053M}, where $W$ is the equivalent width of \MgII at $\lambda\lambda2796$~\AA. 
The \stis spectra of \sne \bl, \bm, \cp and \et are not included in the sample due to the low 
signal-to-noise around the \MgII doublet. Upon visual inspection of Figure~\ref{fig:Mg}, one can see that the \MgII lines are 
deep for all four \sne and the line ratios obtained from fitting a double Gaussian profile further indicate that the lines are not 
optically thin. 
In all \stis spectra, except those of \fe, the \MgII doublet is strongly blended. Further, it cannot be excluded that the entire 
absorption is attributed to the Milky Way. We therefore measure the total equivalent width of the observed features and 
consider $2/3$ of the value to be the upper limit of the component at $\lambda\lambda2796$~\AA. 
The fraction is chosen from the line ratio of the \MgII doublet, if it is optically thin. The limits we obtain 
along with the corresponding $\EBV$ computed from the relation above are shown in Table~\ref{tb:ewnaid}.
The \MgII doublet is resolved better in the \fe~\stis\ spectra and the equivalent width can be determined from the Gaussian 
fits more accurately. The value presented in Table~\ref{tb:ewnaid} is the average equivalent width computed from all available \fe~\stis spectra. 
In comparison to the photometric $\EBV$, the \citet{2008MNRAS.385.1053M} relation clearly 
underestimates the reddening of \jj, whereas no firm conclusions can be drawn for the other \sne due to the large errors and 
since blending with Milky Way features cannot be excluded.
\subsection{The origin of  \sn reddening}\label{sec:origin}
While it can be expected that the observed \snia colour arise from a mixture of extinction by dust and intrinsic 
\snia colour diversity, the latter is typically of order $|\Delta (B-V)|\lesssim 0.1$ \citep[\eg][]{2008A&A...487...19N}. 
Thus, we can assume that the impact of dimming by dust will dominate with increased 
observed reddening and that the fitted extinction laws will be less suceptible to intrinsic \snia variations.  It is therefore
particularly interesting to compare the fits for \sne \cu and \jj.  They are the reddest objects in our sample, and at the 
same time, show very different total-to-selective extinction, $\RV=\RVcu$ and $\RV=\RVjj$ for \sne \cu and \jj respectively.

The evidence for material in the line-of-sight is strengthened by the fact that both \sne show deep \NaID absorption lines in 
their spectra \citep[\eg][for \cu and \jj respectively]{2014MNRAS.443.1849S,2014ApJ...784L..12G}. Since CS dust is 
expected to give rise to steep extinction laws, %
such a scenario may offer an explanation for the low $\RV$ of \snjj.  
However, no time-variability has been found for the \NaID absorption lines for any of the \sne we
studied, although \citet{2015ApJ...801..136G} report a detection of time-varying \KI absorption for \snjj.
When combined with the non-varying \NaID, these findings point to circumstellar dust at radii of 
$r_\mathrm{dust}\gtrsim10^{19}$~cm due to the difference in ionisation energies.  The detection of light echoes
\citep{2015ApJ...804L..37C} also points to the existence of dust at a comparable distance from the site of the explosion, 
$11$ pc, i.e., $r_\mathrm{dust}=3\cdot10^{19}$ cm.  

Further, we have studied the point spread function shape of \snjj in the \hst A14 photometry to search for signs of dust at smaller 
radii but none could be detected, which is consistent with the lack of time-evolution of the \snjj colours with respect to \snfe.   For CS 
dust at $r_\mathrm{dust}\gtrsim10^{19}$~cm the photon time-delay exceeds the typical time-scale of the \snia light curve and 
no colour time-evolution is expected as discussed in \citet{2011ApJ...735...20A}.

Other observations suggesting that the nature of the extinction of \snjj is primarily interstellar, is that the polarisation 
angle of the \sn light is well aligned with the spiral structure of the host galaxy \citep{2015A&A...577A..53P}, the lack 
of thermal emission from the \sn environment in near and mid-IR \citep{2014arXiv1411.3332J}, and that the
velocities of the observed multiple \NaID lines are consistent with measured H\,\textsc{i} velocities along the line-of-sight 
in the host galaxy \citep{2015ApJ...799..197R}.  The latter is also the case for \sne \cu \citep{2000AJ....120..703H}.  
Even if dust does exist at $r_\mathrm{dust}\sim10^{19}$~cm this is not expected to have any major impact on the 
measured $\RV$ for such large distances \citep{2005ApJ...635L..33W}.

If the reddening of both \sne \cu and \jj are dominated by interstellar extinction, the dust properties along the two lines-of-sight 
must be significantly different, and the dust in M82 must be of a nature that has not been observed in the Milky Way 
\citep[although see ][ for a possible explanation]{2015ApJ...805...74B}. M82 is a starburst galaxy, and the dust properties 
have been studied by \citet{2014MNRAS.440..150H}.  They find that a steep Milky Way-like law is preferred over the \
\citet{2001NewAR..45..601C} law, used for starburst galaxies and they derive a wavelength-dependence 
of the scattering within a central projected radius of $<3$~kpc that suggests that "only small grains are entrained in the inner 
\sne-driven wind", which is consistent with both the location (see Table~\ref{tb:snsummary}) and the characteristic of the observed
reddening of \snjj.

For \sneia, \citet{2013A&A...560A..66R} found that when using \halpha emission as a tracer for ongoing star formation, \sne 
with local \halpha emission is on average redder and more homogeneous, resulting in a lower brightness dispersion.  These results
have been confirmed by \citet{2015Sci...347.1459K} and \citet{2015ApJ...802...20R} using UV data from the {\em Galaxy 
Evolution Explorer} (\galex).  If the low $\RV$ originates from small dust grains in regions undergoing intense star formation, we
can expect to observe different local properties between \sne \cu and \jj.  In Table~\ref{tb:galex} we present the surface brightness
of at the position of the \sne using an aperture with a $2$~kpc radius.   From this we cannot draw any firm conclusion based on
the local UV surface brightness of the host galaxies.  Even after we attempt to correct for host galaxy dust, following the recipe
used in \citet{2015ApJ...802...20R}, we see no correlation with UV surface brightness and measured $\RV$.


\begin{table}
	\centering
	\begin{tabular}{l r@{}l r@{}l}
	\hline\hline
	\multicolumn{1}{c}{\sn} & \multicolumn{2}{c}{FUV} &  \multicolumn{2}{c}{NUV} \\
	\hline
	\jj & $26.5$ & $(0.2)$ & $24.9$ & $(0.1)$\\
	\bm & $26.7$ & $(6.1)$ & $25.9$ & $(2.4)$\\
	\cg & $25.1$ & $(0.4)$ & $24.4$ & $(0.2)$\\
	\cp & $24.7$ & $(1.3)$ & $24.6$ & $(0.7)$\\
	\cu & $26.6$ & $(2.3)$ & $26.0$ & $(1.0)$\\
	\et & $24.9$ & $(2.6)$ & $24.4$ & $(1.2)$\\
	\bl & $26.1$ & $(3.3)$ & $26.2$ & $(2.1)$\\
	\hline\hline
	\end{tabular}
	\caption{%
		Surface brightness (mag/arcsec) measured using a $2$~kpc radius centred on the \sn position
		in the \galex FUV and NUV filters.
		\label{tb:galex}}
\end{table}

\begin{figure}
  \centering
    \includegraphics[width=\columnwidth]{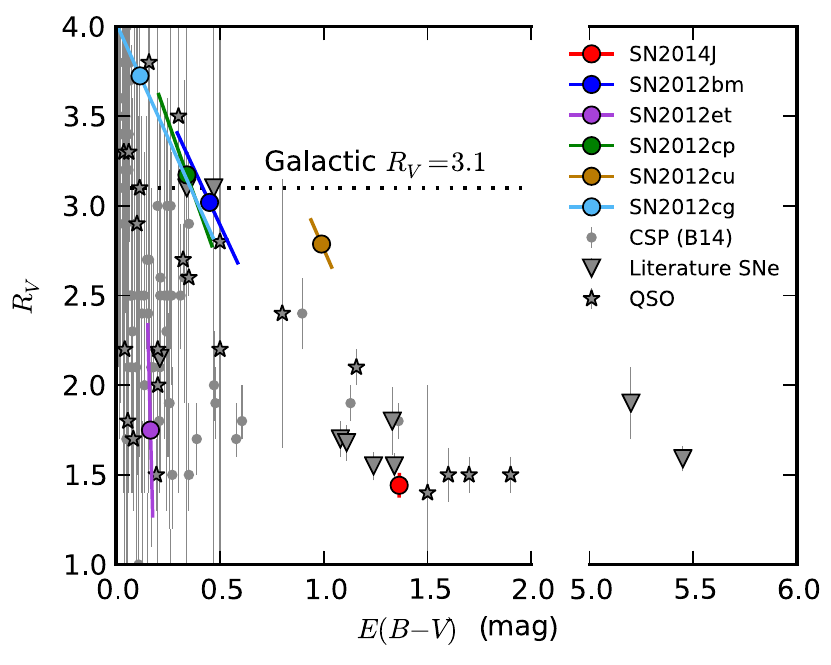}
   \caption{%
	   Fitted $\RV$ and $\EBV$ for the full phase range of $-10$ to $+35$ days, where \snby has been used as reference for 
	   \sne \bm, \cg, and \cp and \snfe was used for the remaining \sne. Also plotted are measurements from the
	   literature, where 9 \sneia from %
	\citet{2000MNRAS.319..223H,2006AJ....131.1639K,2007AJ....133...58K,2004MNRAS.348..261B,2006MNRAS.369.1880E,2008MNRAS.384..107E,2009ApJ...700..331H} and \citet{2010AJ....139..120F} 
	together with three Type~II \sne from \citet{2006MNRAS.368.1169P} and \citet{2009ApJ...694.1067P} are included.  
	The parameters measured by studying reddening of QSO by foreground galaxies are from 
	\citet{2006ApJS..166..443E} and \citet{2006A&A...450..971O}.
    	\label{fig:RVEBV}}
\end{figure}

In Figure~\ref{fig:RVEBV} we show the fitted \ftz parameters from Table~\ref{tb:fitlawresults} for all \sne together with similar 
measurements from the literature.  Low values of $\RV$ are not unique for \sneia, and has also been observed in the 
line-of-sight of other light sources such as gamma-ray burst afterglows \citep{2014A&A...572A..12F} and quasi-stellar objects 
\citep[QSO, \eg][]{2013ApJS..204....6F,2014ApJ...788..123L}.  From the figure it is apparent that there are several observations 
with low $\RV$ values for high extinction \sneia, while \sncu is both showing high extinction and a relatively high $\RV$.  For 
low-extinction \sne it is by nature difficult to measure $\RV$ for individual objects, but there are several statistical studies that 
obtain higher ($\RV\lesssim3$) average values of $\RV$ compared to what is observed for high-extinction \sneia 
\citep[\eg][]{2011ApJ...731..120M,2011A&A...529L...4C}.  The UV observations allow us to improve the sensitivity on $\RV$ for
individual low-reddened \sne, and it is interesting to note that we measure $\RV$ discrepancies similar to what we saw for 
high-extinction objects.  The colours of \sne \cg and \et are however a mixture of intrinsic variation as was shown by the 
comparison of \sncg with \sne \by and \fe in particular. For these objects we therefore cannot draw any firm conclusions to 
what extent the measured $\RV$ is solely a dust property.  Note though that there are strong indications of dust along the 
line-of-sight for these \sne from \NaID absorption \citep[as we have seen for \sncg and for \snet in ][]{2013MNRAS.436..222M}.  
Independent of the origin of the reddening we conclude that 
our sample of reddened \sneia with well observed colours, although limited, suggests that a range of $\RV$ values can be 
expected for both low and high-reddened \sneia.

Given the findings of M13 that all ''NUV-blue'' \sne belong to the photospheric NV class while the ''NUV-red'' 
consists of both NV and HV \sne and the colour-velocity relation for  $\mathrm{\Bband}-\mathrm{\Vband}$
\citep{2009ApJ...699L.139W}, we can also compare the derived extinction parameters with the measured 
\SiII~$\lambda6355$~\AA\ velocities.  From Figures~\ref{fig:velocity} and~\ref{fig:RVEBV} we conclude that 
both \sne for which we measure a low $\RV$ (\jj and \et) are classified as HV \sne while the remaining belong to 
the NV class.  This is consistent with the statistical study of samples of \sneia by \citet{2009ApJ...699L.139W},
although our sample size does not allow us to draw firm conclusions.   However, the fact that we present \sne with
both high and low values of $\RV$ in both high and low extinction environments argues against that the origin
of the discrepancy in observed reddening laws is solely due to intrinsic \sn properties.

One possible explanation to both of these findings could be that HV \sne primarily explode in environments with low 
$\RV$ dust. \citet{2013Sci...340..170W} and \citet{2015MNRAS.446..354P} have shown that HV \sne are, on average, 
both located in more luminous hosts, and in brighter regions, closer to their host nuclei, compared the corresponding 
NV samples.   Further, the results from \citet{2014MNRAS.440..150H} showed that a steep Milky-Way law is
preferred for the central region of M82, and similarly the extinction towards the Galactic bulge also follows a steeper
extinction curve \citep[$\RV\approx2.5$][]{2013ApJ...769...88N} compared to the Milky Way average.  Both the dust and
\snia properties are likely to be affected by the environment where they are produced and if HV \sneia
originate from younger and more metal-rich progenitors than NV \sneia \citep{2013Sci...340..170W}, it is possible that
the average dust properties in the two environments are different as well.

\begin{figure}
  \centering
  \includegraphics[width=\columnwidth]{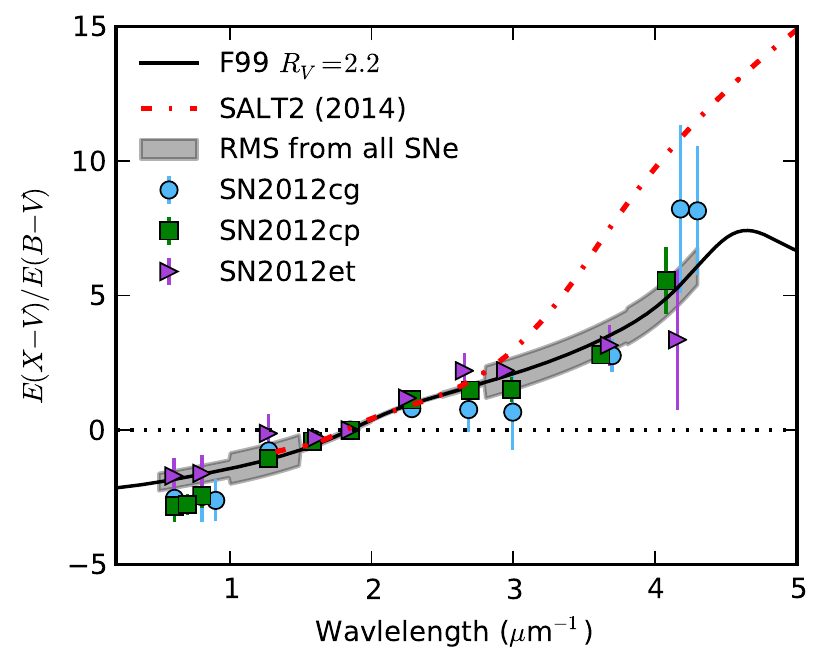}
  \caption{%
    The normalised colour excess $\EXV/\EBV$ is shown using the same approach as in Figure~\ref{fig:extdiversity}
    for the measurements of the three least reddened \sne.  The errorbars are dominated by the adopted intrinsic
    colour uncertainties.  Overplotted is also the average \ftz law (solid black) for the three \sne (fitted by comparing the colours 
    to \snfe for \snet and to \snby for the others) and the version of the SALT2 colour law 
    \citep{2007AA...466...11G} used in \citet{2014A&A...568A..22B}.  The SALT2 law is only defined over the wavelength range 
    $2000$--$8000$~\AA. 
	The grey regions show the RMS of the full sample with respect to the average reddening law for different 
	wavelength bins. 
    \label{fig:exvebv}}
\end{figure}

Studies of high-redshift \sne for cosmological applications are unlikely to include highly reddened \sne like 
\cu and \jj,  but objects with similar reddening to \sne \cg and \et are common.   
In Figure~\ref{fig:exvebv} we show the normalised colour excesses, $\EXV/\EBV$, for the three least 
reddened \sne, where we have used the approach from Figure~\ref{fig:extdiversity} to obtain the plotted wavelengths.  
Figure~\ref{fig:exvebv} also shows the average reddening \ftz law and the SALT2 colour law \citep{2007AA...466...11G} which has been
empirically derived without attempting to discriminate between the different sources of \sn reddening.  Plotted here is 
the version used in the cosmology analysis from \citet{2014A&A...568A..22B}.  Although the reddening laws are consistent 
at wavelengths covered by the \Bband--\Rband bands, the SALT2 law starts to diverge from the data in the UV, wavelengths 
that becomes increasingly important when \sneia at high-redshifts are studied.  The lack of understanding \sn reddening 
is one of the dominating astrophysical systematic uncertainties when \sneia are used for cosmology
\citep[see \eg][]{2010ApJ...716..712A,2011ApJ...737..102S,2012ApJ...746...85S,2014A&A...568A..22B} and further 
UV studies are needed in order to discriminate between different sources of reddening.

\section{Summary and conclusions}\label{sec:conclusions}
In this work we have presented and analysed broadband photometry of seven Type Ia supernovae 
covering the wavelength range $0.2$--$2~\mu$m, obtained mainly from \hst, \swift and the Nordic Optical Telescope.

All \sne are found to be normal with respect to their light curves and optical spectroscopic properties.  The 
\sne have colour excesses in the range $\EBV=0.0$--$1.4$~mag and we have studied the reddening properties 
by comparing them to the normal pristine \sneia, \fe and \by.  In particular, we have explored three different extinction laws: 
\citet{1994ApJ...422..158O},   \citet{1999PASP..111...63F}, and \citet{2008ApJ...686L.103G}. 

We have further studied the time-evolution of the fitted reddening laws and tested the consistency over a broad 
wavelength range.  For the one \sn, \cg, we do find indications of time-evolution of the bluest colours and we have
investigated if this could be explained by circumstellar dust.  Further, we have also searched for time-varying
absorption features, in particular \NaID, for \sncg in multi-epoch high-resolution spectroscopy.  The high-resolution 
spectra were also used to estimate the extinction along the line-of-sight using the absorption of \NaID and \MgII
as dust proxies.  The validity of using \snfe as a colour template was discussed and compared to the resulting
reddening parameters when using \snby.

We summarise the main findings as follows:
\begin{enumerate}
\item We observe a diversity in extinction properties characterised by a total-to-selective extinction, $\RV$,
  ranging from $\RV\approx1.4$ to $\RV\approx3$. 
\item The diversity between the fitted reddening laws for different \sne is significantly greater than the
  discrepancy between the laws for a single \sn.  Although, we do find that the empirical \ftz law provides a better
  fit for \sne with $\RV\sim3$ while a  power-law is preferred for the low-$\RV$ \snjj.
  \item For the two reddest \sne, both with $\EBV>1$~mag, we observe significantly different values of 
  $\RV$ with $\RV=2.8\pm0.1$ and $\RV=1.4\pm0.1$, for \sn \cu and \jj respectively.  Based on their \NaID profiles
  being consistent with the observed H\,\textsc{i} velocity fields in their host galaxies, the lack of thermal emission
  and the observed polarization being aligned with the host galaxy plane for \snjj, we conclude that the reddening
  is likely dominated by extinction in the interstellar medium of the host galaxy.  
  Further, pre-existing circumstellar dust at radii $r_\mathrm{dust}<10^{19}$~cm is expected to give rise to evolution of  
  the colour excesses, $E(\mathrm{\Xband}-\mathrm{\Vband})$, but this is not observed for either 
  \sncu nor \snjj.
\item We compare nebular spectra of \sne \fe, \cg and \jj and conclude that all three are very similar at epochs $+280$~days
  from maximum, with \sne \fe and \cg being remarkably similar.
\item We compare the \stis spectra at maximum of \sne \by, \fe and \cg and conclude that while \sncg is well matched to
  both \sne at $\lambda>3800$~\AA\ it is better matched to \snby at shorter wavelengths.
\item For \sncg we observe time-evolution in $E(\mathrm{\uvmtwo}-\mathrm{\Vband})$ with respect to \snfe, but when 
  the object is compared to \snby the evolution is less significant.  Circumstellar dust at $r_\mathrm{dust}\sim10^{17}$~cm
  is consistent with $E(\mathrm{\uvmtwo}-\mathrm{\Vband})$ but disfavoured by the lack of evolution of the redder colours.
  However, this example illustrates that observations over a wide wavelength range provides a powerful tool for separating 
  the origin of colour excess evolution of \sneia.
\item For low-reddening \sne such as \sncg, adding UV data to optical and near-IR can decrease the statistical
  uncertainty of $\RV$ by $>50\,\%$.
\item The two \sne with low $\RV$ are found to have the highest \SiII velocities at maximum which is consistent with 
  previous studies \citep{2009ApJ...699L.139W}.
\item By measuring the equivalent width from absorption of \NaI, we confirm the results of \citet{2013ApJ...779...38P},  and find
  that the extinction derived for \sne \cg, \cu and \et using the \citet{2012MNRAS.426.1465P} relation is higher than what we 
  observe directly for these objects.
\end{enumerate}

Future studies, including a larger sample of low reddening objects covering a wide range in host galaxy morphologies,
will be necessary to robustly estimate  the impact of the diversity
in the wavelength dependence of reddening on the estimates of cosmological parameters from \sneia. The current
work, clearly showing that there is significant range in the extinction parameters, and possibly also extinction laws,
should motivate further multi-wavelength studies of \sneia, also including the UV.

\section*{Acknowledgement}
We would like to thank Denise Taylor at Space Telescope Science Institute for advising and assisting 
us in carrying out this program.  We would also like to thank Livia Vallini for carrying out observations of \sncg
during the Nordic Millimetre and Optical/NIR Astronomy Summer School 2012.  We are grateful to the anonymous
referee for thoroughly going through the manuscript and providing us with many useful comments.
R.A. and A.G. acknowledge support from the Swedish Research Council 
and the Swedish Space Board.  
P. J. Brown and the \swift Optical/Ultraviolet Supernova Archive (SOUSA) are supported by NASA's Astrophysics 
Data Analysis Program through grant NNX13AF35G.
The work of P.S. is sponsored by FCT - Funda\c{c}\~ao para a Ci\^encia e Tecnologia, under the grant 
SFRH / BD / 62075 / 2009.
V.S. acknowledges support from 
Funda\c{c}\~{a}o para a Ci\^{e}ncia e a Tecnologia (Ci\^{e}ncia 2008) and grant PTDC/CTE-AST/112582/2009.
NER acknowledges the support from the European Union Seventh Framework Programme (FP7/2007-2013) under 
grant agreement n. 267251 ``Astronomy Fellowships in Italy" (AstroFIt).
The Oskar Klein Centre is funded by the Swedish Research Council.
E.Y.H. acknowledge the generous support provided by the Danish Agency for Science and Technology and Innovation through a Sapere Aude Level 2 grant.
The Dark Cosmology Centre is funded by the DNRF. The research leading 
to these results has received funding from the European Research Council under the European Union's 
Seventh Framework Program (FP7/2007-2013)/ERC Grant agreement no. EGGS-278202.
Observations were made with the {\it Hubble Space Telescope}; 
the Nordic Optical Telescope, operated by the Nordic Optical Telescope Scientific Association at the 
Observatorio del Roque de los Muchachos, La Palma, Spain.
The data presented here were obtained in part with ALFOSC, which is provided by the Instituto de 
Astrofisica de Andalucia (IAA) under a joint agreement with the University of Copenhagen and NOTSA.
STSDAS and PyRAF is a product of the Space Telescope Science Institute, which is operated by AURA for NASA. 
\appendix

\section{Summary of \sne studied in this work}\label{sec:snsummary}
The seven \sneia that we study in this work are briefly summarised below:
\begin{itemize}
\item{\bf\snbl} was discovered by the Chilean Automatic Supernova Search (CHASE)  $42\farcs9$ east and 
$14\farcs6$ north of the center of the galaxy ESO~234-19 on Mar.~26.38~UT from an unfiltered image 
\citep{2012CBET.3076....1P}.   A spectrum \citep{2012CBET.3076....2P} was obtained with the 2.5~m du~Pont 
telescope at the Las Campanas Observatory on Mar.~27.28 and object was classified as a \snia using SNID
\citep{2007ApJ...666.1024B}.
\item{\bf\snbm}, located $10\farcs95$ west and $10\farcs4$ north of the centre of the galaxy UGC~8189, was discovered on 
Mar.~27th by \citet{2012CBET.3077....1P}.  A spectrum was obtained on Mar.~28.05~UT and it was classified 
\citep{2012CBET.3077....2C} as a \snia using GELATO \citep{2008A&A...488..383H}.
%
\item{\bf\sncg} was discovered \citep{2012CBET.3111....1K} by the Lick Observatory Supernova Search 
\citep[LOSS,][]{2001ASPC..246..121F}  on May~17.220~UT. The redder than normal SN was located $17\farcs3$ east 
and $1\farcs5$ south \citep{2012CBET.3111....2C} of the center of the Virgo Cluster member NGC~4424 in a region with 
many blue stars and disturbed dust lanes.
%
\citet{2012ATel.4226....1G} used pre-explosion {\em HST} images of this well-studied galaxy to rule-out most supergiants 
as possible binary companions for the progenitor scenario.
\item{\bf\sncp} was discovered $6\farcs7$ east and $1\farcs2$ south of the center of UGC~8713 on May 23.2~UT by the
Puckett Observatory Supernova Search \citep{2012CBET.3130....1C}. \citet{2012CBET.3130....3M} and 
\citet{2012CBET.3130....2Z} obtained spectroscopy of the object on May~25.3 and May~25.7, respectively, and 
reported that it was consistent with a \snia before maximum.
\item{\bf\sncu} was discovered on June 14.6~UT by \citet{2012CBET.3146....1I} $3\farcs1$ east and 
$27\farcs1$ south of the nucleus of the galaxy NGC~4772 and later classified as a \snia by \citet{2012CBET.3146....2M}. 
Pre-explosion {\em Chandra} observations are available of the host galaxy and were used by \citet{2013MNRAS.435..187N} 
to constrain the X-ray emission from the progenitor scenario.
\item{\bf\snet}, located $5\farcs3$ east and $0\farcs8$ north of the center of MCG~+04-55-47, was discovered by
\citet{2012CBET.3226....1R} on Sep.~12.057~UT, and later classified on Sept. 13.80 UT as a high-velocity \sneia
\citep{2012CBET.3226....2D}.
\item{\bf\snjj} was discovered by \cite{2014CBET.3792....1F} in the nearby galaxy M82 on January 21.805~UT, and later 
classified by \citet{ATEL5786}.  The \sn was located $54''$ west and $21''$ south of the ill-defined nucleus of the
host galaxy.
\end{itemize}

\section{A spectral and colour model from \snfe}\label{sec:femodel}
\snfe is a normal \snia with excellent temporal and wavelength coverage with negligible extinction along the 
line-of-sight and is therefore a suitable object for reddening studies.  In \jjpaper we used the available UV-NIR
spectroscopic time-series \citep{2013A&A...554A..27P,2014MNRAS.439.1959M} and the NIR light curves 
\citep{2012ApJ...754...19M} to derive the extinction law of \snjj.  In this work we extend this template with UV
photometry \citep{2012ApJ...753...22B} and optical photometry at $25$ to $40$ days past \Bband-band 
maximum.

The light curves in the UV \wfcone, \wfctwo, \wfcthree and the optical \Bband\Vband\Rband\iband 
bands are first constructed by synthesizing the spectra from \cite{2014MNRAS.439.1959M} and 
\cite{2013A&A...554A..27P}.  A light curve for each filter is then obtained by fitting smoothed splines 
using \snoopy \citep{2011AJ....141...19B}.  For the redder optical \Rband and \iband bands, the sparse temporal 
coverage after 20~days past \Bband-maximum does not allow an accurate spline fit of the second bump.  We 
therefore also use the data from \citet{2013NewA...20...30M} to constrain the spline fit at $>25$~days past 
\Bband-max.  For the \swift/\uvot\ \uvmtwo filter and the NIR \Jband\Hband\Ksband we fit spline models to the 
measured light curves from \cite{2012ApJ...753...22B} and \cite{2012ApJ...754...19M} respectively.  
The light curves are then used to calculate the $(\mathrm{\Xband}-\mathrm{\Vband})_0$ colours shown
in Figure~\ref{fig:colors}.

As explained in \S\,\ref{sec:lawfitting} we also need a template of the spectral energy density of the 
\emph{unreddened} object in order to be able to calculate the expected extinction for each passband for a given 
extinction model.  In \jjpaper the extinction for a given set of reddening law parameters was first calculated for all 
available spectra of \snfe and the extinction for any given epoch was then obtained using spline interpolations.  Similar 
to the light curve spline model, this approach is adequate for the UV, but due to the sparse spectral coverage, it will 
cause the model to deviate from the light curve for the redder optical bands and NIR at $>25$~days past maximum.  
In \jjpaper this did not affect the final results due to the high extinction of \snjj and the conservative intrinsic colour 
dispersions adopted in the analysis.

\begin{figure}
\centering
\includegraphics[width=\columnwidth]{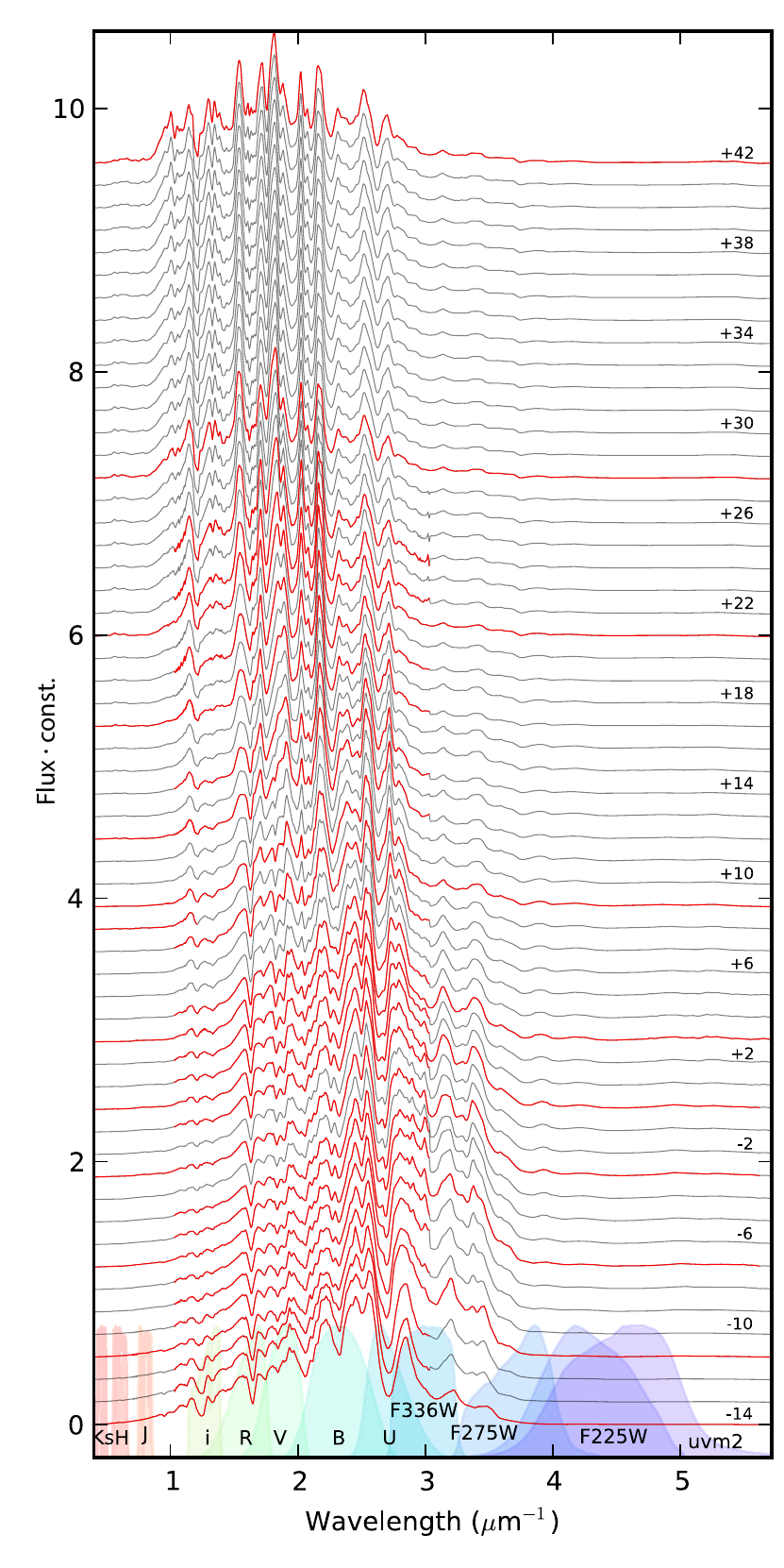}
\caption{%
  The spectral template constructed from measured spectra of \snfe.  Sections based on measured data
  are shown in red, while the interpolated sections, warped to the light curves are shown in 
  grey.  The filters used for the warping are shown at the bottom of the figure.\label{fig:sedmodel}}
\end{figure}

\begin{figure}
\centering
\includegraphics[width=\columnwidth]{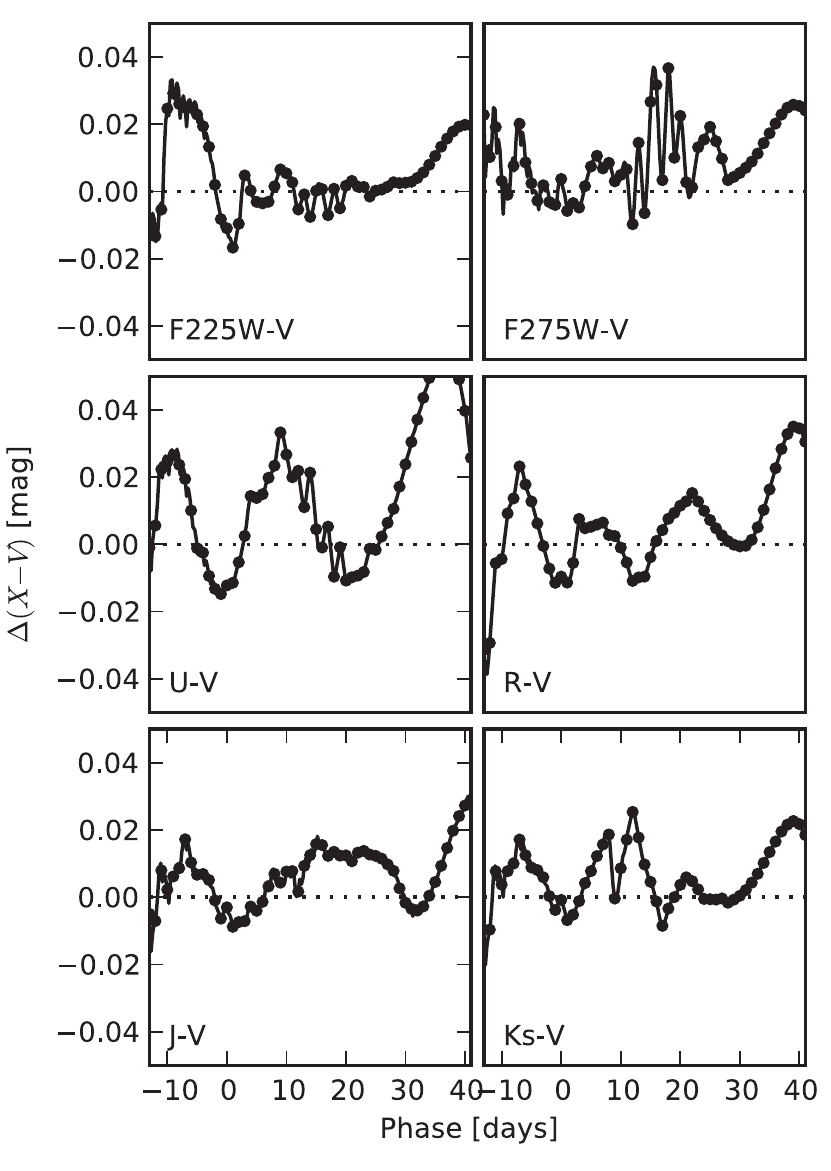}
\caption{%
  Examples of the difference between synthetic colours calculated from the SED template, and the 
  corresponding colours from the light curve spline templates.
\label{fig:modelcomp}}
\end{figure}

In this work we have developed a more accurate SED template by first resampling the spectral time series used 
in \jjpaper to a resolution of $10$~\AA, which are shown as the red spectra in Figure~\ref{fig:sedmodel}. 
These were then used to calculate the missing wavelength and temporal elements through linear interpolation (shown
in grey in Figure~\ref{fig:sedmodel}), and finally the full matrix is mangled to match the light curves described above.
The mangling was carried out by first calculating the ratio between the light curve fluxes and the corresponding synthetic 
SED fluxes.  A function was then setup for each phase where the calculated ratios were adopted for the effective 
wavelengths of the filters.  For all wavelengths shorter (longer) than the bluest (reddest) point, the function was assumed
to take the same value as in the extreme points, while we interpolate the function value for all intermediate wavelengths,
using the calculated ratios.  The SED for the given phase was then multiplied by the function.  The difference between
the synthetic colours calculated from the mangled SED and the spline colour model was always found to be within 
$<0.05$~mag for all epochs, and often even smaller.  In Figure~\ref{fig:modelcomp} these differences are shown
for six different colours.  Since these differences are within the adopted intrinsic colour uncertainties they will not 
have any significant impact on the fitted reddening laws.

\section{A model for the UV of \snby}\label{sec:bytemplate}
The \sne \by and \fe have been found to be remarkably similar in the optical, both around maximum and
at nebular phases, while showing significant discrepancies in the UV 
\citep{2013ApJ...769L...1F,2015MNRAS.446.2073G}.  In Figure~\ref{fig:nuvratio} the ratio 
between the spectra are shown at maximum to illustrate this.  Unfortunately, \snby does not have 
the wavelength and temporal coverage that \snfe has, but we can create a crude template for \snby 
based on the assumption that it is indeed identical to \snfe with the exception of the UV wavelengths.


In Figure~\ref{fig:nuvratio} the, solid, green line is showing the function
\begin{equation}
f(x')  =  c\left[\frac{x'}{\left(x'^4 + 1\right)^{1/4}} - 1\right] + 1\quad\textrm{with}\quad
x'  = \frac{x - x_0}{\alpha}
\label{eq:ratiofunc}
\end{equation}
where $x$ is the wavelength given in microns.  In the figure the function is plotted for the parameters
$x_0 = 0.2495$, $\alpha = 0.0360$ and  $c=0.3359$. 

\begin{figure}
  \begin{center}
    \includegraphics[width=\columnwidth]{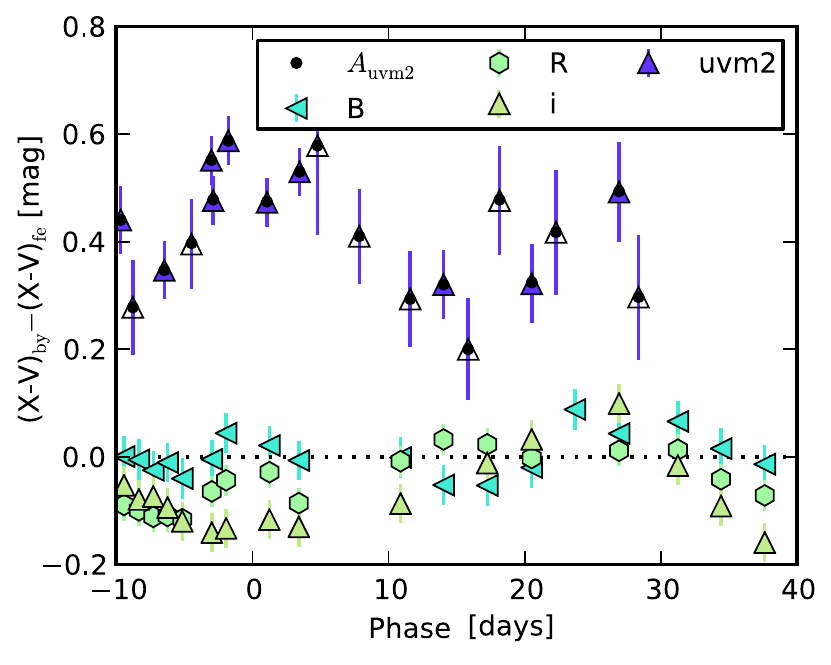}
    \caption{%
    The difference in colour between \sne \by and \fe, where the ground based photometry 
    has been adopted from \citet{2013MNRAS.430.1030S}.    
    Observations for which the colour was obtained 
	from measured \Vband-band data are shown with filled symbols, while values obtained from a \Vband-band template 
	are shown with open symbols.
	The colour difference between the \sne can also be calculated synthetically based on Equation~\eqref{eq:ratiofunc} under
	the assumption that this function describes the flux ratio of the two objects, for all epochs, after adjusting for the difference
	in distance. This is motivated by the observation that these \sne appear to be twins at all wavelengths except for the UV.
	The parameter $c$ in the equation was fitted by matching the synthetic and measured  \uvmtwo$-$\Vband for
	each epoch, where the former is shown by the black points for the best fitted values.
    \label{fig:exvbyfe}}
  \end{center}
\end{figure}

\snby was observed by \swift in the UV \citep{2013ApJ...779...23M} and in the optical 
\citep{2013MNRAS.430.1030S}, and the colour excesses of these measurements with respect to \snfe are 
shown in Figure~\ref{fig:exvbyfe}.  Here we note that $E(\mathrm{\uvmtwo}-\mathrm{\Vband})$
is showing a similar time-evolution as was seen for \sncg in the upper-left panel of Figure~\ref{fig:afits}.

\begin{figure}
  \begin{center}
    \includegraphics[width=\columnwidth]{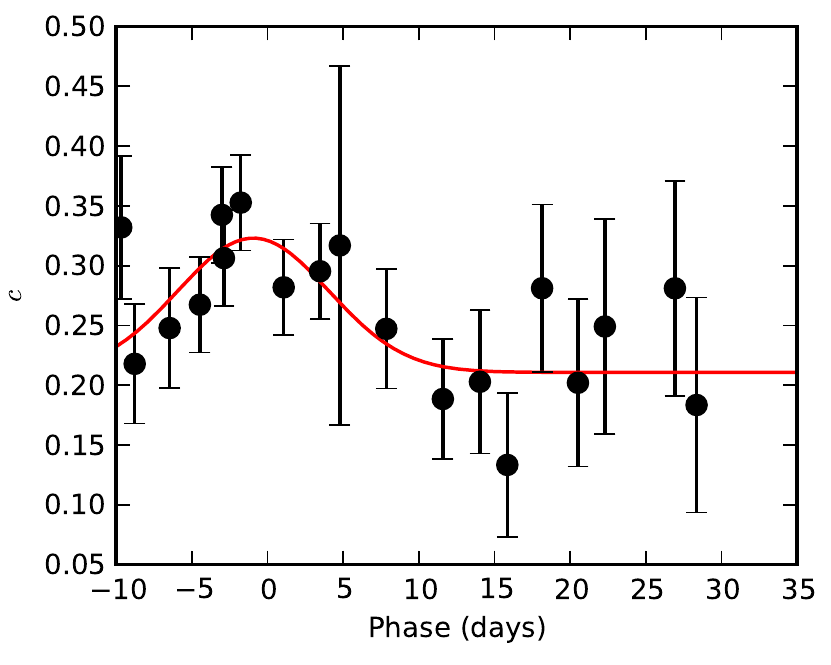}
    \caption{%
    	The fitted values of parameter $c$ in after the SED of \snfe has been fitted together with 
	Equation~\eqref{eq:ratiofunc} to match \snby\ \uvmtwo photometry.  The solid line is showing
	a Gaussian fit.
	\label{fig:constant}}
  \end{center}
\end{figure}

Under the assumption that \sne \by and \fe are twins at wavelengths $\lambda>3000$~\AA\ we can create
an approximate template for \snby by multiplying the SED of \snfe, presented in Appendix~\ref{sec:femodel}
and shown in Figure~\ref{fig:sedmodel},  by Equation~\eqref{eq:ratiofunc}.  Further, if we fix the 
parameters $x_0$ and $\alpha$ to the values given above, we can fit $c$ for each epoch with 
\uvmtwo observations by forcing the synthetically calculated $A_\mathrm{\uvmtwo}$, based on the SED to
match the observed $E(\mathrm{\uvmtwo}-\mathrm{\Vband})$ in Figure~\ref{fig:exvbyfe} 
(again, assuming that $\mathrm{\Vband}^\mathrm{\fe}=\mathrm{\Vband}^\mathrm{\by}$).
The fitted values of $c$ are shown in Figure~\ref{fig:constant} and the synthetic values of $A_\mathrm{\uvmtwo}$ 
that these values correspond to are been plotted as black dots in Figure~\ref{fig:exvbyfe}.

Further, we parametrise $c(p)$ with a Gaussian, 
\begin{displaymath}
c(p) = \frac{a}{\sqrt{2\pi}\sigma}\exp\left[-\frac{1}{2}\left(\frac{p - p_0}{\sigma}\right)^2\right] + b\,,
\end{displaymath}
as shown by the solid line in the Figure~\ref{fig:constant}, with $p_0 = -0.939$, $a=1.406$, $b=0.211$ and $\sigma=4.992$.

To summarise, we create a crude light curve and SED template of \snby by using the template of \snfe from 
\S\,\ref{sec:femodel} and multiply it with Equation~\eqref{eq:ratiofunc}, where the parameters, including the
time-dependent $c$, are given above.

\section{Red tails of the \hst filters}\label{sec:redtail}
\begin{figure}
\centering
    \includegraphics[width=\columnwidth]{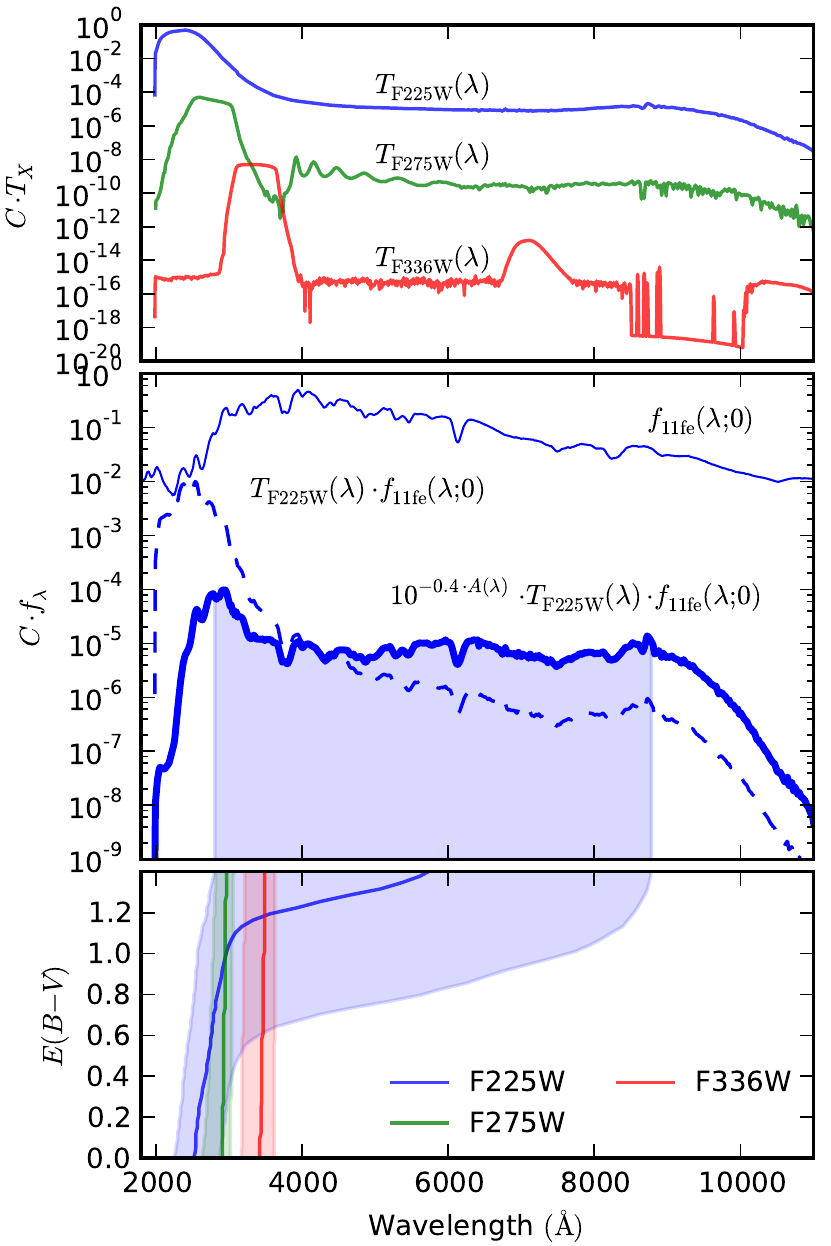}
    \caption{%
    		The  {\it upper} panel shows the filter transmissions, $T_X(\lambda)$, of the \wfcuvis passbands, multiplied with 
		an arbitrary normalisation, $C$.  The thin, blue, line in the {\it middle} panel shows the spectrum of \snfe at \Bband 
		band maximum, while the dashed, blue, line shows the same spectrum after it has been multiplied by the \wfcone 
		throughput. The thick, blue, line shows the same spectrum after it has been reddened with a \ftz extinction law with
		$\EBV=1.4$~mag and $\RV=1.4$, similar to what we observe for \snjj.  All three spectra have arbitrary normalisation, 
		$C$. In the plot we have also marked the wavelength region from which $80\,$\% of the light originates from, defined 
		such that $10\,$\% of light comes from bluer wavelengths, and $10\,$\% comes from redder wavelengths.  The 
		{\em lower} panel shows the median wavelength, \ie where $50\,$\% of the light originates from redder and bluer 
		wavelengths, for three passbands for \snfe at  \Bband maximum, reddened with a \ftz extinction law ($\RV=1.4$) with 
		increasing extinction.  The $80\,$\% region is shown in the same manner as for the middle panel. \label{fig:leak}}
\end{figure}

The \hst filter transmission used in this work, shown in the upper panel of Figure~\ref{fig:leak}, suffer from red 
tails. That is, part of the flux observed for an object in these filters originates from wavelengths that are redder 
than the range where most of the transmission lies.  As mentioned in \jjpaper, the red tails are relatively small
for the \wfcuvis filters,  e.g. the transmission of \wfcone is roughly one part in $100\,000$ at $5000$~\AA\ compared 
to the peak transmission at $\sim2250$~\AA.  Despite this, the red tails will have implications when reddened 
\sneia are studied.   For example, from spectrophotometry of \snfe at maximum (thin, blue line in the middle 
panel of Figure~\ref{fig:leak}),  we conclude that $\sim 0.5\,$\% of the flux (dashed blue line) in the \wfcone filter will 
originate from wavelengths $>4000$~\AA.  However, extinction will suppress the UV-flux significantly and after 
applying a \ftz reddening law with $E(B-V)=1.4$, $R_V=1.4$,  $\sim60\,\%$ of the \wfcone flux (thick, blue line) 
comes from photons with $\lambda>4000$~\AA.

\begin{figure}
  \begin{center}
    \includegraphics[width=0.98\columnwidth]{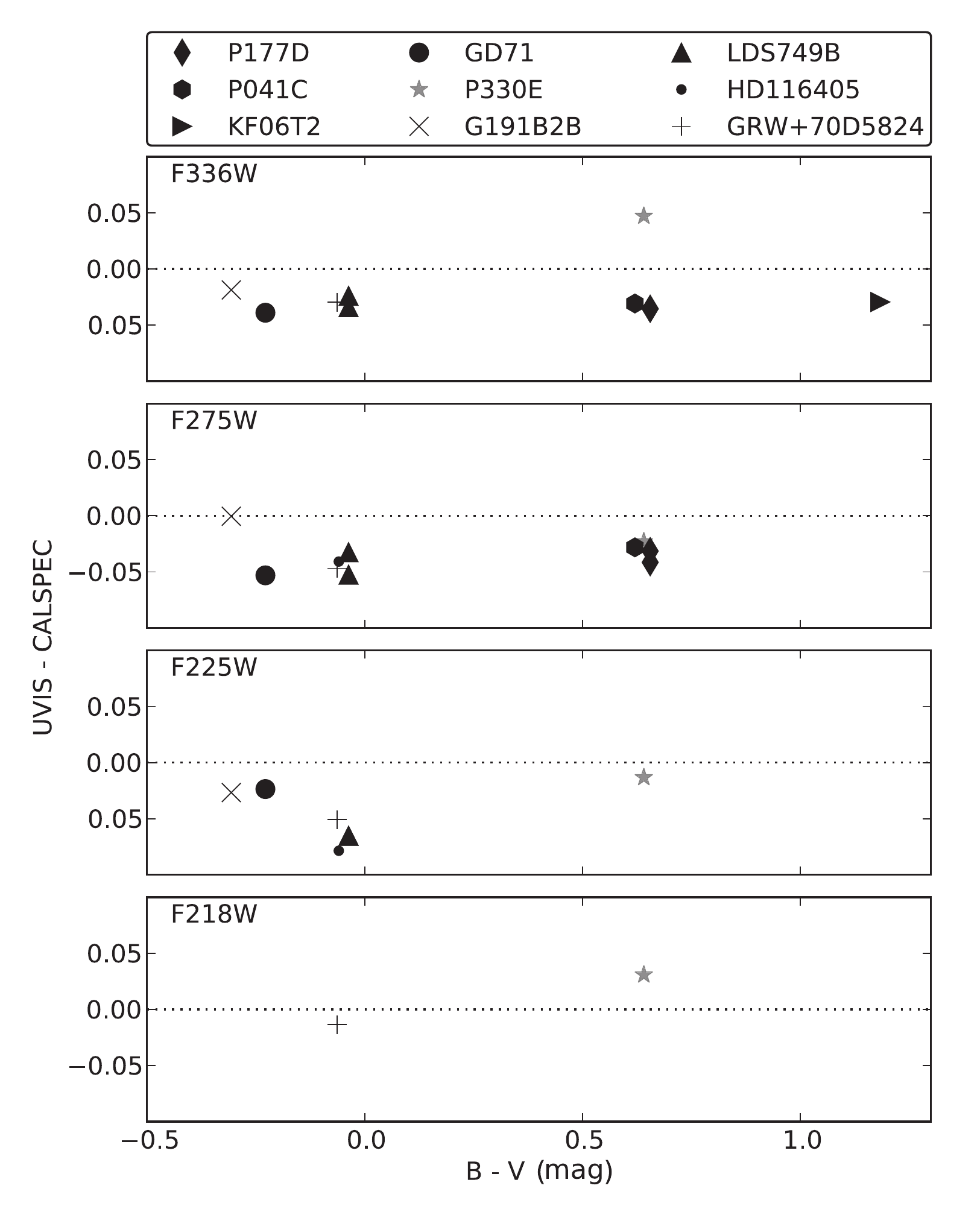}
    \caption{%
    	Comparison between spectrophotometry of \calspec stars using the \wfcuvis filter
	throughputs, with the corresponding \wfcuvis photometry against the synthetically
	calculated \Bband$-$\Vband colour. The star P330E was used for calibrating \wfcuvis.
    \label{fig:uviscalspec}}
  \end{center}
\end{figure}

Another way to illustrate the effect of the red tails is shown in the lower panel of Figure~\ref{fig:leak} where
the solid lines show the effective median wavelengths for an observation of \snfe at maximum through the 
\wfc filters with various amounts of reddening applied.  The effective median wavelength is 
defined such that $50\,$\% of the light originates from redder and bluer wavelengths respectively.  Further, 
the wavelength range defined by the points where $10\,$\% and $90\,$\% or the light comes from bluer
wavelengths (\ie the region that includes $80\,$\% of the light) is shown for increasing reddening.
From the figure we conclude that, \wfctwo and \wfcthree are less plagued by red tails than \wfcone,
where the latter, in fact, will be probing redder effective wavelengths of the \sn spectrum than the two 
former filters for $\EBV\gtrsim1$~mag.

The fact that we are comparing different effective wavelengths when calculating 
the colour excess between a reddened and unreddened source is true for all broad band 
observations.  However, fitting extinction properties based on filters with red tails that span 
a wavelength range where the SED drops rapidly, also puts demands on knowing the passband 
throughput accurately even when the transmission drops several orders of magnitude as for the 
filters in Figure~\ref{fig:leak}.

The expected \wfcuvis passband red tails were tested \citep{2011brown} in the final round of 
thermal vacuum tests of WFC3 prior to launch.  They studied the filter throughputs in
four wavelength bins centered on $450$, $600$, $750$ and $900$~nm and found deviations of up
to $\sim40\,$\% between the measured and expected transmission.  If the deviations are taken
as an estimate of the throughput uncertainty at these wavelengths, we can evaluate its impact 
on the extinction analysis.  

We carried out this test by first applying a \ftz reddening law with  $E(B-V)=1.4$, $R_V=1.4$ to the
SED of \snfe described in  Section~\ref{sec:intrinsic}.  The impact on the calculated extinction for the 
two bands \wfcone and \wfctwo \cite[the test of \wfcthree was not carried out due to 
limitations in the setup as described in][]{2011brown} can then be calculated.  We do this by comparing 
synthetic magnitudes from the original \synphot throughputs, to magnitudes obtained after the filters have 
been multiplied by a spline function derived from the deviations for the four wavelength bins mentioned above.
The impact on the calculated extinction was found to always be $<0.05$~mag.  Further, the worst case 
scenario can be estimated by letting all the measured deviations from \cite{2011brown} maximise 
the amount of red light entering the filter.  The result on the extinction from this was $<0.10$~mag for
$t<20$~days with respect to \Bband max and $0.15$--$0.20$~mag for $t = 20$--$40$~days, which is
still within the dispersion that we have adopted for the intrinsic UV colours.

It is also possible to investigate the accuracy of the full {\em HST/WFC3/UVIS} \synphot throughput by 
comparing spectrophotometry of objects, with known spectra, to \wfcuvis photometry.  This is
in fact the most direct test we can carry out of our method for fitting extinction laws since this is using
the very same procedure. In figure~\ref{fig:uviscalspec} we have plotted stars from %
\calspec\footnote{\url{http://www.stsci.edu/hst/observatory/crds/calspec.html}} with the
corresponding UVIS-photometry using the same analysis path as in Section~\ref{sec:hstdata}
with a 0.4" radius.  The agreement is in general more than sufficient for the purpose of this analysis,
but it is unfortunate that data are not available for redder stars for which the impact of the red tail can 
be expected to be more significant.  The star P330E should be discarded from the comparison
since this star has been used for calibrating the \wfcuvis system.

However, we can carry out the same test for the reddened \sneia that have been observed with both \stis 
and \wfcuvis.  In Table~\ref{tb:stiswfcsynphot} we present spectrophotometry at maximum for \sne \cg and
\jj, using the transmissions of the \wfcuvis filters,  compared to the corresponding \wfcuvis photometry.  For \jj
the spectrum and photometry were obtained on consecutive days, and the magnitudes based on the former
were correct to correspond to the \wfcuvis epoch using the \snfe template.  These corrections are $<0.1$~mag.

For these two \sne, the discrepancies between two instruments are within $0.1$~mag, including 
\wfcone.  This is the filter with the most significant red tails, and the photon wavelength distribution for \wfcone
is significantly different for two \sne.  The consistency between spectrophotometry and \wfcuvis photometry for 
two \sne that define the end-points of the reddening range in our study is showing that any errors of the filter 
transmissions will not affect the conclusions drawn in the paper.

Further, the fact that we also observe a discrepancy of $\sim0.1$~mag for the \wfcthree filter, which has minor red 
tails, suggests that the discrepancies we observe for \wfcone may not necessarily originate from uncertainties of
these.

\begin{table}
\centering
\begin{tabular}{l l l l@{}r}
\hline\hline
\multicolumn{1}{c}{\sn} &
\multicolumn{1}{c}{Filter} & 
\multicolumn{1}{c}{\stis} &
\multicolumn{2}{c}{\wfc}\\
\hline
\multirow{3}{*}{\rotatebox[origin=c]{90}{\small \cg}}
& \wfcone   &  15.89 & 15.98&$\,(0.05)$\\
& \wfctwo    &  13.67 & 13.61&$\,(0.05)$\\
& \wfcthree &  11.85 & 11.74&$\,(0.06)$\\
\hline
\multirow{3}{*}{\rotatebox[origin=c]{90}{\small \jj}}
& \wfcone   &  18.43 & 18.50&$\,(0.04)$\\
& \wfctwo    &  16.55 & 16.55&$\,(0.04)$\\
& \wfcthree &  13.15 & 13.02&$\,(0.04)$\\
\hline\hline
\end{tabular}
\caption{%
	Spectrophotometry of \stis spectra at maximum of \sne \cg and \jj at maximum using the filter
	transmissions shown in Figure~\ref{fig:leak} compared to \wfcuvis photometry of the same objects.
	For \sncg the observations were obtained in consecutive \hst orbits, while the data for \snjj was
	taken 1~day apart.  For this \sn, we have calculated the magnitude difference synthetically from the
	\snfe spectra described in Appendix~\ref{sec:femodel}.  Since the colour evolution is slow 
	between the two observations, any discrepancy in the filter transmissions will be of second order for 
	this correction. 
\label{tb:stiswfcsynphot}}
\end{table}

%

\bibliography{hstred}

\begin{thebibliography}{185}
\expandafter\ifx\csname natexlab\endcsname\relax\def\natexlab#1{#1}\fi

\bibitem[{{Amanullah} \& {Goobar}(2011)}]{2011ApJ...735...20A}
{Amanullah} R., {Goobar} A., 2011, \apj, 735, 20

\bibitem[{{Amanullah} {et~al}\mbox{.}(2014){Amanullah}, {Goobar}, {Johansson},
  {Banerjee}, {Venkataraman}, {Joshi}, {Ashok}, {Cao}, {Kasliwal}, {Kulkarni},
  {Nugent}, {Petrushevska}, \& {Stanishev}}]{2014ApJ...788L..21A}
{Amanullah} R. {et~al.}, 2014, \apjl, 788, L21

\bibitem[{{Amanullah} {et~al}\mbox{.}(2010){Amanullah}, {Lidman}, {Rubin},
  {Aldering}, {Astier}, {Barbary}, {Burns}, {Conley}, {Dawson}, {Deustua},
  {Doi}, {Fabbro}, {Faccioli}, {Fakhouri}, {Folatelli}, {Fruchter}, {Furusawa},
  {Garavini}, {Goldhaber}, {Goobar}, {Groom}, {Hook}, {Howell}, {Kashikawa},
  {Kim}, {Knop}, {Kowalski}, {Linder}, {Meyers}, {Morokuma}, {Nobili},
  {Nordin}, {Nugent}, {{\"O}stman}, {Pain}, {Panagia}, {Perlmutter}, {Raux},
  {Ruiz-Lapuente}, {Spadafora}, {Strovink}, {Suzuki}, {Wang}, {Wood-Vasey},
  {Yasuda}, \& {Supernova Cosmology Project}}]{2010ApJ...716..712A}
{Amanullah} R. {et~al.}, 2010, \apj, 716, 712

\bibitem[{{Astier} {et~al}\mbox{.}(2006){Astier}, {Guy}, {Regnault}, {Pain},
  {Aubourg}, {Balam}, {Basa}, {Carlberg}, {Fabbro}, {Fouchez}, {Hook},
  {Howell}, {Lafoux}, {Neill}, {Palanque-Delabrouille}, {Perrett}, {Pritchet},
  {Rich}, {Sullivan}, {Taillet}, {Aldering}, {Antilogus}, {Arsenijevic},
  {Balland}, {Baumont}, {Bronder}, {Courtois}, {Ellis}, {Filiol}, {Gon{\c
  c}alves}, {Goobar}, {Guide}, {Hardin}, {Lusset}, {Lidman}, {McMahon},
  {Mouchet}, {Mourao}, {Perlmutter}, {Ripoche}, {Tao}, \&
  {Walton}}]{2006A&A...447...31A}
{Astier} P. {et~al.}, 2006, \aap, 447, 31

\bibitem[{{Benetti} {et~al}\mbox{.}(2005){Benetti}, {Cappellaro}, {Mazzali},
  {Turatto}, {Altavilla}, {Bufano}, {Elias-Rosa}, {Kotak}, {Pignata}, {Salvo},
  \& {Stanishev}}]{2005ApJ...623.1011B}
{Benetti} S. {et~al.}, 2005, \apj, 623, 1011

\bibitem[{{Benetti} {et~al}\mbox{.}(2004){Benetti}, {Meikle}, {Stehle},
  {Altavilla}, {Desidera}, {Folatelli}, {Goobar}, {Mattila}, {Mendez},
  {Navasardyan}, {Pastorello}, {Patat}, {Riello}, {Ruiz-Lapuente}, {Tsvetkov},
  {Turatto}, {Mazzali}, \& {Hillebrandt}}]{2004MNRAS.348..261B}
{Benetti} S. {et~al.}, 2004, \mnras, 348, 261

\bibitem[{{Betoule} {et~al}\mbox{.}(2014){Betoule}, {Kessler}, {Guy}, {Mosher},
  {Hardin}, {Biswas}, {Astier}, {El-Hage}, {Konig}, {Kuhlmann}, {Marriner},
  {Pain}, {Regnault}, {Balland}, {Bassett}, {Brown}, {Campbell}, {Carlberg},
  {Cellier-Holzem}, {Cinabro}, {Conley}, {D'Andrea}, {DePoy}, {Doi}, {Ellis},
  {Fabbro}, {Filippenko}, {Foley}, {Frieman}, {Fouchez}, {Galbany}, {Goobar},
  {Gupta}, {Hill}, {Hlozek}, {Hogan}, {Hook}, {Howell}, {Jha}, {Le Guillou},
  {Leloudas}, {Lidman}, {Marshall}, {M{\"o}ller}, {Mour{\~a}o}, {Neveu},
  {Nichol}, {Olmstead}, {Palanque-Delabrouille}, {Perlmutter}, {Prieto},
  {Pritchet}, {Richmond}, {Riess}, {Ruhlmann-Kleider}, {Sako}, {Schahmaneche},
  {Schneider}, {Smith}, {Sollerman}, {Sullivan}, {Walton}, \&
  {Wheeler}}]{2014A&A...568A..22B}
{Betoule} M. {et~al.}, 2014, \aap, 568, A22

\bibitem[{{Blondin} {et~al}\mbox{.}(2012){Blondin}, {Matheson}, {Kirshner},
  {Mandel}, {Berlind}, {Calkins}, {Challis}, {Garnavich}, {Jha}, {Modjaz},
  {Riess}, \& {Schmidt}}]{2012AJ....143..126B}
{Blondin} S. {et~al.}, 2012, \aj, 143, 126

\bibitem[{{Blondin} {et~al}\mbox{.}(2009){Blondin}, {Prieto}, {Patat},
  {Challis}, {Hicken}, {Kirshner}, {Matheson}, \&
  {Modjaz}}]{2009ApJ...693..207B}
{Blondin} S., {Prieto} J.~L., {Patat} F., {Challis} P., {Hicken} M., {Kirshner}
  R.~P., {Matheson} T., {Modjaz} M., 2009, \apj, 693, 207

\bibitem[{{Blondin} \& {Tonry}(2007)}]{2007ApJ...666.1024B}
{Blondin} S., {Tonry} J.~L., 2007, \apj, 666, 1024

\bibitem[{{Borkowski}, {Blondin} \& {Reynolds}(2009){Borkowski}, {Blondin}, \&
  {Reynolds}}]{2009ApJ...699L..64B}
{Borkowski} K.~J., {Blondin} J.~M., {Reynolds} S.~P., 2009, \apjl, 699, L64

\bibitem[{{Bourque} \& {Anderson}(2014)}]{2014wfcSTAN}
{Bourque} M., {Anderson} J., 2014, {CTE Correction for UVIS Subarrays Without
  Pre-Scan Bias Pixels}. Tech. rep., STScI

\bibitem[{{Branch} {et~al}\mbox{.}(2006){Branch}, {Dang}, {Hall}, {Ketchum},
  {Melakayil}, {Parrent}, {Troxel}, {Casebeer}, {Jeffery}, \&
  {Baron}}]{2006PASP..118..560B}
{Branch} D. {et~al.}, 2006, \pasp, 118, 560

\bibitem[{{Branch} {et~al}\mbox{.}(2003){Branch}, {Garnavich}, {Matheson},
  {Baron}, {Thomas}, {Hatano}, {Challis}, {Jha}, \&
  {Kirshner}}]{2003AJ....126.1489B}
{Branch} D. {et~al.}, 2003, \aj, 126, 1489

\bibitem[{{Breeveld} {et~al}\mbox{.}(2011){Breeveld}, {Landsman}, {Holland},
  {Roming}, {Kuin}, \& {Page}}]{2011AIPC.1358..373B}
{Breeveld} A.~A., {Landsman} W., {Holland} S.~T., {Roming} P., {Kuin} N.~P.~M.,
  {Page} M.~J., 2011, in American Institute of Physics Conference Series, Vol.
  1358, American Institute of Physics Conference Series, {McEnery} J.~E.,
  {Racusin} J.~L., {Gehrels} N., eds., pp. 373--376

\bibitem[{{Brown}(2014)}]{2014ApJ...796L..18B}
{Brown} P.~J., 2014, \apjl, 796, L18

\bibitem[{{Brown} {et~al}\mbox{.}(2014){Brown}, {Breeveld}, {Holland}, {Kuin},
  \& {Pritchard}}]{2014Ap&SS.354...89B}
{Brown} P.~J., {Breeveld} A.~A., {Holland} S., {Kuin} P., {Pritchard} T., 2014,
  \apss, 354, 89

\bibitem[{{Brown} {et~al}\mbox{.}(2012){Brown}, {Dawson}, {de Pasquale},
  {Gronwall}, {Holland}, {Immler}, {Kuin}, {Mazzali}, {Milne}, {Oates}, \&
  {Siegel}}]{2012ApJ...753...22B}
{Brown} P.~J. {et~al.}, 2012, \apj, 753, 22

\bibitem[{{Brown} {et~al}\mbox{.}(2010){Brown}, {Roming}, {Milne}, {Bufano},
  {Ciardullo}, {Elias-Rosa}, {Filippenko}, {Foley}, {Gehrels}, {Gronwall},
  {Hicken}, {Holland}, {Hoversten}, {Immler}, {Kirshner}, {Li}, {Mazzali},
  {Phillips}, {Pritchard}, {Still}, {Turatto}, \& {Vanden
  Berk}}]{2010ApJ...721.1608B}
{Brown} P.~J. {et~al.}, 2010, \apj, 721, 1608

\bibitem[{{Brown} {et~al}\mbox{.}(2015){Brown}, {Smitka}, {Wang}, {Breeveld},
  {de Pasquale}, {Hartmann}, {Krisciunas}, {Kuin}, {Milne}, {Page}, \&
  {Siegel}}]{2015ApJ...805...74B}
{Brown} P.~J. {et~al.}, 2015, \apj, 805, 74

\bibitem[{{Brown}(2011)}]{2011brown}
{Brown} T., 2011, {Instrument Science Report WFC3}, 2008-49

\bibitem[{{Burns} {et~al}\mbox{.}(2014){Burns}, {Stritzinger}, {Phillips},
  {Hsiao}, {Contreras}, {Persson}, {Folatelli}, {Boldt}, {Campillay},
  {Castell{\'o}n}, {Freedman}, {Madore}, {Morrell}, {Salgado}, \&
  {Suntzeff}}]{2014ApJ...789...32B}
{Burns} C.~R. {et~al.}, 2014, \apj, 789, 32

\bibitem[{{Burns} {et~al}\mbox{.}(2011){Burns}, {Stritzinger}, {Phillips},
  {Kattner}, {Persson}, {Madore}, {Freedman}, {Boldt}, {Campillay},
  {Contreras}, {Folatelli}, {Gonzalez}, {Krzeminski}, {Morrell}, {Salgado}, \&
  {Suntzeff}}]{2011AJ....141...19B}
{Burns} C.~R. {et~al.}, 2011, \aj, 141, 19

\bibitem[{{Calzetti}(2001)}]{2001NewAR..45..601C}
{Calzetti} D., 2001, \nar, 45, 601

\bibitem[{{Cao}, {Kasliwal} \& {McKay}(2014){Cao}, {Kasliwal}, \&
  {McKay}}]{ATEL5786}
{Cao} Y., {Kasliwal} M.~M., {McKay} A., 2014, The Astronomer's Telegram, 5786,
  1

\bibitem[{{Cappellaro} {et~al}\mbox{.}(2012){Cappellaro}, {Pastorello},
  {Tomasella}, {Benetti}, {Fiaschi}, {Ochner}, {Turatto}, \&
  {Valenti}}]{2012CBET.3077....2C}
{Cappellaro} E., {Pastorello} A., {Tomasella} L., {Benetti} S., {Fiaschi} M.,
  {Ochner} P., {Turatto} M., {Valenti} S., 2012, Central Bureau Electronic
  Telegrams, 3077, 2

\bibitem[{{Cardelli}, {Clayton} \& {Mathis}(1989){Cardelli}, {Clayton}, \&
  {Mathis}}]{1989ApJ...345..245C}
{Cardelli} J.~A., {Clayton} G.~C., {Mathis} J.~S., 1989, \apj, 345, 245

\bibitem[{{Cenko} {et~al}\mbox{.}(2012){Cenko}, {Filippenko}, {Silverman},
  {Gal-Yam}, {Pei}, {Nguyen}, {Carson}, \& {Barth}}]{2012CBET.3111....2C}
{Cenko} S.~B., {Filippenko} A.~V., {Silverman} J.~M., {Gal-Yam} A., {Pei} L.,
  {Nguyen} M., {Carson} D., {Barth} A.~J., 2012, Central Bureau Electronic
  Telegrams, 3111, 2

\bibitem[{{Chomiuk} {et~al}\mbox{.}(2012{\natexlab{a}}){Chomiuk}, {Soderberg},
  {Simon}, \& {Foley}}]{2012ATel.4453....1C}
{Chomiuk} L., {Soderberg} A., {Simon} J., {Foley} R., 2012{\natexlab{a}}, The
  Astronomer's Telegram, 4453, 1

\bibitem[{{Chomiuk} {et~al}\mbox{.}(2012{\natexlab{b}}){Chomiuk}, {Soderberg},
  {Moe}, {Chevalier}, {Rupen}, {Badenes}, {Margutti}, {Fransson}, {Fong}, \&
  {Dittmann}}]{2012ApJ...750..164C}
{Chomiuk} L. {et~al.}, 2012{\natexlab{b}}, \apj, 750, 164

\bibitem[{{Chotard} {et~al}\mbox{.}(2011){Chotard}, {Gangler}, {Aldering},
  {Antilogus}, {Aragon}, {Bailey}, {Baltay}, {Bongard}, {Buton}, {Canto},
  {Childress}, {Copin}, {Fakhouri}, {Hsiao}, {Kerschhaggl}, {Kowalski},
  {Loken}, {Nugent}, {Paech}, {Pain}, {Pecontal}, {Pereira}, {Perlmutter},
  {Rabinowitz}, {Runge}, {Scalzo}, {Smadja}, {Tao}, {Thomas}, {Weaver}, {Wu},
  \& {Nearby Supernova Factory}}]{2011A&A...529L...4C}
{Chotard} N. {et~al.}, 2011, \aap, 529, L4

\bibitem[{{Conley} {et~al}\mbox{.}(2011){Conley}, {Guy}, {Sullivan},
  {Regnault}, {Astier}, {Balland}, {Basa}, {Carlberg}, {Fouchez}, {Hardin},
  {Hook}, {Howell}, {Pain}, {Palanque-Delabrouille}, {Perrett}, {Pritchet},
  {Rich}, {Ruhlmann-Kleider}, {Balam}, {Baumont}, {Ellis}, {Fabbro},
  {Fakhouri}, {Fourmanoit}, {Gonz{\'a}lez-Gait{\'a}n}, {Graham}, {Hudson},
  {Hsiao}, {Kronborg}, {Lidman}, {Mourao}, {Neill}, {Perlmutter}, {Ripoche},
  {Suzuki}, \& {Walker}}]{2011ApJS..192....1C}
{Conley} A. {et~al.}, 2011, \apjs, 192, 1

\bibitem[{{Conley} {et~al}\mbox{.}(2008){Conley}, {Sullivan}, {Hsiao}, {Guy},
  {Astier}, {Balam}, {Balland}, {Basa}, {Carlberg}, {Fouchez}, {Hardin},
  {Howell}, {Hook}, {Pain}, {Perrett}, {Pritchet}, \&
  {Regnault}}]{2008ApJ...681..482C}
{Conley} A. {et~al.}, 2008, \apj, 681, 482

\bibitem[{{Contreras} {et~al}\mbox{.}(2010){Contreras}, {Hamuy}, {Phillips},
  {Folatelli}, {Suntzeff}, {Persson}, {Stritzinger}, {Boldt}, {Gonz{\'a}lez},
  {Krzeminski}, {Morrell}, {Roth}, {Salgado}, {Jos{\'e} Maureira}, {Burns},
  {Freedman}, {Madore}, {Murphy}, {Wyatt}, {Li}, \&
  {Filippenko}}]{2010AJ....139..519C}
{Contreras} C. {et~al.}, 2010, \aj, 139, 519

\bibitem[{{Cort{\'e}s}, {Kenney} \& {Hardy}(2006){Cort{\'e}s}, {Kenney}, \&
  {Hardy}}]{2006AJ....131..747C}
{Cort{\'e}s} J.~R., {Kenney} J.~D.~P., {Hardy} E., 2006, \aj, 131, 747

\bibitem[{{Cort{\'e}s}, {Kenney} \& {Hardy}(2008){Cort{\'e}s}, {Kenney}, \&
  {Hardy}}]{2008ApJ...683...78C}
{Cort{\'e}s} J.~R., {Kenney} J.~D.~P., {Hardy} E., 2008, \apj, 683, 78

\bibitem[{{Cox} {et~al}\mbox{.}(2012){Cox}, {Newton}, {Puckett}, {Post},
  {Brimacombe}, {Cortini}, {Martignoni}, {Zhang}, {Wu}, {Wang}, \&
  {Milisavljevic}}]{2012CBET.3130....1C}
{Cox} L. {et~al.}, 2012, Central Bureau Electronic Telegrams, 3130, 1

\bibitem[{{Crotts}(2015)}]{2015ApJ...804L..37C}
{Crotts} A.~P.~S., 2015, \apjl, 804, L37

\bibitem[{{Dalcanton} {et~al}\mbox{.}(2009){Dalcanton}, {Williams}, {Seth},
  {Dolphin}, {Holtzman}, {Rosema}, {Skillman}, {Cole}, {Girardi}, {Gogarten},
  {Karachentsev}, {Olsen}, {Weisz}, {Christensen}, {Freeman}, {Gilbert},
  {Gallart}, {Harris}, {Hodge}, {de Jong}, {Karachentseva}, {Mateo}, {Stetson},
  {Tavarez}, {Zaritsky}, {Governato}, \& {Quinn}}]{2009ApJS..183...67D}
{Dalcanton} J.~J. {et~al.}, 2009, \apjs, 183, 67

\bibitem[{{Dennefeld} {et~al}\mbox{.}(2012){Dennefeld}, {Pizzella},
  {Valentini}, {Tomasella}, {Benetti}, {Cortes}, {Ozel}, \&
  {Rajpurohit}}]{2012CBET.3226....2D}
{Dennefeld} M., {Pizzella} A., {Valentini} M., {Tomasella} L., {Benetti} S.,
  {Cortes} M., {Ozel} N., {Rajpurohit} A., 2012, Central Bureau Electronic
  Telegrams, 3226, 2

\bibitem[{{Dilday} {et~al}\mbox{.}(2012){Dilday}, {Howell}, {Cenko},
  {Silverman}, {Nugent}, {Sullivan}, {Ben-Ami}, {Bildsten}, {Bolte}, {Endl},
  {Filippenko}, {Gnat}, {Horesh}, {Hsiao}, {Kasliwal}, {Kirkman}, {Maguire},
  {Marcy}, {Moore}, {Pan}, {Parrent}, {Podsiadlowski}, {Quimby}, {Sternberg},
  {Suzuki}, {Tytler}, {Xu}, {Bloom}, {Gal-Yam}, {Hook}, {Kulkarni}, {Law},
  {Ofek}, {Polishook}, \& {Poznanski}}]{2012Sci...337..942D}
{Dilday} B. {et~al.}, 2012, Science, 337, 942

\bibitem[{{Draine}(2003)}]{2003ApJ...598.1017D}
{Draine} B.~T., 2003, \apj, 598, 1017

\bibitem[{{Elias-Rosa} {et~al}\mbox{.}(2006){Elias-Rosa}, {Benetti},
  {Cappellaro}, {Turatto}, {Mazzali}, {Patat}, {Meikle}, {Stehle},
  {Pastorello}, {Pignata}, {Kotak}, {Harutyunyan}, {Altavilla}, {Navasardyan},
  {Qiu}, {Salvo}, \& {Hillebrandt}}]{2006MNRAS.369.1880E}
{Elias-Rosa} N. {et~al.}, 2006, \mnras, 369, 1880

\bibitem[{{Elias-Rosa} {et~al}\mbox{.}(2008){Elias-Rosa}, {Benetti}, {Turatto},
  {Cappellaro}, {Valenti}, {Arkharov}, {Beckman}, {di Paola}, {Dolci},
  {Filippenko}, {Foley}, {Krisciunas}, {Larionov}, {Li}, {Meikle},
  {Pastorello}, {Valentini}, \& {Hillebrandt}}]{2008MNRAS.384..107E}
{Elias-Rosa} N. {et~al.}, 2008, \mnras, 384, 107

\bibitem[{{El{\'{\i}}asd{\'o}ttir}
  {et~al}\mbox{.}(2006){El{\'{\i}}asd{\'o}ttir}, {Hjorth}, {Toft}, {Burud}, \&
  {Paraficz}}]{2006ApJS..166..443E}
{El{\'{\i}}asd{\'o}ttir} {\'A}., {Hjorth} J., {Toft} S., {Burud} I., {Paraficz}
  D., 2006, \apjs, 166, 443

\bibitem[{{Filippenko} {et~al}\mbox{.}(2001){Filippenko}, {Li}, {Treffers}, \&
  {Modjaz}}]{2001ASPC..246..121F}
{Filippenko} A.~V., {Li} W.~D., {Treffers} R.~R., {Modjaz} M., 2001, in
  Astronomical Society of the Pacific Conference Series, Vol. 246, IAU Colloq.
  183: Small Telescope Astronomy on Global Scales, {Paczynski} B., {Chen}
  W.-P., {Lemme} C., eds., p. 121

\bibitem[{{Firth} {et~al}\mbox{.}(2015){Firth}, {Sullivan}, {Gal-Yam},
  {Howell}, {Maguire}, {Nugent}, {Piro}, {Baltay}, {Feindt}, {Hadjiyksta},
  {McKinnon}, {Ofek}, {Rabinowitz}, \& {Walker}}]{2015MNRAS.446.3895F}
{Firth} R.~E. {et~al.}, 2015, \mnras, 446, 3895

\bibitem[{{Fitzpatrick}(1999)}]{1999PASP..111...63F}
{Fitzpatrick} E.~L., 1999, \pasp, 111, 63

\bibitem[{{Folatelli} {et~al}\mbox{.}(2013){Folatelli}, {Morrell}, {Phillips},
  {Hsiao}, {Campillay}, {Contreras}, {Castell{\'o}n}, {Hamuy}, {Krzeminski},
  {Roth}, {Stritzinger}, {Burns}, {Freedman}, {Madore}, {Murphy}, {Persson},
  {Prieto}, {Suntzeff}, {Krisciunas}, {Anderson}, {F{\"o}rster}, {Maza},
  {Pignata}, {Rojas}, {Boldt}, {Salgado}, {Wyatt}, {Olivares E.}, {Gal-Yam}, \&
  {Sako}}]{2013ApJ...773...53F}
{Folatelli} G. {et~al.}, 2013, \apj, 773, 53

\bibitem[{{Folatelli} {et~al}\mbox{.}(2010){Folatelli}, {Phillips}, {Burns},
  {Contreras}, {Hamuy}, {Freedman}, {Persson}, {Stritzinger}, {Suntzeff},
  {Krisciunas}, {Boldt}, {Gonz{\'a}lez}, {Krzeminski}, {Morrell}, {Roth},
  {Salgado}, {Madore}, {Murphy}, {Wyatt}, {Li}, {Filippenko}, \&
  {Miller}}]{2010AJ....139..120F}
{Folatelli} G. {et~al.}, 2010, \aj, 139, 120

\bibitem[{{Folatelli} {et~al}\mbox{.}(2012){Folatelli}, {Phillips}, {Morrell},
  {Tanaka}, {Maeda}, {Nomoto}, {Stritzinger}, {Burns}, {Hamuy}, {Mazzali},
  {Boldt}, {Campillay}, {Contreras}, {Gonz{\'a}lez}, {Roth}, {Salgado},
  {Freedman}, {Madore}, {Persson}, \& {Suntzeff}}]{2012ApJ...745...74F}
{Folatelli} G. {et~al.}, 2012, \apj, 745, 74

\bibitem[{{Foley} {et~al}\mbox{.}(2014){Foley}, {Fox}, {McCully}, {Phillips},
  {Sand}, {Zheng}, {Challis}, {Filippenko}, {Folatelli}, {Hillebrandt},
  {Hsiao}, {Jha}, {Kirshner}, {Kromer}, {Marion}, {Nelson}, {Pakmor},
  {Pignata}, {R{\"o}pke}, {Seitenzahl}, {Silverman}, {Skrutskie}, \&
  {Stritzinger}}]{2014MNRAS.443.2887F}
{Foley} R.~J. {et~al.}, 2014, \mnras, 443, 2887

\bibitem[{{Foley} \& {Kasen}(2011)}]{2011ApJ...729...55F}
{Foley} R.~J., {Kasen} D., 2011, \apj, 729, 55

\bibitem[{{Foley} \& {Kirshner}(2013)}]{2013ApJ...769L...1F}
{Foley} R.~J., {Kirshner} R.~P., 2013, \apjl, 769, L1

\bibitem[{{F{\"o}rster} {et~al}\mbox{.}(2013){F{\"o}rster},
  {Gonz{\'a}lez-Gait{\'a}n}, {Folatelli}, \& {Morrell}}]{2013ApJ...772...19F}
{F{\"o}rster} F., {Gonz{\'a}lez-Gait{\'a}n} S., {Folatelli} G., {Morrell} N.,
  2013, \apj, 772, 19

\bibitem[{{Fossey} {et~al}\mbox{.}(2014){Fossey}, {Cooke}, {Pollack}, {Wilde},
  \& {Wright}}]{2014CBET.3792....1F}
{Fossey} J., {Cooke} B., {Pollack} G., {Wilde} M., {Wright} T., 2014, Central
  Bureau Electronic Telegrams, 3792, 1

\bibitem[{{Fox} {et~al}\mbox{.}(2011){Fox}, {Chevalier}, {Skrutskie},
  {Soderberg}, {Filippenko}, {Ganeshalingam}, {Silverman}, {Smith}, \&
  {Steele}}]{2011ApJ...741....7F}
{Fox} O.~D. {et~al.}, 2011, \apj, 741, 7

\bibitem[{{Fox} \& {Filippenko}(2013)}]{2013ApJ...772L...6F}
{Fox} O.~D., {Filippenko} A.~V., 2013, \apjl, 772, L6

\bibitem[{{Friedman}(2012)}]{2012PhDT........42F}
{Friedman} A.~S., 2012, PhD thesis, Harvard University

\bibitem[{{Friedman} {et~al}\mbox{.}(2014){Friedman}, {Wood-Vasey}, {Marion},
  {Challis}, {Mandel}, {Bloom}, {Modjaz}, {Narayan}, {Hicken}, {Foley},
  {Klein}, {Starr}, {Morgan}, {Rest}, {Blake}, {Miller}, {Falco}, {Wyatt},
  {Mink}, {Skrutskie}, \& {Kirshner}}]{2014arXiv1408.0465F}
{Friedman} A.~S. {et~al.}, 2014, accepted to ApJS, \texttt{arXiv:1408.0465}

\bibitem[{{Fruchter} \& {et al.}(2010)}]{2010bdrz.conf..382F}
{Fruchter} A.~S., {et al.}, 2010, in 2010 Space Telescope Science Institute
  Calibration Workshop, p. 382-387, pp. 382--387

\bibitem[{{Fynbo} {et~al}\mbox{.}(2013){Fynbo}, {Krogager}, {Venemans},
  {Noterdaeme}, {Vestergaard}, {M{\o}ller}, {Ledoux}, \&
  {Geier}}]{2013ApJS..204....6F}
{Fynbo} J.~P.~U., {Krogager} J.-K., {Venemans} B., {Noterdaeme} P.,
  {Vestergaard} M., {M{\o}ller} P., {Ledoux} C., {Geier} S., 2013, \apjs, 204,
  6

\bibitem[{{Fynbo} {et~al}\mbox{.}(2014){Fynbo}, {Kr{\"u}hler}, {Leighly},
  {Ledoux}, {Vreeswijk}, {Schulze}, {Noterdaeme}, {Watson}, {Wijers}, {Bolmer},
  {Cano}, {Christensen}, {Covino}, {D'Elia}, {Flores}, {Friis}, {Goldoni},
  {Greiner}, {Hammer}, {Hjorth}, {Jakobsson}, {Japelj}, {Kaper}, {Klose},
  {Knust}, {Leloudas}, {Levan}, {Malesani}, {Milvang-Jensen}, {M{\o}ller},
  {Nicuesa Guelbenzu}, {Oates}, {Pian}, {Schady}, {Sparre}, {Tagliaferri},
  {Tanvir}, {Th{\"o}ne}, {de Ugarte Postigo}, {Vergani}, {Wiersema}, {Xu}, \&
  {Zafar}}]{2014A&A...572A..12F}
{Fynbo} J.~P.~U. {et~al.}, 2014, \aap, 572, A12

\bibitem[{{Garavini} {et~al}\mbox{.}(2007){Garavini}, {Folatelli}, {Nobili},
  {Aldering}, {Amanullah}, {Antilogus}, {Astier}, {Blanc}, {Bronder}, {Burns},
  {Conley}, {Deustua}, {Doi}, {Fabbro}, {Fadeyev}, {Gibbons}, {Goldhaber},
  {Goobar}, {Groom}, {Hook}, {Howell}, {Kashikawa}, {Kim}, {Kowalski},
  {Kuznetsova}, {Lee}, {Lidman}, {Mendez}, {Morokuma}, {Motohara}, {Nugent},
  {Pain}, {Perlmutter}, {Quimby}, {Raux}, {Regnault}, {Ruiz-Lapuente},
  {Sainton}, {Schahmaneche}, {Smith}, {Spadafora}, {Stanishev}, {Thomas},
  {Walton}, {Wang}, {Wood-Vasey}, \& {Yasuda}}]{2007A&A...470..411G}
{Garavini} G. {et~al.}, 2007, \aap, 470, 411

\bibitem[{{Gehrels} {et~al}\mbox{.}(2004){Gehrels}, {Chincarini}, {Giommi},
  {Mason}, {Nousek}, {Wells}, {White}, {Barthelmy}, {Burrows}, {Cominsky},
  {Hurley}, {Marshall}, {M{\'e}sz{\'a}ros}, {Roming}, {Angelini}, {Barbier},
  {Belloni}, {Campana}, {Caraveo}, {Chester}, {Citterio}, {Cline}, {Cropper},
  {Cummings}, {Dean}, {Feigelson}, {Fenimore}, {Frail}, {Fruchter}, {Garmire},
  {Gendreau}, {Ghisellini}, {Greiner}, {Hill}, {Hunsberger}, {Krimm},
  {Kulkarni}, {Kumar}, {Lebrun}, {Lloyd-Ronning}, {Markwardt}, {Mattson},
  {Mushotzky}, {Norris}, {Osborne}, {Paczynski}, {Palmer}, {Park}, {Parsons},
  {Paul}, {Rees}, {Reynolds}, {Rhoads}, {Sasseen}, {Schaefer}, {Short},
  {Smale}, {Smith}, {Stella}, {Tagliaferri}, {Takahashi}, {Tashiro},
  {Townsley}, {Tueller}, {Turner}, {Vietri}, {Voges}, {Ward}, {Willingale},
  {Zerbi}, \& {Zhang}}]{2004ApJ...611.1005G}
{Gehrels} N. {et~al.}, 2004, \apj, 611, 1005

\bibitem[{{Goobar}(2008)}]{2008ApJ...686L.103G}
{Goobar} A., 2008, \apjl, 686, L103

\bibitem[{{Goobar} {et~al}\mbox{.}(2014){Goobar}, {Johansson}, {Amanullah},
  {Cao}, {Perley}, {Kasliwal}, {Ferretti}, {Nugent}, {Harris}, {Gal-Yam},
  {Ofek}, {Tendulkar}, {Dennefeld}, {Valenti}, {Arcavi}, {Banerjee},
  {Venkataraman}, {Joshi}, {Ashok}, {Cenko}, {Diaz}, {Fremling}, {Horesh},
  {Howell}, {Kulkarni}, {Papadogiannakis}, {Petrushevska}, {Sand}, {Sollerman},
  {Stanishev}, {Bloom}, {Surace}, {Dupuy}, \& {Liu}}]{2014ApJ...784L..12G}
{Goobar} A. {et~al.}, 2014, \apjl, 784, L12

\bibitem[{{Goobar} {et~al}\mbox{.}(2015){Goobar}, {Kromer}, {Siverd},
  {Stassun}, {Pepper}, {Amanullah}, {Kasliwal}, {Sollerman}, \&
  {Taddia}}]{2015ApJ...799..106G}
{Goobar} A. {et~al.}, 2015, \apj, 799, 106

\bibitem[{{Goobar} \& {Leibundgut}(2011)}]{2011ARNPS..61..251G}
{Goobar} A., {Leibundgut} B., 2011, Annual Review of Nuclear and Particle
  Science, 61, 251

\bibitem[{{Graham} {et~al}\mbox{.}(2015{\natexlab{a}}){Graham}, {Foley},
  {Zheng}, {Kelly}, {Shivvers}, {Silverman}, {Filippenko}, {Clubb}, \&
  {Ganeshalingam}}]{2015MNRAS.446.2073G}
{Graham} M.~L. {et~al.}, 2015{\natexlab{a}}, \mnras, 446, 2073

\bibitem[{{Graham} {et~al}\mbox{.}(2015{\natexlab{b}}){Graham}, {Valenti},
  {Fulton}, {Weiss}, {Shen}, {Kelly}, {Zheng}, {Filippenko}, {Marcy}, {Howell},
  {Burt}, \& {Rivera}}]{2015ApJ...801..136G}
{Graham} M.~L. {et~al.}, 2015{\natexlab{b}}, \apj, 801, 136

\bibitem[{{Graur} \& {Maoz}(2012)}]{2012ATel.4226....1G}
{Graur} O., {Maoz} D., 2012, The Astronomer's Telegram, 4226, 1

\bibitem[{Guy {et~al}\mbox{.}(2007)Guy, Astier, Baumont, Hardin, Pain,
  Regnault, Basa, Carlberg, Conley, Fabbro, Fouchez, Hook, Howell, Perrett,
  Pritchet, Rich, Sullivan, Antilogus, Aubourg, Bazin, Bronder, Filiol,
  Palanque-Delabrouille, Ripoche, \& Ruhlmann-Kleider}]{2007A&A...466...11G}
Guy J. {et~al.}, 2007, \aap, 466, 11

\bibitem[{{Guy} {et~al}\mbox{.}(2007){Guy}, {Astier}, {Baumont}, {Hardin},
  {Pain}, {Regnault}, {Basa}, {Carlberg}, {Conley}, {Fabbro}, {Fouchez},
  {Hook}, {Howell}, {Perrett}, {Pritchet}, {Rich}, {Sullivan}, {Antilogus},
  {Aubourg}, {Bazin}, {Bronder}, {Filiol}, {Palanque-Delabrouille}, {Ripoche},
  \& {Ruhlmann-Kleider}}]{2007AA...466...11G}
{Guy} J. {et~al.}, 2007, \aap, 466, 11

\bibitem[{{Hamuy} {et~al}\mbox{.}(2006){Hamuy}, {Folatelli}, {Morrell},
  {Phillips}, {Suntzeff}, {Persson}, {Roth}, {Gonzalez}, {Krzeminski},
  {Contreras}, {Freedman}, {Murphy}, {Madore}, {Wyatt}, {Maza}, {Filippenko},
  {Li}, \& {Pinto}}]{2006PASP..118....2H}
{Hamuy} M. {et~al.}, 2006, \pasp, 118, 2

\bibitem[{{Harutyunyan} {et~al}\mbox{.}(2008){Harutyunyan}, {Pfahler},
  {Pastorello}, {Taubenberger}, {Turatto}, {Cappellaro}, {Benetti},
  {Elias-Rosa}, {Navasardyan}, {Valenti}, {Stanishev}, {Patat}, {Riello},
  {Pignata}, \& {Hillebrandt}}]{2008A&A...488..383H}
{Harutyunyan} A.~H. {et~al.}, 2008, \aap, 488, 383

\bibitem[{{Haynes} {et~al}\mbox{.}(2000){Haynes}, {Jore}, {Barrett}, {Broeils},
  \& {Murray}}]{2000AJ....120..703H}
{Haynes} M.~P., {Jore} K.~P., {Barrett} E.~A., {Broeils} A.~H., {Murray} B.~M.,
  2000, \aj, 120, 703

\bibitem[{{Hernandez} {et~al}\mbox{.}(2000){Hernandez}, {Meikle}, {Aparicio},
  {Benn}, {Burleigh}, {Chrysostomou}, {Fernandes}, {Geballe}, {Hammersley},
  {Iglesias-Paramo}, {James}, {James}, {Kemp}, {Lister}, {Martinez-Delgado},
  {Oscoz}, {Pollacco}, {Rozas}, {Smartt}, {Sorensen}, {Swaters}, {Telting},
  {Vacca}, {Walton}, \& {Zapatero-Osorio}}]{2000MNRAS.319..223H}
{Hernandez} M. {et~al.}, 2000, \mnras, 319, 223

\bibitem[{{Hicken} {et~al}\mbox{.}(2009){Hicken}, {Challis}, {Jha}, {Kirshner},
  {Matheson}, {Modjaz}, {Rest}, {Wood-Vasey}, {Bakos}, {Barton}, {Berlind},
  {Bragg}, {Brice{\~n}o}, {Brown}, {Caldwell}, {Calkins}, {Cho}, {Ciupik},
  {Contreras}, {Dendy}, {Dosaj}, {Durham}, {Eriksen}, {Esquerdo}, {Everett},
  {Falco}, {Fernandez}, {Gaba}, {Garnavich}, {Graves}, {Green}, {Groner},
  {Hergenrother}, {Holman}, {Hradecky}, {Huchra}, {Hutchison}, {Jerius},
  {Jordan}, {Kilgard}, {Krauss}, {Luhman}, {Macri}, {Marrone}, {McDowell},
  {McIntosh}, {McNamara}, {Megeath}, {Mochejska}, {Munoz}, {Muzerolle},
  {Naranjo}, {Narayan}, {Pahre}, {Peters}, {Peterson}, {Rines}, {Ripman},
  {Roussanova}, {Schild}, {Sicilia-Aguilar}, {Sokoloski}, {Smalley}, {Smith},
  {Spahr}, {Stanek}, {Barmby}, {Blondin}, {Stubbs}, {Szentgyorgyi}, {Torres},
  {Vaz}, {Vikhlinin}, {Wang}, {Westover}, {Woods}, \&
  {Zhao}}]{2009ApJ...700..331H}
{Hicken} M. {et~al.}, 2009, \apj, 700, 331

\bibitem[{{Hsiao} {et~al}\mbox{.}(2007){Hsiao}, {Conley}, {Howell}, {Sullivan},
  {Pritchet}, {Carlberg}, {Nugent}, \& {Phillips}}]{2007ApJ...663.1187H}
{Hsiao} E.~Y., {Conley} A., {Howell} D.~A., {Sullivan} M., {Pritchet} C.~J.,
  {Carlberg} R.~G., {Nugent} P.~E., {Phillips} M.~M., 2007, \apj, 663, 1187

\bibitem[{{Hsiao} {et~al}\mbox{.}(2013){Hsiao}, {Marion}, {Phillips}, {Burns},
  {Winge}, {Morrell}, {Contreras}, {Freedman}, {Kromer}, {Gall}, {Gerardy},
  {H{\"o}flich}, {Im}, {Jeon}, {Kirshner}, {Nugent}, {Persson}, {Pignata},
  {Roth}, {Stanishev}, {Stritzinger}, \& {Suntzeff}}]{2013ApJ...766...72H}
{Hsiao} E.~Y. {et~al.}, 2013, \apj, 766, 72

\bibitem[{{Hutton} {et~al}\mbox{.}(2014){Hutton}, {Ferreras}, {Wu}, {Kuin},
  {Breeveld}, {Yershov}, {Cropper}, \& {Page}}]{2014MNRAS.440..150H}
{Hutton} S., {Ferreras} I., {Wu} K., {Kuin} P., {Breeveld} A., {Yershov} V.,
  {Cropper} M., {Page} M., 2014, \mnras, 440, 150

\bibitem[{{Itagaki} {et~al}\mbox{.}(2012){Itagaki}, {Howerton}, {Noguchi},
  {Nakano}, {Elenin}, {Molotov}, {Marion}, {Milisavljevic}, {Rines},
  {Wilhelmy}, {Zhang}, {Lin}, \& {Wang}}]{2012CBET.3146....1I}
{Itagaki} K. {et~al.}, 2012, Central Bureau Electronic Telegrams, 3146, 1

\bibitem[{Jha, Riess \& Kirshner(2007)Jha, Riess, \&
  Kirshner}]{2007ApJ...659..122J}
Jha S., Riess A.~G., Kirshner R.~P., 2007, \apj, 659, 122

\bibitem[{{Jin} \& {Gao}(2011)}]{2011CBET.2708....1J}
{Jin} Z., {Gao} X., 2011, Central Bureau Electronic Telegrams, 2708, 1

\bibitem[{{Johansson}, {Amanullah} \& {Goobar}(2013){Johansson}, {Amanullah},
  \& {Goobar}}]{2013MNRAS.431L..43J}
{Johansson} J., {Amanullah} R., {Goobar} A., 2013, \mnras, 431, L43

\bibitem[{{Johansson} {et~al}\mbox{.}(2014){Johansson}, {Goobar}, {Kasliwal},
  {Helou}, {Masci}, {Tinyanont}, {Jencson}, {Cao}, {Fox}, {Kromer},
  {Amanullah}, {Banerjee}, {Joshi}, {Jerkstrand}, {Kankare}, \&
  {Prince}}]{2014arXiv1411.3332J}
{Johansson} J. {et~al.}, 2014, submitted to MNRAS, \texttt{arXiv:1411.3332}

\bibitem[{{Kandrashoff} {et~al}\mbox{.}(2012){Kandrashoff}, {Cenko}, {Li},
  {Filippenko}, {Cortini}, {Silverman}, {Gal-Yam}, {Pei}, {Nguyen}, {Carson},
  {Barth}, {Marion}, {Kirshner}, {Foley}, {Challis}, \&
  {Irwin}}]{2012CBET.3111....1K}
{Kandrashoff} M. {et~al.}, 2012, Central Bureau Electronic Telegrams, 3111, 1

\bibitem[{{Kelly} {et~al}\mbox{.}(2015){Kelly}, {Filippenko}, {Burke},
  {Hicken}, {Ganeshalingam}, \& {Zheng}}]{2015Sci...347.1459K}
{Kelly} P.~L., {Filippenko} A.~V., {Burke} D.~L., {Hicken} M., {Ganeshalingam}
  M., {Zheng} W., 2015, Science, 347, 1459

\bibitem[{{Kirshner} {et~al}\mbox{.}(1993){Kirshner}, {Jeffery}, {Leibundgut},
  {Challis}, {Sonneborn}, {Phillips}, {Suntzeff}, {Smith}, {Winkler}, {Winge},
  {Hamuy}, {Hunter}, {Roth}, {Blades}, {Branch}, {Chevalier}, {Fransson},
  {Panagia}, {Wagoner}, {Wheeler}, \& {Harkness}}]{1993ApJ...415..589K}
{Kirshner} R.~P. {et~al.}, 1993, \apj, 415, 589

\bibitem[{{Krisciunas} {et~al}\mbox{.}(2007){Krisciunas}, {Garnavich},
  {Stanishev}, {Suntzeff}, {Prieto}, {Espinoza}, {Gonzalez}, {Salvo}, {Elias de
  la Rosa}, {Smartt}, {Maund}, \& {Kudritzki}}]{2007AJ....133...58K}
{Krisciunas} K. {et~al.}, 2007, \aj, 133, 58

\bibitem[{{Krisciunas} {et~al}\mbox{.}(2006){Krisciunas}, {Prieto},
  {Garnavich}, {Riley}, {Rest}, {Stubbs}, \& {McMillan}}]{2006AJ....131.1639K}
{Krisciunas} K., {Prieto} J.~L., {Garnavich} P.~M., {Riley} J.-L.~G., {Rest}
  A., {Stubbs} C., {McMillan} R., 2006, \aj, 131, 1639

\bibitem[{{Krisciunas} {et~al}\mbox{.}(2003){Krisciunas}, {Suntzeff}, {Candia},
  {Arenas}, {Espinoza}, {Gonzalez}, {Gonzalez}, {H{\"o}flich}, {Landolt},
  {Phillips}, \& {Pizarro}}]{2003AJ....125..166K}
{Krisciunas} K. {et~al.}, 2003, \aj, 125, 166

\bibitem[{{Landolt}(1992)}]{1992AJ....104..340L}
{Landolt} A.~U., 1992, \aj, 104, 340

\bibitem[{{Leibundgut}(1988)}]{1988PhDT.......171L}
{Leibundgut} B., 1988, PhD thesis, PhD thesis.~Univ.~Basel.137 pp.~, (1988)

\bibitem[{{Leighly} {et~al}\mbox{.}(2014){Leighly}, {Terndrup}, {Baron},
  {Lucy}, {Dietrich}, \& {Gallagher}}]{2014ApJ...788..123L}
{Leighly} K.~M., {Terndrup} D.~M., {Baron} E., {Lucy} A.~B., {Dietrich} M.,
  {Gallagher} S.~C., 2014, \apj, 788, 123

\bibitem[{{Maeda} {et~al}\mbox{.}(2010{\natexlab{a}}){Maeda}, {Benetti},
  {Stritzinger}, {R{\"o}pke}, {Folatelli}, {Sollerman}, {Taubenberger},
  {Nomoto}, {Leloudas}, {Hamuy}, {Tanaka}, {Mazzali}, \&
  {Elias-Rosa}}]{2010Natur.466...82M}
{Maeda} K. {et~al.}, 2010{\natexlab{a}}, \nat, 466, 82

\bibitem[{{Maeda}, {Nozawa} \& {Motohara}(2014){Maeda}, {Nozawa}, \&
  {Motohara}}]{2014arXiv1411.3778M}
{Maeda} K., {Nozawa} T., {Motohara} K., 2014, submitted to MNRAS,
  \texttt{arXiv:1411.3778}

\bibitem[{{Maeda} {et~al}\mbox{.}(2010{\natexlab{b}}){Maeda}, {Taubenberger},
  {Sollerman}, {Mazzali}, {Leloudas}, {Nomoto}, \&
  {Motohara}}]{2010ApJ...708.1703M}
{Maeda} K., {Taubenberger} S., {Sollerman} J., {Mazzali} P.~A., {Leloudas} G.,
  {Nomoto} K., {Motohara} K., 2010{\natexlab{b}}, \apj, 708, 1703

\bibitem[{{Maguire} {et~al}\mbox{.}(2012){Maguire}, {Sullivan}, {Ellis},
  {Nugent}, {Howell}, {Gal-Yam}, {Cooke}, {Mazzali}, {Pan}, {Dilday}, {Thomas},
  {Arcavi}, {Ben-Ami}, {Bersier}, {Bianco}, {Fulton}, {Hook}, {Horesh},
  {Hsiao}, {James}, {Podsiadlowski}, {Walker}, {Yaron}, {Kasliwal}, {Laher},
  {Law}, {Ofek}, {Poznanski}, \& {Surace}}]{2012MNRAS.426.2359M}
{Maguire} K. {et~al.}, 2012, \mnras, 426, 2359

\bibitem[{{Maguire} {et~al}\mbox{.}(2013){Maguire}, {Sullivan}, {Patat},
  {Gal-Yam}, {Hook}, {Dhawan}, {Howell}, {Mazzali}, {Nugent}, {Pan},
  {Podsiadlowski}, {Simon}, {Sternberg}, {Valenti}, {Baltay}, {Bersier},
  {Blagorodnova}, {Chen}, {Ellman}, {Feindt}, {F{\"o}rster}, {Fraser},
  {Gonz{\'a}lez-Gait{\'a}n}, {Graham}, {Guti{\'e}rrez}, {Hachinger},
  {Hadjiyska}, {Inserra}, {Knapic}, {Laher}, {Leloudas}, {Margheim},
  {McKinnon}, {Molinaro}, {Morrell}, {Ofek}, {Rabinowitz}, {Rest}, {Sand},
  {Smareglia}, {Smartt}, {Taddia}, {Walker}, {Walton}, \&
  {Young}}]{2013MNRAS.436..222M}
{Maguire} K. {et~al.}, 2013, \mnras, 436, 222

\bibitem[{{Mandel}, {Foley} \& {Kirshner}(2014){Mandel}, {Foley}, \&
  {Kirshner}}]{2014ApJ...797...75M}
{Mandel} K.~S., {Foley} R.~J., {Kirshner} R.~P., 2014, \apj, 797, 75

\bibitem[{{Mandel}, {Narayan} \& {Kirshner}(2011){Mandel}, {Narayan}, \&
  {Kirshner}}]{2011ApJ...731..120M}
{Mandel} K.~S., {Narayan} G., {Kirshner} R.~P., 2011, \apj, 731, 120

\bibitem[{{Marion} {et~al}\mbox{.}(2012){Marion}, {Milisavljevic}, {Rines}, \&
  {Wilhelmy}}]{2012CBET.3146....2M}
{Marion} G.~H., {Milisavljevic} D., {Rines} K., {Wilhelmy} S., 2012, Central
  Bureau Electronic Telegrams, 3146, 2

\bibitem[{{Marion} {et~al}\mbox{.}(2015){Marion}, {Sand}, {Hsiao}, {Banerjee},
  {Valenti}, {Stritzinger}, {Vink{\'o}}, {Joshi}, {Venkataraman}, {Ashok},
  {Amanullah}, {Binzel}, {Bochanski}, {Bryngelson}, {Burns}, {Drozdov},
  {Fieber-Beyer}, {Graham}, {Howell}, {Johansson}, {Kirshner}, {Milne},
  {Parrent}, {Silverman}, {Vervack}, \& {Wheeler}}]{2015ApJ...798...39M}
{Marion} G.~H. {et~al.}, 2015, \apj, 798, 39

\bibitem[{{Matheson} {et~al}\mbox{.}(2012){Matheson}, {Joyce}, {Allen}, {Saha},
  {Silva}, {Wood-Vasey}, {Adams}, {Anderson}, {Beck}, {Bentz}, {Bershady},
  {Binkert}, {Butler}, {Camarata}, {Eigenbrot}, {Everett}, {Gallagher},
  {Garnavich}, {Glikman}, {Harbeck}, {Hargis}, {Herbst}, {Horch}, {Howell},
  {Jha}, {Kaczmarek}, {Knezek}, {Manne-Nicholas}, {Mathieu}, {Meixner},
  {Milliman}, {Power}, {Rajagopal}, {Reetz}, {Rhode}, {Schechtman-Rook},
  {Schwamb}, {Schweiker}, {Simmons}, {Simon}, {Summers}, {Young}, {Weyant},
  {Wilcots}, {Will}, \& {Williams}}]{2012ApJ...754...19M}
{Matheson} T. {et~al.}, 2012, \apj, 754, 19

\bibitem[{{Mazzali} {et~al}\mbox{.}(2014){Mazzali}, {Sullivan}, {Hachinger},
  {Ellis}, {Nugent}, {Howell}, {Gal-Yam}, {Maguire}, {Cooke}, {Thomas},
  {Nomoto}, \& {Walker}}]{2014MNRAS.439.1959M}
{Mazzali} P.~A. {et~al.}, 2014, \mnras, 439, 1959

\bibitem[{{McClelland} {et~al}\mbox{.}(2013){McClelland}, {Garnavich}, {Milne},
  {Shappee}, \& {Pogge}}]{2013ApJ...767..119M}
{McClelland} C.~M., {Garnavich} P.~M., {Milne} P.~A., {Shappee} B.~J., {Pogge}
  R.~W., 2013, \apj, 767, 119

\bibitem[{{Meikle}(2000)}]{2000MNRAS.314..782M}
{Meikle} W.~P.~S., 2000, \mnras, 314, 782

\bibitem[{{M{\'e}nard} {et~al}\mbox{.}(2008){M{\'e}nard}, {Nestor}, {Turnshek},
  {Quider}, {Richards}, {Chelouche}, \& {Rao}}]{2008MNRAS.385.1053M}
{M{\'e}nard} B., {Nestor} D., {Turnshek} D., {Quider} A., {Richards} G.,
  {Chelouche} D., {Rao} S., 2008, \mnras, 385, 1053

\bibitem[{{Milisavljevic}(2012)}]{2012CBET.3130....3M}
{Milisavljevic} D., 2012, Central Bureau Electronic Telegrams, 3130, 3

\bibitem[{{Milne} {et~al}\mbox{.}(2013){Milne}, {Brown}, {Roming}, {Bufano}, \&
  {Gehrels}}]{2013ApJ...779...23M}
{Milne} P.~A., {Brown} P.~J., {Roming} P.~W.~A., {Bufano} F., {Gehrels} N.,
  2013, \apj, 779, 23

\bibitem[{{Milne} {et~al}\mbox{.}(2010){Milne}, {Brown}, {Roming}, {Holland},
  {Immler}, {Filippenko}, {Ganeshalingam}, {Li}, {Stritzinger}, {Phillips},
  {Hicken}, {Kirshner}, {Challis}, {Mazzali}, {Schmidt}, {Bufano}, {Gehrels},
  \& {Vanden Berk}}]{2010ApJ...721.1627M}
{Milne} P.~A. {et~al.}, 2010, \apj, 721, 1627

\bibitem[{{Munari} {et~al}\mbox{.}(2013){Munari}, {Henden}, {Belligoli},
  {Castellani}, {Cherini}, {Righetti}, \& {Vagnozzi}}]{2013NewA...20...30M}
{Munari} U., {Henden} A., {Belligoli} R., {Castellani} F., {Cherini} G.,
  {Righetti} G.~L., {Vagnozzi} A., 2013, \na, 20, 30

\bibitem[{{Munari} \& {Zwitter}(1997)}]{1997A&A...318..269M}
{Munari} U., {Zwitter} T., 1997, \aap, 318, 269

\bibitem[{{Murga} {et~al}\mbox{.}(2015){Murga}, {Zhu}, {M{\'e}nard}, \&
  {Lan}}]{2015arXiv150302697M}
{Murga} M., {Zhu} G., {M{\'e}nard} B., {Lan} T.-W., 2015, submitted to MNRAS,
  \texttt{arXiv:1503.02697}

\bibitem[{{Nataf} {et~al}\mbox{.}(2013){Nataf}, {Gould}, {Fouqu{\'e}},
  {Gonzalez}, {Johnson}, {Skowron}, {Udalski}, {Szyma{\'n}ski}, {Kubiak},
  {Pietrzy{\'n}ski}, {Soszy{\'n}ski}, {Ulaczyk}, {Wyrzykowski}, \&
  {Poleski}}]{2013ApJ...769...88N}
{Nataf} D.~M. {et~al.}, 2013, \apj, 769, 88

\bibitem[{{Nielsen}, {Voss} \& {Nelemans}(2013){Nielsen}, {Voss}, \&
  {Nelemans}}]{2013MNRAS.435..187N}
{Nielsen} M.~T.~B., {Voss} R., {Nelemans} G., 2013, \mnras, 435, 187

\bibitem[{{Nobili} \& {Goobar}(2008)}]{2008A&A...487...19N}
{Nobili} S., {Goobar} A., 2008, \aap, 487, 19

\bibitem[{{Nordin} {et~al}\mbox{.}(2011){Nordin}, {{\"O}stman}, {Goobar},
  {Amanullah}, {Nichol}, {Smith}, {Sollerman}, {Bassett}, {Frieman},
  {Garnavich}, {Leloudas}, {Sako}, \& {Schneider}}]{2011A&A...526A.119N}
{Nordin} J. {et~al.}, 2011, \aap, 526, A119

\bibitem[{{Nugent}, {Kim} \& {Perlmutter}(2002){Nugent}, {Kim}, \&
  {Perlmutter}}]{2002PASP..114..803N}
{Nugent} P., {Kim} A., {Perlmutter} S., 2002, \pasp, 114, 803

\bibitem[{{Nugent} {et~al}\mbox{.}(2011){Nugent}, {Sullivan}, {Cenko},
  {Thomas}, {Kasen}, {Howell}, {Bersier}, {Bloom}, {Kulkarni}, {Kandrashoff},
  {Filippenko}, {Silverman}, {Marcy}, {Howard}, {Isaacson}, {Maguire},
  {Suzuki}, {Tarlton}, {Pan}, {Bildsten}, {Fulton}, {Parrent}, {Sand},
  {Podsiadlowski}, {Bianco}, {Dilday}, {Graham}, {Lyman}, {James}, {Kasliwal},
  {Law}, {Quimby}, {Hook}, {Walker}, {Mazzali}, {Pian}, {Ofek}, {Gal-Yam}, \&
  {Poznanski}}]{2011Natur.480..344N}
{Nugent} P.~E. {et~al.}, 2011, \nat, 480, 344

\bibitem[{{O'Donnell}(1994)}]{1994ApJ...422..158O}
{O'Donnell} J.~E., 1994, \apj, 422, 158

\bibitem[{{Ofek} {et~al}\mbox{.}(2012){Ofek}, {Laher}, {Law}, {Surace},
  {Levitan}, {Sesar}, {Horesh}, {Poznanski}, {van Eyken}, {Kulkarni}, {Nugent},
  {Zolkower}, {Walters}, {Sullivan}, {Ag{\"u}eros}, {Bildsten}, {Bloom},
  {Cenko}, {Gal-Yam}, {Grillmair}, {Helou}, {Kasliwal}, \&
  {Quimby}}]{2012PASP..124...62O}
{Ofek} E.~O. {et~al.}, 2012, \pasp, 124, 62

\bibitem[{{{\"O}stman}, {Goobar} \& {M{\"o}rtsell}(2006){{\"O}stman}, {Goobar},
  \& {M{\"o}rtsell}}]{2006A&A...450..971O}
{{\"O}stman} L., {Goobar} A., {M{\"o}rtsell} E., 2006, \aap, 450, 971

\bibitem[{{Pan} {et~al}\mbox{.}(2015){Pan}, {Sullivan}, {Maguire}, {Gal-Yam},
  {Hook}, {Howell}, {Nugent}, \& {Mazzali}}]{2015MNRAS.446..354P}
{Pan} Y.-C., {Sullivan} M., {Maguire} K., {Gal-Yam} A., {Hook} I.~M., {Howell}
  D.~A., {Nugent} P.~E., {Mazzali} P.~A., 2015, \mnras, 446, 354

\bibitem[{{Parnovsky} \& {Parnowski}(2010)}]{2010Ap&SS.325..163P}
{Parnovsky} S.~L., {Parnowski} A.~S., 2010, \apss, 325, 163

\bibitem[{{Patat} {et~al}\mbox{.}(2007){Patat}, {Chandra}, {Chevalier},
  {Justham}, {Podsiadlowski}, {Wolf}, {Gal-Yam}, {Pasquini}, {Crawford},
  {Mazzali}, {Pauldrach}, {Nomoto}, {Benetti}, {Cappellaro}, {Elias-Rosa},
  {Hillebrandt}, {Leonard}, {Pastorello}, {Renzini}, {Sabbadin}, {Simon}, \&
  {Turatto}}]{2007Sci...317..924P}
{Patat} F. {et~al.}, 2007, Science, 317, 924

\bibitem[{{Patat} {et~al}\mbox{.}(2013){Patat}, {Cordiner}, {Cox}, {Anderson},
  {Harutyunyan}, {Kotak}, {Palaversa}, {Stanishev}, {Tomasella}, {Benetti},
  {Goobar}, {Pastorello}, \& {Sollerman}}]{2013A&A...549A..62P}
{Patat} F. {et~al.}, 2013, \aap, 549, A62

\bibitem[{{Patat} {et~al}\mbox{.}(2010){Patat}, {Cox}, {Parrent}, \&
  {Branch}}]{2010A&A...514A..78P}
{Patat} F., {Cox} N.~L.~J., {Parrent} J., {Branch} D., 2010, \aap, 514, A78

\bibitem[{{Patat} {et~al}\mbox{.}(2015){Patat}, {Taubenberger}, {Cox}, {Baade},
  {Clocchiatti}, {H{\"o}flich}, {Maund}, {Reilly}, {Spyromilio}, {Wang},
  {Wheeler}, \& {Zelaya}}]{2015A&A...577A..53P}
{Patat} F. {et~al.}, 2015, \aap, 577, A53

\bibitem[{{Pereira} {et~al}\mbox{.}(2013){Pereira}, {Thomas}, {Aldering},
  {Antilogus}, {Baltay}, {Benitez-Herrera}, {Bongard}, {Buton}, {Canto},
  {Cellier-Holzem}, {Chen}, {Childress}, {Chotard}, {Copin}, {Fakhouri},
  {Fink}, {Fouchez}, {Gangler}, {Guy}, {Hillebrandt}, {Hsiao}, {Kerschhaggl},
  {Kowalski}, {Kromer}, {Nordin}, {Nugent}, {Paech}, {Pain}, {P{\'e}contal},
  {Perlmutter}, {Rabinowitz}, {Rigault}, {Runge}, {Saunders}, {Smadja}, {Tao},
  {Taubenberger}, {Tilquin}, \& {Wu}}]{2013A&A...554A..27P}
{Pereira} R. {et~al.}, 2013, \aap, 554, A27

\bibitem[{{Perlmutter} {et~al}\mbox{.}(1999){Perlmutter}, {Aldering},
  {Goldhaber}, {Knop}, {Nugent}, {Castro}, {Deustua}, {Fabbro}, {Goobar},
  {Groom}, {Hook}, {Kim}, {Kim}, {Lee}, {Nunes}, {Pain}, {Pennypacker},
  {Quimby}, {Lidman}, {Ellis}, {Irwin}, {McMahon}, {Ruiz-Lapuente}, {Walton},
  {Schaefer}, {Boyle}, {Filippenko}, {Matheson}, {Fruchter}, {Panagia},
  {Newberg}, {Couch}, \& {Supernova Cosmology Project}}]{1999ApJ...517..565P}
{Perlmutter} S. {et~al.}, 1999, \apj, 517, 565

\bibitem[{{Perlmutter} {et~al}\mbox{.}(1997){Perlmutter}, {Gabi}, {Goldhaber},
  {Goobar}, {Groom}, {Hook}, {Kim}, {Kim}, {Lee}, {Pain}, {Pennypacker},
  {Small}, {Ellis}, {McMahon}, {Boyle}, {Bunclark}, {Carter}, {Irwin},
  {Glazebrook}, {Newberg}, {Filippenko}, {Matheson}, {Dopita}, \&
  {Couch}}]{1997ApJ...483..565P}
{Perlmutter} S. {et~al.}, 1997, \apj, 483, 565

\bibitem[{{Persson} {et~al}\mbox{.}(1998){Persson}, {Murphy}, {Krzeminski},
  {Roth}, \& {Rieke}}]{1998AJ....116.2475P}
{Persson} S.~E., {Murphy} D.~C., {Krzeminski} W., {Roth} M., {Rieke} M.~J.,
  1998, \aj, 116, 2475

\bibitem[{Phillips(1993)}]{1993ApJ...413L.105P}
Phillips M.~M., 1993, \apjl, 413, L105

\bibitem[{{Phillips} {et~al}\mbox{.}(2013){Phillips}, {Simon}, {Morrell},
  {Burns}, {Cox}, {Foley}, {Karakas}, {Patat}, {Sternberg}, {Williams},
  {Gal-Yam}, {Hsiao}, {Leonard}, {Persson}, {Stritzinger}, {Thompson},
  {Campillay}, {Contreras}, {Folatelli}, {Freedman}, {Hamuy}, {Roth},
  {Shields}, {Suntzeff}, {Chomiuk}, {Ivans}, {Madore}, {Penprase}, {Perley},
  {Pignata}, {Preston}, \& {Soderberg}}]{2013ApJ...779...38P}
{Phillips} M.~M. {et~al.}, 2013, \apj, 779, 38

\bibitem[{{Pignata} {et~al}\mbox{.}(2012){Pignata}, {Cifuentes},
  {Apostolovski}, {Vidal}, {Maza}, {Hamuy}, {Antezana}, {Gonzalez}, {Cartier},
  {Forster}, {Silva}, {Carrasco}, {Sanchez}, {Hervias}, {Ramirez}, {Aros},
  {Conuel}, {Folatelli}, {Reichart}, {Ivarsen}, {Haislip}, {Crain}, {Foster},
  {Nysewander}, {LaCluyze}, \& {Prieto}}]{2012CBET.3076....1P}
{Pignata} G. {et~al.}, 2012, Central Bureau Electronic Telegrams, 3076, 1

\bibitem[{{Poznanski} {et~al}\mbox{.}(2009){Poznanski}, {Butler}, {Filippenko},
  {Ganeshalingam}, {Li}, {Bloom}, {Chornock}, {Foley}, {Nugent}, {Silverman},
  {Cenko}, {Gates}, {Leonard}, {Miller}, {Modjaz}, {Serduke}, {Smith}, {Swift},
  \& {Wong}}]{2009ApJ...694.1067P}
{Poznanski} D. {et~al.}, 2009, \apj, 694, 1067

\bibitem[{{Poznanski} {et~al}\mbox{.}(2011){Poznanski}, {Ganeshalingam},
  {Silverman}, \& {Filippenko}}]{2011MNRAS.415L..81P}
{Poznanski} D., {Ganeshalingam} M., {Silverman} J.~M., {Filippenko} A.~V.,
  2011, \mnras, 415, L81

\bibitem[{{Poznanski}, {Prochaska} \& {Bloom}(2012){Poznanski}, {Prochaska}, \&
  {Bloom}}]{2012MNRAS.426.1465P}
{Poznanski} D., {Prochaska} J.~X., {Bloom} J.~S., 2012, \mnras, 426, 1465

\bibitem[{{Pozzo} {et~al}\mbox{.}(2006){Pozzo}, {Meikle}, {Rayner}, {Joseph},
  {Filippenko}, {Foley}, {Li}, {Mattila}, \& {Sollerman}}]{2006MNRAS.368.1169P}
{Pozzo} M. {et~al.}, 2006, \mnras, 368, 1169

\bibitem[{{Prieto}(2012)}]{2012CBET.3076....2P}
{Prieto} J.~L., 2012, Central Bureau Electronic Telegrams, 3076, 2

\bibitem[{{Puckett} {et~al}\mbox{.}(2012){Puckett}, {Newton}, {Cappellaro},
  {Pastorello}, {Tomasella}, {Benetti}, {Fiaschi}, {Ochner}, {Turatto}, \&
  {Valenti}}]{2012CBET.3077....1P}
{Puckett} T. {et~al.}, 2012, Central Bureau Electronic Telegrams, 3077, 1

\bibitem[{{Rau} {et~al}\mbox{.}(2009){Rau}, {Kulkarni}, {Law}, {Bloom},
  {Ciardi}, {Djorgovski}, {Fox}, {Gal-Yam}, {Grillmair}, {Kasliwal}, {Nugent},
  {Ofek}, {Quimby}, {Reach}, {Shara}, {Bildsten}, {Cenko}, {Drake},
  {Filippenko}, {Helfand}, {Helou}, {Howell}, {Poznanski}, \&
  {Sullivan}}]{2009PASP..121.1334R}
{Rau} A. {et~al.}, 2009, \pasp, 121, 1334

\bibitem[{{Rich} {et~al}\mbox{.}(2012){Rich}, {Dennefeld}, {Pizzella},
  {Valentini}, {Tomasella}, {Benetti}, {Cortes}, {Ozel}, \&
  {Rajpurohit}}]{2012CBET.3226....1R}
{Rich} D. {et~al.}, 2012, Central Bureau Electronic Telegrams, 3226, 1

\bibitem[{{Riess} {et~al}\mbox{.}(1998){Riess}, {Filippenko}, {Challis},
  {Clocchiatti}, {Diercks}, {Garnavich}, {Gilliland}, {Hogan}, {Jha},
  {Kirshner}, {Leibundgut}, {Phillips}, {Reiss}, {Schmidt}, {Schommer},
  {Smith}, {Spyromilio}, {Stubbs}, {Suntzeff}, \&
  {Tonry}}]{1998AJ....116.1009R}
{Riess} A.~G. {et~al.}, 1998, \aj, 116, 1009

\bibitem[{Riess, {Press} \& {Kirshner}(1996)Riess, {Press}, \&
  {Kirshner}}]{1996ApJ...473...88R}
Riess A.~G., {Press} W.~H., {Kirshner} R.~P., 1996, \apj, 473, 88

\bibitem[{{Rigault} {et~al}\mbox{.}(2015){Rigault}, {Aldering}, {Kowalski},
  {Copin}, {Antilogus}, {Aragon}, {Bailey}, {Baltay}, {Baugh}, {Bongard},
  {Boone}, {Buton}, {Chen}, {Chotard}, {Fakhouri}, {Feindt}, {Fagrelius},
  {Fleury}, {Fouchez}, {Gangler}, {Hayden}, {Kim}, {Leget}, {Lombardo},
  {Nordin}, {Pain}, {Pecontal}, {Pereira}, {Perlmutter}, {Rabinowitz}, {Runge},
  {Rubin}, {Saunders}, {Smadja}, {Sofiatti}, {Suzuki}, {Tao}, \&
  {Weaver}}]{2015ApJ...802...20R}
{Rigault} M. {et~al.}, 2015, \apj, 802, 20

\bibitem[{{Rigault} {et~al}\mbox{.}(2013){Rigault}, {Copin}, {Aldering},
  {Antilogus}, {Aragon}, {Bailey}, {Baltay}, {Bongard}, {Buton}, {Canto},
  {Cellier-Holzem}, {Childress}, {Chotard}, {Fakhouri}, {Feindt}, {Fleury},
  {Gangler}, {Greskovic}, {Guy}, {Kim}, {Kowalski}, {Lombardo}, {Nordin},
  {Nugent}, {Pain}, {P{\'e}contal}, {Pereira}, {Perlmutter}, {Rabinowitz},
  {Runge}, {Saunders}, {Scalzo}, {Smadja}, {Tao}, {Thomas}, \&
  {Weaver}}]{2013A&A...560A..66R}
{Rigault} M. {et~al.}, 2013, \aap, 560, A66

\bibitem[{{Ritchey} {et~al}\mbox{.}(2015){Ritchey}, {Welty}, {Dahlstrom}, \&
  {York}}]{2015ApJ...799..197R}
{Ritchey} A.~M., {Welty} D.~E., {Dahlstrom} J.~A., {York} D.~G., 2015, \apj,
  799, 197

\bibitem[{{Roming} {et~al}\mbox{.}(2005){Roming}, {Kennedy}, {Mason}, {Nousek},
  {Ahr}, {Bingham}, {Broos}, {Carter}, {Hancock}, {Huckle}, {Hunsberger},
  {Kawakami}, {Killough}, {Koch}, {McLelland}, {Smith}, {Smith}, {Soto},
  {Boyd}, {Breeveld}, {Holland}, {Ivanushkina}, {Pryzby}, {Still}, \&
  {Stock}}]{2005SSRv..120...95R}
{Roming} P.~W.~A. {et~al.}, 2005, \ssr, 120, 95

\bibitem[{{Schlafly} \& {Finkbeiner}(2011)}]{2011ApJ...737..103S}
{Schlafly} E.~F., {Finkbeiner} D.~P., 2011, \apj, 737, 103

\bibitem[{{Schlegel}, {Finkbeiner} \& {Davis}(1998){Schlegel}, {Finkbeiner}, \&
  {Davis}}]{1998ApJ...500..525S}
{Schlegel} D.~J., {Finkbeiner} D.~P., {Davis} M., 1998, \apj, 500, 525

\bibitem[{{Scolnic} {et~al}\mbox{.}(2014){Scolnic}, {Riess}, {Foley}, {Rest},
  {Rodney}, {Brout}, \& {Jones}}]{2014ApJ...780...37S}
{Scolnic} D.~M., {Riess} A.~G., {Foley} R.~J., {Rest} A., {Rodney} S.~A.,
  {Brout} D.~J., {Jones} D.~O., 2014, \apj, 780, 37

\bibitem[{{Silverman} \& {Filippenko}(2012)}]{2012MNRAS.425.1917S}
{Silverman} J.~M., {Filippenko} A.~V., 2012, \mnras, 425, 1917

\bibitem[{{Silverman} {et~al}\mbox{.}(2012){Silverman}, {Ganeshalingam},
  {Cenko}, {Filippenko}, {Li}, {Barth}, {Carson}, {Childress}, {Clubb},
  {Cucchiara}, {Graham}, {Marion}, {Nguyen}, {Pei}, {Tucker}, {Vinko},
  {Wheeler}, \& {Worseck}}]{2012ApJ...756L...7S}
{Silverman} J.~M. {et~al.}, 2012, \apjl, 756, L7

\bibitem[{{Silverman}, {Ganeshalingam} \& {Filippenko}(2013){Silverman},
  {Ganeshalingam}, \& {Filippenko}}]{2013MNRAS.430.1030S}
{Silverman} J.~M., {Ganeshalingam} M., {Filippenko} A.~V., 2013, \mnras, 430,
  1030

\bibitem[{{Silverman}, {Kong} \& {Filippenko}(2012){Silverman}, {Kong}, \&
  {Filippenko}}]{2012MNRAS.425.1819S}
{Silverman} J.~M., {Kong} J.~J., {Filippenko} A.~V., 2012, \mnras, 425, 1819

\bibitem[{{Silverman} {et~al}\mbox{.}(2013){Silverman}, {Nugent}, {Gal-Yam},
  {Sullivan}, {Howell}, {Filippenko}, {Arcavi}, {Ben-Ami}, {Bloom}, {Cenko},
  {Cao}, {Chornock}, {Clubb}, {Coil}, {Foley}, {Graham}, {Griffith}, {Horesh},
  {Kasliwal}, {Kulkarni}, {Leonard}, {Li}, {Matheson}, {Miller}, {Modjaz},
  {Ofek}, {Pan}, {Perley}, {Poznanski}, {Quimby}, {Steele}, {Sternberg}, {Xu},
  \& {Yaron}}]{2013ApJS..207....3S}
{Silverman} J.~M. {et~al.}, 2013, \apjs, 207, 3

\bibitem[{{Simon} {et~al}\mbox{.}(2009){Simon}, {Gal-Yam}, {Gnat}, {Quimby},
  {Ganeshalingam}, {Silverman}, {Blondin}, {Li}, {Filippenko}, {Wheeler},
  {Kirshner}, {Patat}, {Nugent}, {Foley}, {Vogt}, {Butler}, {Peek},
  {Rosolowsky}, {Herczeg}, {Sauer}, \& {Mazzali}}]{2009ApJ...702.1157S}
{Simon} J.~D. {et~al.}, 2009, \apj, 702, 1157

\bibitem[{{Smith} {et~al}\mbox{.}(2002){Smith}, {Tucker}, {Kent}, {Richmond},
  {Fukugita}, {Ichikawa}, {Ichikawa}, {Jorgensen}, {Uomoto}, {Gunn}, {Hamabe},
  {Watanabe}, {Tolea}, {Henden}, {Annis}, {Pier}, {McKay}, {Brinkmann}, {Chen},
  {Holtzman}, {Shimasaku}, \& {York}}]{2002AJ....123.2121S}
{Smith} J.~A. {et~al.}, 2002, \aj, 123, 2121

\bibitem[{{Soker}(2014)}]{2014MNRAS.444L..73S}
{Soker} N., 2014, \mnras, 444, L73

\bibitem[{{Sollerman} {et~al}\mbox{.}(2005){Sollerman}, {Cox}, {Mattila},
  {Ehrenfreund}, {Kaper}, {Leibundgut}, \& {Lundqvist}}]{2005A&A...429..559S}
{Sollerman} J., {Cox} N., {Mattila} S., {Ehrenfreund} P., {Kaper} L.,
  {Leibundgut} B., {Lundqvist} P., 2005, \aap, 429, 559

\bibitem[{{Springob} {et~al}\mbox{.}(2009){Springob}, {Masters}, {Haynes},
  {Giovanelli}, \& {Marinoni}}]{2009ApJS..182..474S}
{Springob} C.~M., {Masters} K.~L., {Haynes} M.~P., {Giovanelli} R., {Marinoni}
  C., 2009, \apjs, 182, 474

\bibitem[{{Stanishev} {et~al}\mbox{.}(2015){Stanishev}, {Goobar}, {Amanullah},
  {Bassett}, {Fantaye}, {Garnavich}, {Hlozek}, {Nordin}, {Okouma}, {Ostman},
  {Sako}, {Scalzo}, \& {Smith}}]{2015arXiv150507707S}
{Stanishev} V. {et~al.}, 2015, submitted to A\&A, \texttt{arXiv:1505.07707}

\bibitem[{{Sternberg} {et~al}\mbox{.}(2011){Sternberg}, {Gal-Yam}, {Simon},
  {Leonard}, {Quimby}, {Phillips}, {Morrell}, {Thompson}, {Ivans}, {Marshall},
  {Filippenko}, {Marcy}, {Bloom}, {Patat}, {Foley}, {Yong}, {Penprase},
  {Beeler}, {Allende Prieto}, \& {Stringfellow}}]{2011Sci...333..856S}
{Sternberg} A. {et~al.}, 2011, Science, 333, 856

\bibitem[{{Sternberg} {et~al}\mbox{.}(2014){Sternberg}, {Gal-Yam}, {Simon},
  {Patat}, {Hillebrandt}, {Phillips}, {Foley}, {Thompson}, {Morrell},
  {Chomiuk}, {Soderberg}, {Yong}, {Kraus}, {Herczeg}, {Hsiao}, {Raskutti},
  {Cohen}, {Mazzali}, \& {Nomoto}}]{2014MNRAS.443.1849S}
{Sternberg} A. {et~al.}, 2014, \mnras, 443, 1849

\bibitem[{{Stetson}(1987)}]{1987PASP...99..191S}
{Stetson} P.~B., 1987, \pasp, 99, 191

\bibitem[{{Stritzinger} {et~al}\mbox{.}(2002){Stritzinger}, {Hamuy},
  {Suntzeff}, {Smith}, {Phillips}, {Maza}, {Strolger}, {Antezana},
  {Gonz{\'a}lez}, {Wischnjewsky}, {Candia}, {Espinoza}, {Gonz{\'a}lez},
  {Stubbs}, {Becker}, {Rubenstein}, \& {Galaz}}]{2002AJ....124.2100S}
{Stritzinger} M. {et~al.}, 2002, \aj, 124, 2100

\bibitem[{{Sullivan} {et~al}\mbox{.}(2011){Sullivan}, {Guy}, {Conley},
  {Regnault}, {Astier}, {Balland}, {Basa}, {Carlberg}, {Fouchez}, {Hardin},
  {Hook}, {Howell}, {Pain}, {Palanque-Delabrouille}, {Perrett}, {Pritchet},
  {Rich}, {Ruhlmann-Kleider}, {Balam}, {Baumont}, {Ellis}, {Fabbro},
  {Fakhouri}, {Fourmanoit}, {Gonz{\'a}lez-Gait{\'a}n}, {Graham}, {Hudson},
  {Hsiao}, {Kronborg}, {Lidman}, {Mourao}, {Neill}, {Perlmutter}, {Ripoche},
  {Suzuki}, \& {Walker}}]{2011ApJ...737..102S}
{Sullivan} M. {et~al.}, 2011, \apj, 737, 102

\bibitem[{{Suntzeff}(2000)}]{2000AIPC..522...65S}
{Suntzeff} N.~B., 2000, in American Institute of Physics Conference Series,
  Vol. 522, American Institute of Physics Conference Series, {Holt} S.~S.,
  {Zhang} W.~W., eds., pp. 65--74

\bibitem[{{Suzuki} {et~al}\mbox{.}(2012){Suzuki}, {Rubin}, {Lidman},
  {Aldering}, {Amanullah}, {Barbary}, {Barrientos}, {Botyanszki}, {Brodwin},
  {Connolly}, {Dawson}, {Dey}, {Doi}, {Donahue}, {Deustua}, {Eisenhardt},
  {Ellingson}, {Faccioli}, {Fadeyev}, {Fakhouri}, {Fruchter}, {Gilbank},
  {Gladders}, {Goldhaber}, {Gonzalez}, {Goobar}, {Gude}, {Hattori}, {Hoekstra},
  {Hsiao}, {Huang}, {Ihara}, {Jee}, {Johnston}, {Kashikawa}, {Koester},
  {Konishi}, {Kowalski}, {Linder}, {Lubin}, {Melbourne}, {Meyers}, {Morokuma},
  {Munshi}, {Mullis}, {Oda}, {Panagia}, {Perlmutter}, {Postman}, {Pritchard},
  {Rhodes}, {Ripoche}, {Rosati}, {Schlegel}, {Spadafora}, {Stanford},
  {Stanishev}, {Stern}, {Strovink}, {Takanashi}, {Tokita}, {Wagner}, {Wang},
  {Yasuda}, {Yee}, \& {Supernova Cosmology Project}}]{2012ApJ...746...85S}
{Suzuki} N. {et~al.}, 2012, \apj, 746, 85

\bibitem[{{Taddia} {et~al}\mbox{.}(2012){Taddia}, {Stritzinger}, {Phillips},
  {Burns}, {Heinrich-Josties}, {Morrell}, {Sollerman}, {Valenti}, {Anderson},
  {Boldt}, {Campillay}, {Castellon}, {Contreras}, {Folatelli}, {Freedman},
  {Hamuy}, {Krzeminski}, {Leloudas}, {Maeda}, {Persson}, {Roth}, \&
  {Suntzeff}}]{2012A&A...545L...7T}
{Taddia} F. {et~al.}, 2012, \aap, 545, L7

\bibitem[{{Telting} {et~al}\mbox{.}(2014){Telting}, {Avila}, {Buchhave},
  {Frandsen}, {Gandolfi}, {Lindberg}, {Stempels}, {Prins}, \& {NOT
  staff}}]{2014AN....335...41T}
{Telting} J.~H. {et~al.}, 2014, Astronomische Nachrichten, 335, 41

\bibitem[{{Thomas} {et~al}\mbox{.}(2011){Thomas}, {Aldering}, {Antilogus},
  {Aragon}, {Bailey}, {Baltay}, {Bongard}, {Buton}, {Canto}, {Childress},
  {Chotard}, {Copin}, {Fakhouri}, {Gangler}, {Hsiao}, {Kerschhaggl},
  {Kowalski}, {Loken}, {Nugent}, {Paech}, {Pain}, {Pecontal}, {Pereira},
  {Perlmutter}, {Rabinowitz}, {Rigault}, {Rubin}, {Runge}, {Scalzo}, {Smadja},
  {Tao}, {Weaver}, {Wu}, {Brown}, {Milne}, \& {Nearby Supernova
  Factory}}]{2011ApJ...743...27T}
{Thomas} R.~C. {et~al.}, 2011, \apj, 743, 27

\bibitem[{{Tomasella } {et~al}\mbox{.}(2014){Tomasella }, {Benetti},
  {Cappellaro}, {Pastorello}, {Turatto}, {Barbon}, {Elias-Rosa}, {Harutyunyan},
  {Ochner}, {Tartaglia}, \& {Valenti}}]{2014AN....335..841T}
{Tomasella } L. {et~al.}, 2014, Astronomische Nachrichten, 335, 841

\bibitem[{{Tripp}(1998)}]{1998A&A...331..815T}
{Tripp} R., 1998, \aap, 331, 815

\bibitem[{{Tully} {et~al}\mbox{.}(2009){Tully}, {Rizzi}, {Shaya}, {Courtois},
  {Makarov}, \& {Jacobs}}]{2009AJ....138..323T}
{Tully} R.~B., {Rizzi} L., {Shaya} E.~J., {Courtois} H.~M., {Makarov} D.~I.,
  {Jacobs} B.~A., 2009, \aj, 138, 323

\bibitem[{{Valenti} {et~al}\mbox{.}(2011){Valenti}, {Fraser}, {Benetti},
  {Pignata}, {Sollerman}, {Inserra}, {Cappellaro}, {Pastorello}, {Smartt},
  {Ergon}, {Botticella}, {Brimacombe}, {Bufano}, {Crockett}, {Eder}, {Fugazza},
  {Haislip}, {Hamuy}, {Harutyunyan}, {Ivarsen}, {Kankare}, {Kotak}, {Lacluyze},
  {Magill}, {Mattila}, {Maza}, {Mazzali}, {Reichart}, {Taubenberger},
  {Turatto}, \& {Zampieri}}]{2011MNRAS.416.3138V}
{Valenti} S. {et~al.}, 2011, \mnras, 416, 3138

\bibitem[{{Wang}(2005)}]{2005ApJ...635L..33W}
{Wang} L., 2005, \apjl, 635, L33

\bibitem[{Wang {et~al}\mbox{.}(2009)Wang, {Filippenko}, {Ganeshalingam}, {Li},
  {Silverman}, {Wang}, {Chornock}, {Foley}, {Gates}, {Macomber}, {Serduke},
  {Steele}, \& {Wong}}]{2009ApJ...699L.139W}
Wang X. {et~al.}, 2009, \apjl, 699, L139

\bibitem[{{Wang} {et~al}\mbox{.}(2012){Wang}, {Wang}, {Filippenko}, {Baron},
  {Kromer}, {Jack}, {Zhang}, {Aldering}, {Antilogus}, {Arnett}, {Baade},
  {Barris}, {Benetti}, {Bouchet}, {Burrows}, {Canal}, {Cappellaro}, {Carlberg},
  {di Carlo}, {Challis}, {Crotts}, {Danziger}, {Della Valle}, {Fink}, {Foley},
  {Fransson}, {Gal-Yam}, {Garnavich}, {Gerardy}, {Goldhaber}, {Hamuy},
  {Hillebrandt}, {H{\"o}flich}, {Holland}, {Holz}, {Hughes}, {Jeffery}, {Jha},
  {Kasen}, {Khokhlov}, {Kirshner}, {Knop}, {Kozma}, {Krisciunas}, {Lee},
  {Leibundgut}, {Lentz}, {Leonard}, {Lewin}, {Li}, {Livio}, {Lundqvist},
  {Maoz}, {Matheson}, {Mazzali}, {Meikle}, {Miknaitis}, {Milne}, {Mochnacki},
  {Nomoto}, {Nugent}, {Oran}, {Panagia}, {Perlmutter}, {Phillips}, {Pinto},
  {Poznanski}, {Pritchet}, {Reinecke}, {Riess}, {Ruiz-Lapuente}, {Scalzo},
  {Schlegel}, {Schmidt}, {Siegrist}, {Soderberg}, {Sollerman}, {Sonneborn},
  {Spadafora}, {Spyromilio}, {Sramek}, {Starrfield}, {Strolger}, {Suntzeff},
  {Thomas}, {Tonry}, {Tornambe}, {Truran}, {Turatto}, {Turner}, {Van Dyk},
  {Weiler}, {Wheeler}, {Wood-Vasey}, {Woosley}, \&
  {Yamaoka}}]{2012ApJ...749..126W}
{Wang} X. {et~al.}, 2012, \apj, 749, 126

\bibitem[{{Wang} {et~al}\mbox{.}(2013){Wang}, {Wang}, {Filippenko}, {Zhang}, \&
  {Zhao}}]{2013Sci...340..170W}
{Wang} X., {Wang} L., {Filippenko} A.~V., {Zhang} T., {Zhao} X., 2013, Science,
  340, 170

\bibitem[{{Zhang}, {Wu} \& {Wang}(2012){Zhang}, {Wu}, \&
  {Wang}}]{2012CBET.3130....2Z}
{Zhang} T.-M., {Wu} C.-J., {Wang} X.-F., 2012, Central Bureau Electronic
  Telegrams, 3130, 2

\end{thebibliography}

\section*{Supporting information}
Additional Supporting Information may be found in the online version of this article.

\vspace{2cm}
\noindent{\it%
  $^1$Oskar Klein Centre, Physics Department, Stockholm\\
  \phantom{$^1$}University, SE 106 91 Stockholm, Sweden\\
  $^2$George P. and Cynthia Woods Mitchell Institute for\\
  \phantom{$^2$}Fundamental Physics \& Astronomy, Texas A. \& M.\\  
  \phantom{$^2$}University, Department of Physics and Astronomy, 4242\\
  \phantom{$^2$}TAMU, College Station, TX 77843, USA\\
  $^3$Cahill Center for Astrophysics, California Institute of\\
  \phantom{$^3$}Technology, Pasadena, CA 91125, USA\\
  $^4$Carnegie Observatories, Las Campanas Observatory,\\
  \phantom{$^4$}Colina El Pino, Casilla 601, La Serena, Chile\\
  $^5$Institute of Theoretical Astrophysics, University of Oslo,\\
  \phantom{$^5$}P.O. Box 1029, Blindern, N-0315 Oslo, Norway\\
  $^6$INAF - Osservatorio Astronomico di Padova, vicolo\\
  \phantom{$^6$}dell'Osservatorio 5, I-35122 Padova, Italy
  $^7$Dark Cosmology Centre, Niels Bohr Institute, University
  \phantom{$^7$}of Copenhagen, Juliane Maries Vej 30, 2100\\
  \phantom{$^7$}Copenhagen~O, Denmark\\
  $^8$Instituto de Astrof\'isica de Andaluc\'ia (IAA-CSIC),\\
  \phantom{$^8$}Glorieta de la Astronom\'ia s/n, E-18008, Granada, Spain\\
  $^9$Unidad Asociada Grupo Ciencias Planetarias\\
  \phantom{$^9$}UPV/EHU-IAA/CSIC, Departamento de F\'isica\\
  \phantom{$^9$}Aplicada~I, E.T.S., Ingenier\'ia,\\
  \phantom{$^8$}Universidad del Pa\'is Vasco UPV/EHU, Bilbao, Spain\\
  $^{10}$INAF- Osservatorio Astronomico di Roma via\\
  \phantom{$^{10}$}Frascati 33, 00040, Monte Porzio Catone, Roma\\
  $^{11}$Las Cumbres Observatory Global Telescope Network,\\
  \phantom{$^{11}$}6740 Cortona Dr., Suite 102, Goleta, CA 93117, USA\\
  $^{12}$Department of Physics, University of California, Santa\\
  \phantom{$^{12}$}Barbara, Broida Hall, Mail Code 9530, Santa Barbara,\\
  \phantom{$^{12}$}CA 93106-9530, USA\\
  $^{13}$Department of Physics and Astronomy, Aarhus\\
  \phantom{$^{13}$}University, Ny Munkegade 120, DK-8000 Aarhus C,\\
  \phantom{$^{13}$}Denmark\\
  $^{14}$Astrophysics Research Centre, School of Mathematics and\\
  \phantom{$^{14}$}Physics, Queen's University Belfast, BT7 1NN, UK\\
  $^{15}$Observatories of the Carnegie Institution for Science,\\
  \phantom{$^{15}$}813 Santa Barbara Street, Pasadena, CA 91101, USA\\
  $^{16}$Benoziyo Center for Astrophysics, Faculty of Physics,\\
  \phantom{$^{16}$}Weizmann Institute of Science, 76100 Rehovot, Israel\\
  $^{17}$Oskar Klein Centre, Astronomy Department, Stockholm\\
  \phantom{$^{16}$}University, SE 106 91 Stockholm, Sweden\\
  $^{18}$Finnish Centre for Astronomy with ESO (FINCA),\\
  \phantom{$^{18}$}University of Turku, V\"ais\"al\"antie 20, FI-21500 Piikki\"o,\\
  \phantom{$^{18}$}Finland\\
  $^{19}$Department of Astronomy, University of California,\\
  \phantom{$^{19}$}Berkeley, CA 94720-3411, USA\\
  $^{20}$Computational Cosmology Center, Lawrence Berkeley\\
  \phantom{$^{20}$}National Laboratory, 1 Cyclotron Road, Berkeley,\\
  \phantom{$^{20}$}CA 94720, USA\\
  $^{21}$CENTRA - Centro Multidisciplinar de Astrof\'isica,\\
  \phantom{$^{21}$}Instituto Superior T\'ecnico, Av. Rovisco Pais 1,\\
  \phantom{$^{21}$}1049-001 Lisbon, Portugal\\
  $^{22}$School of Physics and Astronomy, University of\\
  \phantom{$^{22}$}Southampton, Southampton, SO17 1BJ, UK\\
  $^{23}$Tuorla Observatory, Department of Physics and
  \phantom{$^{23}$}Astronomy, University of Turku, V\"ais\"al\"antie 20,\\
  \phantom{$^{23}$}FI-21500 Piikki\"o, Finland\\
  $^{24}$Department of Astronomy, Faculty of Mathematics,\\
  \phantom{$^{24}$}University of Belgrade, Serbia\\
  $^{25}$Department of Earth and Space Sciences, Chalmers\\
  \phantom{$^{25}$}University of Technology, Onsala Space Observatory\\
  \phantom{$^{25}$}SE 439-92, Onsala, Sweden\\
  $^{26}$ASTRON, Postbus 2, 7990 AA Dwingeloo,\\
  \phantom{$^{26}$}The Netherlands\\
  $^{27}$Kapteyn Astronomical Institute, Postbus 800, 9700 AV,\\
  \phantom{$^{27}$}Groningen, The Netherlands\\
}

\label{lastpage}
\end{document}